\documentclass[aps,prx,twocolumn,superscriptaddress,balancelastpage,showpacs,reprint]{revtex4-2}

\usepackage[T1]{fontenc}
\usepackage{blindtext}
\usepackage{centernot}
\usepackage{graphicx}
\usepackage{amsmath,bbold}
\usepackage{times}
\usepackage{amssymb}
\usepackage{mathrsfs}
\usepackage{chemarr}
\usepackage{color}
\usepackage{url}
\usepackage{version}
\usepackage[hidelinks]{hyperref}
\usepackage{mwe,tikz}
\usepackage[percent]{overpic}
\usepackage{bm}
\usepackage[export]{adjustbox}
\definecolor{linkcolor}{rgb}{0,0,0.6} 
\definecolor{forestgreen}{rgb}{0.13, 0.55, 0.13}
\definecolor{frenchblue}{rgb}{0.0, 0.45, 0.73}
\definecolor{burntsienna}{rgb}{0.91, 0.45, 0.32}
\usepackage{algorithm}
\usepackage[noend]{algpseudocode}

\usepackage{array}
\usepackage{multirow}

\usepackage{enumitem}
\usepackage{cases}

\usepackage{transparent}

\newcommand{\mL}{\mathcal{L}}
\newcommand{\om}{\Omega}

\newcommand{\vphiq}{\varphi_q}
\newcommand{\dtau}{\Delta \tau}

\newcommand{\RR}{\mathbb{R}}
\newcommand{\NN}{\mathbb{N}}
\newcommand{\calS}{\mathbb{S}}
\newcommand{\PP}{\mathbb{P}}
\newcommand{\EE}{\mathbb{E}}
\newcommand{\<}{\langle}
\renewcommand{\>}{\rangle}

\newcommand{\hphi}{{\hat{\phi}}}
\newcommand{\htheta}{{\hat{\theta}}}

\DeclareMathOperator*{\argmax}{argmax}

\usepackage{lipsum}
\usetikzlibrary{patterns}

\newcolumntype{P}[1]{>{\centering\arraybackslash}p{#1}}
\newcolumntype{M}[1]{>{\centering\arraybackslash}m{#1}}

\begin{document}
  

\title{Minimum Action Method for Nonequilibrium Phase Transitions}

\author{Ruben Zakine}
\affiliation{Courant Institute, New York University, 251 Mercer Street, New York, New York 10012, USA}

\author{Eric Vanden-Eijnden}
\affiliation{Courant Institute, New York University, 251 Mercer Street, New York, New York 10012, USA}

\date{\today}

\begin{abstract}
First-order nonequilibrium phase transitions observed in active matter, fluid dynamics, biology, climate science, and other systems with irreversible dynamics are challenging to analyze because they cannot be inferred from a simple free energy minimization principle. Rather the mechanism of these transitions depends crucially on the system's dynamics, which requires us to analyze them in trajectory space rather than in phase space. Here we consider situations where the path of these transitions can be characterized as the minimizer of an action, whose minimum value can be used in a nonequilibrium generalization of the Arrhenius law to calculate the system's phase diagram. We also develop efficient numerical tools for the minimization of this action. These tools are general enough to be transportable to many situations of interest, in particular when the fluctuations present in the microscopic system are non-Gaussian and its dynamics is not governed by the standard Langevin equation. As an illustration, first-order phase transitions in two spatially-extended nonequilibrium systems are analyzed: a modified Ginzburg-Landau equation with a chemical potential which is non-gradient, and a reaction-diffusion network based on the Schl\"ogl model. The phase diagrams of both systems are calculated as a function of their control parameters, and the paths of the transitions, including their critical nuclei, are identified. These results clearly demonstrate the nonequilibrium nature of the transitions, with differing forward and backward paths. 
\end{abstract}

\maketitle 

\section{Introduction}
\label{sec:intro}

Materials with identical microscopic constitutions can be found in very different macroscopic states when external conditions, such as temperature or pressure, vary.
A major achievement of equilibrium statistical mechanics is to give a first-principle explanation of these phase transitions. The theory posits the existence of a distribution, for example of Boltzmann-Gibbs type, that gives the probability of finding the microscopic system in any of its possible configurations.  
Macroscopic properties like the system's density, magnetization~\cite{weiss1907}, population number in the ground state~\cite{einstein1925}, etc., can then be deduced by enumerating all the microscopic configurations consistent with a given value of the chosen macroscopic observable, and identifying which of these values is most likely. 
More formally, if we denote by~$\phi$ the variable used to characterize the macroscopic state of the system, its statistical weight is obtained by summing the system's probability distribution over all microscopic states consistent with a given realization of~$\phi$.  
Performing this calculation typically require sophisticated tools such as renormalization group theory~\cite{wilson1975}, the replica method or the cavity method~\cite{edwards1975replica, mezard1987cavity}, along with tools from large deviation theory~\cite{ellis2006entropy}. It generically shows that the statistical weight of $\phi$ is asymptotically given by $\exp(-V(\phi)/\epsilon)$, where $V(\phi)$ is some free energy to be calculated, and~$\epsilon$ is a small parameter that tends to zero in the thermodynamic limit when the number of microscopic constituents tends to infinity: as a result, the theory predicts that the system will be found in the macroscopic state~$\phi$ of minimum free energy with probability one in this limit. This also explains phase transitions: they take place when the topology of the free energy $V(\phi)$ changes as a control parameter, like the temperature or some applied external field, is varied. For example, if $V(\phi)$ has two wells whose relative depths change with the control parameter, a first-order phase transition occurs when the deepest well becomes more shallow than the other well, and as a result, the macroscopic state of the system changes from the first to the second. 

The statistical mechanics approach to phase transitions rests on the assumption that the probability distribution of the microscopic system is known. This information is available for equilibrium systems, whose microscopic dynamics is time-reversible. In this work, however, we are primarily interested in nonequilibrium systems, whose dynamics is irreversible. Except for some special situations where it can be computed exactly~\cite{derrida1992, derrida1993} or asymptotically e.g. via some thermodynamic mapping~\cite{tailleur_mapping_2007,tailleur_mapping_2008,tailleur2008,obyrne2020}, the invariant distribution of these systems is not known in general. Yet, these systems too can undergo phase transitions. Examples include  driven systems arising from active matter~\cite{solon2013,buttinoni2013_speck, cates2015, grafke_cates2017, geyer2019, martin2021}, fluid dynamics~\cite{grafke_instanton_2013, grafke_instanton_2015}, biology~\cite{liu2019}, neuroscience~\cite{amit1997, mazzucato2015, jercog2017}, climate science~\cite{ragone2018_bouchet, simonnet2021_bouchet}, etc.  The  description of such nonequilibrium phase transitions requires a generalization of the equilibrium statistical mechanics approach, in which we must consider the probability of trajectories rather than configurations.

\subsection{Minimum Action Principle}
\label{sec:MAP}

\renewcommand{\arraystretch}{1.5}
\begin{table*}
\caption{\label{tab:formalism_comparison}Correspondence of formalisms}
\begin{ruledtabular}
\begin{tabular}{p{1.5cm} p{7.cm} p{7.cm} }
 
& Equilibrium approach in phase space
& Nonequilibrium approach in trajectory space\\
\hline
Microscopic weights 
& Probability distribution (e.g. Boltzmann-Gibbs distribution) of the microscopic variables in state space.
& Probability distribution (e.g. path integral) of the microscopic trajectories.\\
Macroscopic variable 
& Map from microscopic state space to coarse-grained macroscopic space (e.g. spin values to magnetization).
& Map from microscopic trajectories in phase space to macroscopic trajectories. \\
Macroscopic weights
& Marginal distribution of the macroscopic variables, and associated free energy
& Marginal distribution of the macroscopic trajectories, and associated action.  \\
Macroscopic predictions
& Free energy  minimization   
& Action minimization
\end{tabular}
\end{ruledtabular}
\end{table*}
\renewcommand{\arraystretch}{1}

Even if the stationary distribution of  nonequilibrium systems is not known in general, we can often write down the probability distribution of their trajectories, using e.g. path integral approaches such as the Martin-Siggia-Rose-Janssen-De Dominicis~\cite{martin1973,*janssen1976,*de_dominicis1976} or the Doi-Peliti formalism~\cite{doi1976, *peliti1985}, or Girsanov theorem. 
This offers the possibility to generalize the micro-to-macro mapping to trajectory space rather than phase space: that is, enumerate all the microscopic trajectories leading to the same evolution of a macroscopic variable, and thereby deduce the probability weight of these macroscopic trajectories. While these calculations are again to be performed on a case-by-case basis, by analogy with the equilibrium setup we can deduce some of the generic features of the result. Let us discuss those next.

Assuming again that the macroscopic state of the system can be described by a variable or field $\phi$ in some differentiable manifold $\mathcal{M}$ (for example $ \RR^d$, $\calS^d$, or $L_2(\RR^d)$), working in trajectory space amounts to calculating the probability weight of a macroscopic path $\{\phi(t)\}_{t\in[0,T]}$ by enumerating the microscopic trajectories consistent with $\{\phi(t)\}_{t\in[0,T]}$ and summing over their probability distribution. Generically we expect the result of this sum to indicate that the weight of the macroscopic path  $\{\phi(t)\}_{t\in[0,T]}$ is asymptotically given by the factor $\exp(-S_T[\phi]/\epsilon)$, where  $\epsilon$ is again a small parameter that goes to zero as the number of microscopic constituents in the system goes to infinity and  $S_T[\phi]$ is an action, that is, a functional of $\{\phi(t)\}_{t\in[0,T]}$. The specific form of this action  depends on the problem under consideration (examples will be given below),  but it typically takes the form of an integral over a Lagrangian
\begin{equation}
    \label{eq:action:L}
    S_T[\phi] = \int_0^T L(\phi,\dot \phi) dt.
\end{equation}
where $\dot \phi = d\phi/dt\in T_\phi\mathcal{M}$.  The action $S_T[\phi]$ is the non-equilibrium generalization of the free energy $V(\phi)$ and minimization of $S_T[\phi]$ allows us to quantify the probability and  mechanism of various macroscopic events in the limit as $\epsilon\to0$. In particular:

1. The probability that the system started in state $\phi_a$ at time $t=0$ ends up in state $\phi_b$ at time $T$, is obtained by summing $\exp(-S_T[\phi]/\epsilon)$ over all paths with these end points. When $\epsilon\ll1$, the path with minimum action dominates this sum, which means that the aforementioned probability is asymptotically given by
\begin{equation}
    \label{eq:prob}
    \PP(\phi(T)=\phi_b|\phi(0)=\phi_a) \asymp \exp\left(-\inf S_T[\phi]/\epsilon\right)
\end{equation}
where the minimization is taken over all paths $\{\phi(t)\}_{t\in[0,T]}$ such that $\phi(0)=\phi_a$ and $\phi(T) = \phi_b$ and $\asymp$ means exponential asymptotics, i.e  the ratio of the logarithm of both sides in~\eqref{eq:prob} tends to 1 as $\epsilon\to0$. The minimizer of the action also gives the pathway by which the macroscopic transition event occurs with probability one in this limit. 

2. The non-equilibrium invariant distribution of the system can be characterized similarly via the quasipotential defined as 
\begin{equation}
    \label{eq:QP}
    V_{\phi_a}(\phi_b) = \inf_{T>0} \inf S_T[\phi]
\end{equation}
where the inner minimization is again taken over all paths $\{\phi(t)\}_{t\in[0,T]}$ such that $\phi(0)=\phi_a$ and $\phi(T) = \phi_b$. The quasipotential $V_{\phi_a}(\phi_b)$ plays a role analogous to the free energy barrier from state $\phi_a$ to $\phi_b$, and it can be used to identify the possible phases and formulate an equivalent of Arrhenius law. More precisely: $\phi_a$ is a metastable phase if $V_{\phi_a}(\phi) \le V_{\phi}(\phi_a)$ for all $\phi$ in a vicinity of $\phi_a$ (i.e. $\phi_a$ is the non-equilibrium equivalent of a local minimum on the free energy); and if $\phi_a$ and $\phi_b$ are the only two metastable phases in the system, the asymptotic rates of transition from $\phi_a$ to $\phi_b$ and $\phi_b$ to $\phi_a$ are respectively given by
\begin{equation}
    \label{eq:rates:01}
    k_{a,b} \asymp e^{-V_{\phi_a}(\phi_b)/\epsilon}, \quad
    k_{b,a} \asymp e^{-V_{\phi_b}(\phi_a)/\epsilon}
\end{equation} 

3. Eq.~\eqref{eq:rates:01} is a non-equilibrium generalization of Arrhenius law. It implies that the relative probability to find the system in states $\phi_a$ or $\phi_b$ on its non-equilibrium invariant distribution is asymptotically given by
\begin{equation}
    \label{eq:P:01}
    \PP(\phi_b)/\PP(\phi_a) \asymp e^{-[V_{\phi_a}(\phi_b)-V_{\phi_b}(\phi_a)]/\epsilon}.
\end{equation}
As a result, with probability 1 as $\epsilon\to0$, the system is in state $\phi_a$ if $V_{\phi_a}(\phi_b)>V_{\phi_b}(\phi_a)$ and state $\phi_b$ if $V_{\phi_a}(\phi_b)<V_{\phi_b}(\phi_a)$. 
By analyzing how the quasipotential varies in terms of the system's control parameters we can thereby identify non-equilibrium phase transitions that arise when $V_{\phi_b}(\phi_a)=V_{\phi_a}(\phi_b)$ and characterize their mechanism---the details of these calculations will be given below.
We also refer the reader to Fig.~\ref{fig:simple_example} for a graphical illustration in a toy system of non-equilibrium phase transition whose detection requires the formalism above.
\begin{figure}
    \centering
    \includegraphics[width=0.75\columnwidth]{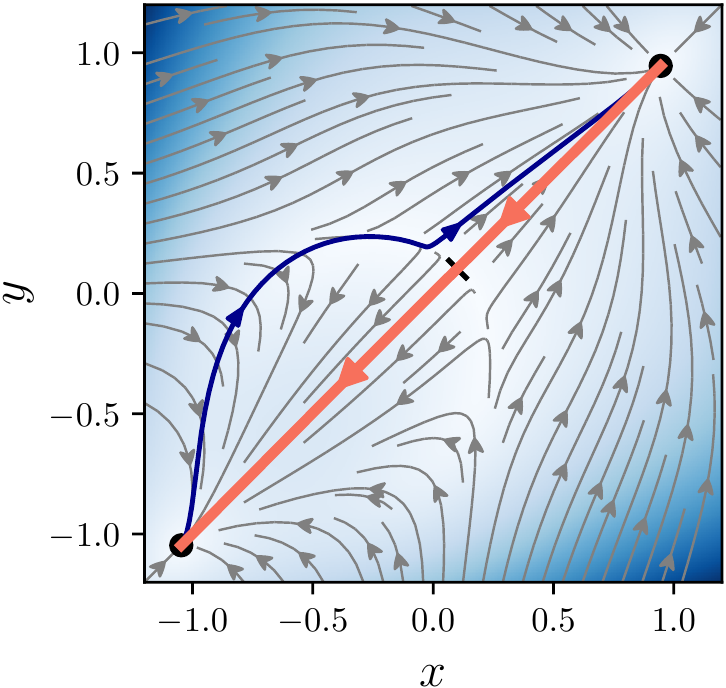}
    \includegraphics[width=0.75\columnwidth]{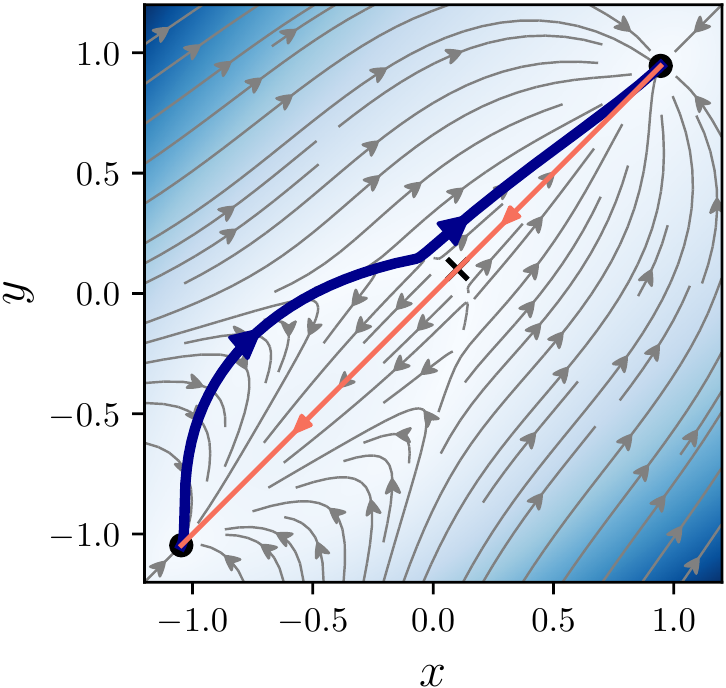}
    \caption{Example of nonequilibrium phase transition in a toy two-dimensional  system with two metastable states (black disks) and action $S_T[x,y] = \frac12\int_0^T  \left(|\dot x-f(x,y)|^2 +|\dot y -f(y,x)|^2 \right)dt$, where $f(x,y)=D(y-x)+x-x^3+h+\nu(x-y)^2$. The parameters $D$ and $h$ are fixed to $0.5$ and $-0.1$, respectively, while $\nu$ is used as a control parameter. The flow of the vector field $f$ for $\nu=0.5$ is shown in the top panel and $\nu=1.5$ in the bottom panel. This flow is non-gradient (i.e. the deterministic dynamics is not steepest descent over an energy) has two stable fixed points, $\phi_a$ (bottom left) and $\phi_b$ (top right), which  solve $f(x,y)=f(y,x)=0$, which are the possible phases in this toy example. The blue line represents the most probable transition path (i.e. the minimizer of the action) from $\phi_a$ to $\phi_b$ and the orange line is the most probable path from $\phi_b$ to $\phi_a$. The thickest line indicates the path with larger rate: that is, $\phi_a$ is the stable phase when $\nu=0.5$ (top panel), while $\phi_b$ is the stable phase when $\nu=1.5$ (bottom panel). The transition paths were calculated with the method developed in this paper, and cross-checked using GMAM~\cite{VE2008gMAM}.  }
    \label{fig:simple_example}
\end{figure}

\subsection{Hamiltonian formalism}
\label{sec:hamilton}

The minimum action principle described in the last section offers a way to study transition events and phase transitions in non-equilibrium systems. Concrete predictions however rest on our ability to: (i) derive the Lagrangian used in the action~\eqref{eq:action:L} and (ii) minimize this action as needed in~\eqref{eq:prob} and \eqref{eq:QP}.

Like in the equilibrium case, resolving the first issue is again complicated in general and requires to be handled on a case-by-case basis. When these calculations can be done (see Sec.~\ref{sec:interpr} for a list of examples), one often deduces that $L(\phi,\dot \phi)$ is given as the Legendre-Fenchel transform of a Hamiltonian $H(\phi,\theta)$:
\begin{equation}
    \label{eq:Ham:Leg}
    L(\phi,\dot \phi) = \sup_{\theta\in T_\phi\mathcal{M}} \left(\< \theta,\dot \phi\> - H(\phi,\theta)\right),
\end{equation}
where $\<\cdot, \cdot\>$ denotes the scalar product in $T_\phi\mathcal{M}$ and $\theta$ is a field conjugate to $\phi$ whose physical meaning will be explained below. The form of the Hamiltonian $H(\phi,\theta)$ is also problem-dependent but it is known in some instances, see Sec.~\ref{sec:collect}. 

Using  \eqref{eq:Ham:Leg}, the minimization of the action can then be formulated as a min-max problem:
\begin{equation}
    \label{eq:min:max}
    \inf_\phi S_T(\phi) = \inf_\phi \sup_\theta \int_0^T \left( \< \dot \phi,\theta\> - H(\phi,\theta)\right)dt.
\end{equation} 
where the supremum is taken over all $\{\theta(t)\}_{t\in[0,T]}$ and the infimum over all $\{\phi(t)\}_{t\in[0,T]}$ such that $\phi(0)=\phi_a$ and $\phi(T)=\phi_b$. To get the quasipotential, we must also consider an extra minimization over all $T>0$ to~\eqref{eq:min:max}, while other applications may require adding terms to~\eqref{eq:min:max} or modifying the boundary conditions for this min-max problem.  In most cases, these calculations must be performed numerically. One of the main goal of this paper is to develop robust numerical methods for these computations. These methods aim to be general enough to be applicable to a wide variety of systems that fit the framework above; here we will also use them to solve some non-trivial examples involving spatially-extended systems undergoing non-equilibrium phase transitions.


\subsection{Related Works} 
\label{sec:related}

The Euler-Lagrange equations associated with the min-max problem~\eqref{eq:min:max} are Hamilton's classical equations:
\begin{equation}
    \label{eq:H1}
    \dot \phi = \partial_\theta H, \qquad 
    \dot \theta = -\partial_\phi H.
\end{equation}
What makes the problem non-standard, however, are the  boundary conditions imposed on $\phi(t)$ at $t=0$ and $t=T$. The nature of these boundary conditions suggests to use shooting methods~\cite{keller2018numerical}, as was proposed e.g. in~\cite{maier1996}, but such methods scale badly with dimension, or can even be ill-posed  for the problems we are interested in, for which the equation for $\theta$ in~\eqref{eq:H1} cannot be integrated forward in time.  Shooting methods are also hard to use when $T=\infty$, which typically arises when we consider $\inf_{T>0} \inf S_T[\phi]$.

To get around this difficulty, the minimum action method (MAM) proposed in \cite{weinan2004} evolves the whole trajectory $\{\phi(t)\}_{t\in[0,T]}$ while keeping $\phi(0)=\phi_a$ and $\phi(T)=\phi_b$ fixed. This amounts to performing gradient descent (GD) on the action in the landscape of all authorized paths satisfying these boundary conditions. Introducing the artificial optimization time~$\tau$, GD results in the following evolution equation for $\{\phi(\tau,t)\}_{\tau\ge0,t\in[0,T]}$
\begin{align}
    \partial_\tau \phi = -\frac{\delta S_T[\phi]}{\delta\phi(t)}, \quad \phi(\tau,0) = \phi_a, \quad \phi(\tau,T) = \phi_b,
\end{align}
or using the Lagrangian formulation of the action
\begin{align}
    \partial_\tau \phi = -\left[ \frac{\partial L}{\partial \phi} - \frac{d}{dt}\left(\frac{\partial L}{\partial \dot \phi}\right)\right],
    \label{eq:euler_lagrange}
\end{align}
with the same boundary conditions at $t=0,T$. The main drawback of MAM is that it involves the Lagrangian $L(\phi,\dot\phi)$ rather than the  Hamiltonian $H(\phi,\theta)$, and there are many problems of interest where the latter is explicitly available but the former is not. Solving \eqref{eq:euler_lagrange} then requires one to perform $\max_\theta[\<\dot \phi,\theta\> - H(\phi,\theta)]$ for all $t\in[0,T]$ at each iteration step in $\tau$ to numerically get an estimate of the function $\vartheta(\phi,\dot \phi)$ such that
\begin{equation}
    \label{eq:min}
    L(\phi,\dot\phi)=  \< \dot \phi,\vartheta(\phi,\dot \phi)\> - H(\phi,\vartheta(\phi,\dot \phi))
\end{equation} 
Proceeding similarly, we can also obtain numerical estimates for the derivative $\partial L/\partial\phi$ and $(d/dt) \partial L/\partial\dot\phi $ appearing on the right hand-side of~\eqref{eq:euler_lagrange}. This approach was used in~\cite{grafke2017}. The downside is that~\eqref{eq:euler_lagrange} is a partial differential equation in physical time~$t$, optimization time $\tau$, and possibly space as well when $\phi$ and $\theta$ are fields; writing efficient numerical solvers for such equations typically requires one to use implicit schemes for numerical stability and/or efficiency, and such schemes are hard to design without explicit knowledge of  $\vartheta(\phi,\dot \phi)$. This is why here we want to bypass the computation of $\vartheta(\phi,\dot\phi)$ and solve the min-max problem in~\eqref{eq:min:max} concurrently by treating the minimization over $\phi$ and  the maximization over $\theta$ on equal footings.

If we now turn our attention to the quasipotential~\eqref{eq:QP}, an extra minimization over all $T>0$ must be added to the min-max problem~\eqref{eq:min:max}. The MAM can be generalized to handle this problem by using a reparametrization of the path $\{\phi(t)\}_{t\in[0,T]}$ by arclength rather than physical time. This formulation leads to the geometric minimum action method (GMAM), in which the minimization over $T$ is performed explicitly beforehand~\cite{VE2008gMAM}. GMAM was further developed and used in~\cite{heymann2008prl,grafke2017} and a variant of it was also recently proposed in~\cite{kikuchi2020} to compute the quasipotential by a spectral decomposition of paths and an optimization of basis coefficients. 
However, these methods are Lagrangian based, with the issues discussed before. We will show below how to use the ideas behind GMAM in a Hamiltonian approach and thereby get an efficient method to solve $\inf_{T>0} \inf_\phi S_T[\phi]$.

On the analytical side, the quasipotential is related to the solution $V(\phi)$ of the Hamilton-Jacobi equation~\cite{freidlinWentzell1998,graham1986} 
\begin{equation}
    \label{eq:HJ}
    0 = H(\phi,\partial V/\partial\phi).
\end{equation}
This equation can be used to deduce some structural properties of the quasipotential. It is a central object in problems that can be tackled through macroscopic fluctuation theory (MFT), and the review~\cite{bertini2015} provides numerous examples and useful insights. Ref.~\cite{bouchet_kuramoto2016} also discusses how to perturbatively solve this equation in systems that are close to equilibrium. In general, however, \eqref{eq:HJ} needs to be solved numerically, which is nontrivial since it is a complicated partial differential equation (or even a functional equation when $\phi$ is a field).   In dimension 2 or 3 this can be done globally using fast marching methods like the one discussed in~\cite{cameron2012,*gan_cameron_2017,*dahiya_cameron_2018}. In dimension higher than 3, these methods become inapplicable, and \eqref{eq:HJ} must be solved locally by the method of characteristics using the variational formulation of this equation: this brings us back to solving $\inf_{T>0} \inf_\phi S_T[\phi]$.

\subsection{Organization and Main Contributions }
\label{sec:main:res}

The remainder of this paper is organized as follows: In Sec.~\ref{sec:interpr}, we give more details about the minimum action principle that is at the core of our approach. To this end,  in Sec.~\ref{sec:collect} we first present typical classes of stochastic dynamical systems that display metastability and non-equilibrium phase transitions, and are amenable to analysis via action minimization. For illustration, we also use our method to compute the transition paths between metastable states in two low-dimensional benchmark models, namely the Maier-Stein model~\cite{maier1996} and the Schl\"ogl model~\cite{schlogl1972}. In Sec.~\ref{sec:general:setup} we discuss the features of the minimum action framework in a general setup, and summarize the main outputs of the approach.

In Sec.~\ref{sec:theo:comp} we present our method for solving the min-max problem~\eqref{eq:min:max}.  The scheme is based on performing gradient descent-ascent (GDA) on the objective, which has the advantage that it can be formulated directly in the Hamiltonian setup. The main issue we need to address is that, in our context, the GDA equations are partial differential equations of hyperbolic type in some optimization time $\tau$ and physical time $t$, to be solved with boundary conditions at $t=0$ and $t=T$. In Sec.~\ref{sec:minmax:T} we show that a simple linear change of variables in these equations  allows us to reformulate them in a way that is convenient for numerical solution via Strang splitting. In Sec.~\ref{sec:minmax:G} we then generalize this scheme to compute the quasipotential when an extra minimization over $T$ is added to~\eqref{eq:min:max}. This is done by using a geometric formulation in which the paths $\{\phi,\theta\}_{t\in[0,T]}$ are reparametrized using normalized arclength. This allows us to handle the minimization of $T$ efficiently and calculate paths whose duration in physical time is infinite but whose length remains finite.

In Secs.~\ref{sec:modified_GL_mainSection} and~\ref{sec:schlogl} we then use the methods we propose to analyze two spatially extended nonequilibrium systems that display first-order phase transitions. More specifically, in Sec.~\ref{sec:modified_GL_mainSection}  we study a modified Ginzburg-Landau (GL) dynamics subject to an additive Gaussian white noise. The noiseless evolution of the field is non-gradient with two stable fixed points; the noise makes these points metastable and we must resort to minimum action algorithms to compute the nonequilibrium transition pathways between them. The action along these paths allows us to estimate the relative probability of the metastable states. We use this procedure to compute the phase diagram of the system in function of two control parameters.
In Sec.~\ref{sec:schlogl}, we study a spatially extended version of the Schl\"ogl model, in which the fluctuations are driven both by diffusion of the microscopic molecules and reactions between them. This system  displays a first-order non-equilibrium phase transition in terms of the diffusivity of the molecules, which we characterize. We also show that the predictions of the minimum action approach explain the transition events observed in the microscopic system.

Concluding remarks are given in Sec.~\ref{sec:conclusion} and some technical developments are deferred to several Appendices.

\section{Problem Setup and Interpretation}
\label{sec:interpr}

The aim of this section is to provide a better motivation of the minimum action principle introduced in the introduction and pinpoint some of its key predictive features. For the reader's convenience we begin by listing a collection of motivating problems where the formalism applies. In Sec.~\ref{sec:general:setup} we will then put the approach in a broader context and explain how to use it to analyze metastability.

\subsection{A collection of motivating problems}
\label{sec:collect}

The first two examples a and b involve no coarse-graining from micro-to-macro and are included because they are simple and transparent; the last three examples c, d, and e requires one to define proper macroscopic variables to derive the minimum action principle and its Hamiltonian.

\paragraph{Diffusion in detailed balance:} Consider the motion of a particle $x(t)\in \RR^d$ whose evolution is governed by the overdamped Langevin equation 
\begin{equation}
\label{eq:sde_equilibrium}
    \dot x = -\nabla U(x) + \sqrt{2 kT} \,  \eta(t)
\end{equation}
where $U(x)$ is some potential, $kT$ is the product of  Boltzmann constant $k$ and the temperature $T$, and $\eta(t)$ is a white-noise. The dynamics~\eqref{eq:sde_equilibrium} is in detailed balance with respect to the Boltzmann-Gibbs probability density function $\rho(x) = Z^{-1}e^{-U(x)/kT}$ where $Z = \int_{\RR^d} e^{-U(x)/kT} dx$. If the potential $U(x)$ has multiple local minima, and the temperature $kT$ is much smaller than the barriers between them, \eqref{eq:sde_equilibrium} displays metastability: the system stays confined for a long time in the well around a minimum of $U(x)$ before finally hopping to another well where the process repeats. In this example, we can use a WKB expansion to analyze the Fokker-Planck equation associated with~\eqref{eq:sde_equilibrium}. The eikonal equation obtained at  leading order in $kT$ is a Hamilton-Jacobi equation whose Hamiltonian is given by
\begin{equation}
    \label{eq:H:pot}
    H(x,\theta) = -\< \theta,\nabla U(x)\> + |\theta|^2.
\end{equation}
This Hamiltonian is the one to be used in the min-max problem~\eqref{eq:min:max}, as can also be proven rigorously using Freidlin-Wentzell LDT~\cite{freidlinWentzell1998}.
In this example, the quasipotential $V_{x_a}(x_b)$ can be calculated explicitly. If $x_a$ and $x_b$ are the locations of two local minima of $U(x)$ with adjacent wells, the path minimizing the action~\eqref{eq:QP} is the minimum energy path between these two points, and $V_{x_a}(x_b)$ is given by
\begin{equation}
    \label{eq:QP:grad}
    V_{x_a} (x_b) = U(x_s) - U(x_a)
\end{equation}
where $U(x_s)$ is the energy of the saddle point of minimum height (aka mountain pass) between $x_a$ ad $x_b$. Thus we recover the Arrhenius law for the rate of transition from $x_a$ to $x_b$,
\begin{equation}
    \label{eq:arrh:grad}
    k_{a,b} \asymp e^{-V_{x_a} (x_b)/kT} = e^{-[U(x_s) - U(x_a)]/kT}.
\end{equation}

\begin{figure}[t]
    \centering
     \begin{tikzpicture}
    \path (0,0) node {    \includegraphics[width=0.8\columnwidth]{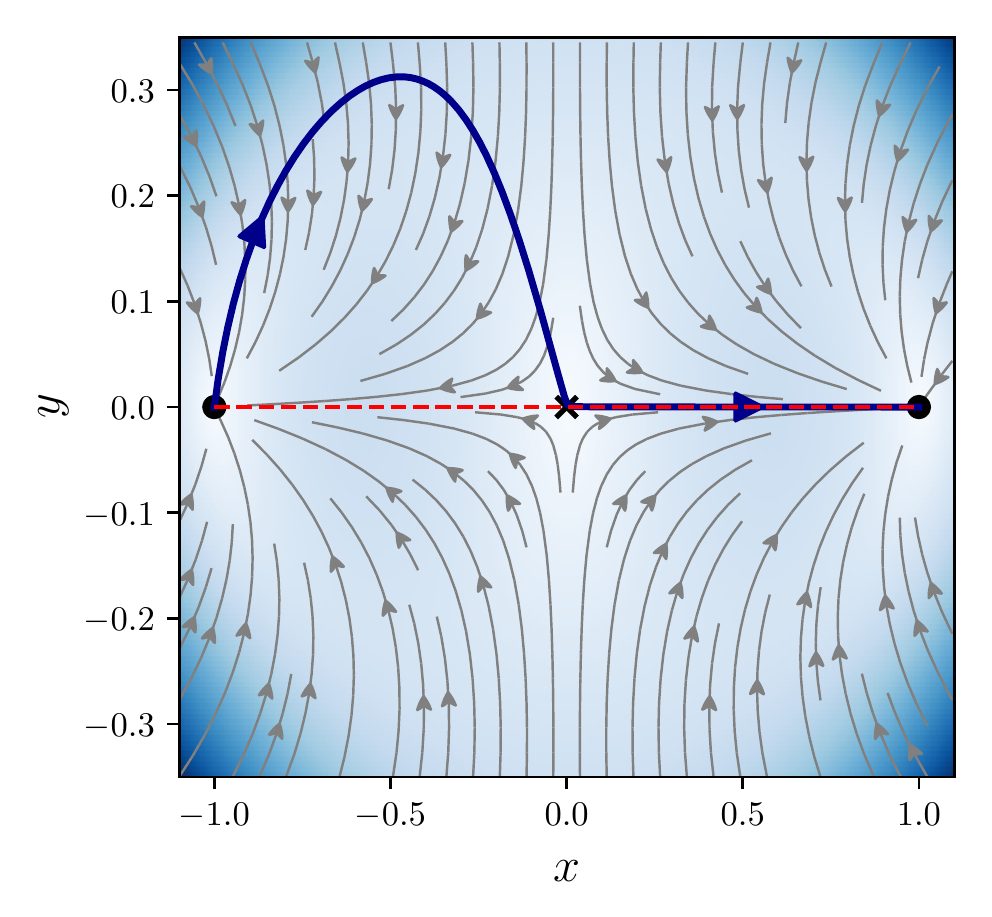}
    };
    \draw (-3.2,2.5) node[anchor=south west] {\bf a)};
  \end{tikzpicture}
     \begin{tikzpicture}
    \path (0,0) node {    \includegraphics[width=0.8\columnwidth]{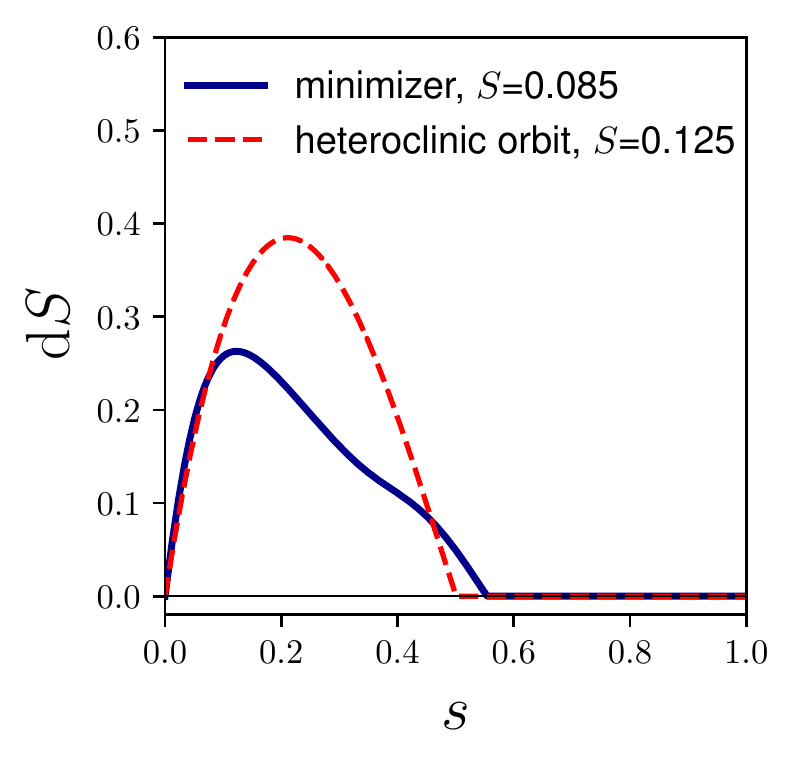}
    };
    \draw (-3.2,2.5) node[anchor=south west] {\bf b)};
  \end{tikzpicture}
    \caption{Optimal reaction path in the Maier-Stein model~\cite{maier1996}: the evolution of $(x,y)\in\RR^2$ is governed by the SDE $\dot x=x-x^3-\beta x y^2+\sqrt{\epsilon}\eta_1$, $\dot y=-(1+x^2)y+\sqrt{\epsilon}\eta_2$, where $\eta_{1,2}$ are independent Gaussian white noises, and $\beta$ some parameter (here $\beta=10$). The flow lines of the noiseless model ($\epsilon=0$)  are shown as grey lines in panel a), and the background color indicates the magnitude of the drift (darker means larger): there are two stable fixed points $x_a=(-1,0)$ and $x_b=(1,0)$ (black disks), and one unstable fixed point $x_c=(0,0)$ (black cross). When $\epsilon$ is small but finite, these two fixed points become metastable, and the noise induces  transitions between them:  the most likely path from $x_a$ to $x_b$ is the  minimum action path, which is shown as the blue line. Also shown as an red dashed line is the heteroclinic orbit between $x_a$ and $x_b$: this orbit is different from the minimum action path, indicative of a nonequilibrium transition. Panel b) shows the increment of the action  along the minimum action path and the heteroclinic orbit, confirming that the former is more likely.  }
    \label{fig:maier_stein}
\end{figure}

\begin{figure}
    \centering
     \begin{tikzpicture}
    \path (0,0) node {    \includegraphics[width=0.75\columnwidth]{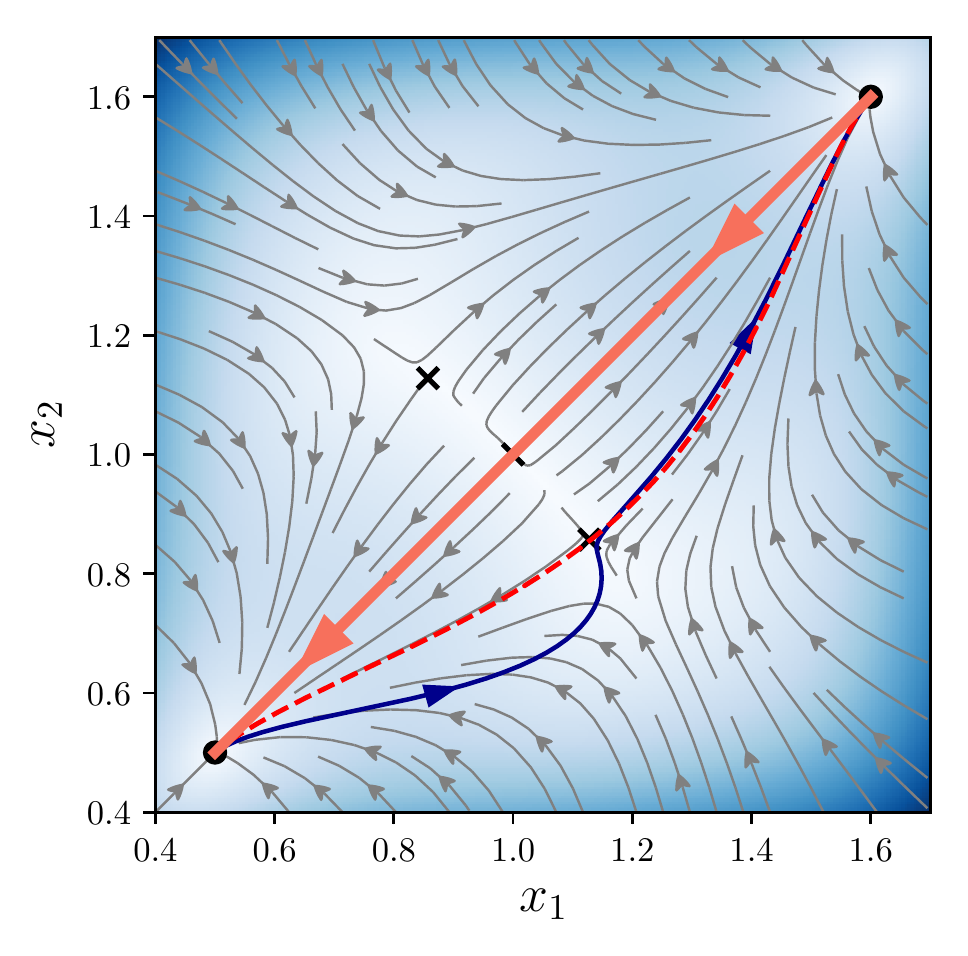}
    };
    \draw (-3.2,2.5) node[anchor=south west] {\bf a)};
  \end{tikzpicture}
       \begin{tikzpicture}
    \path (0,0) node {    \includegraphics[width=0.8\columnwidth]{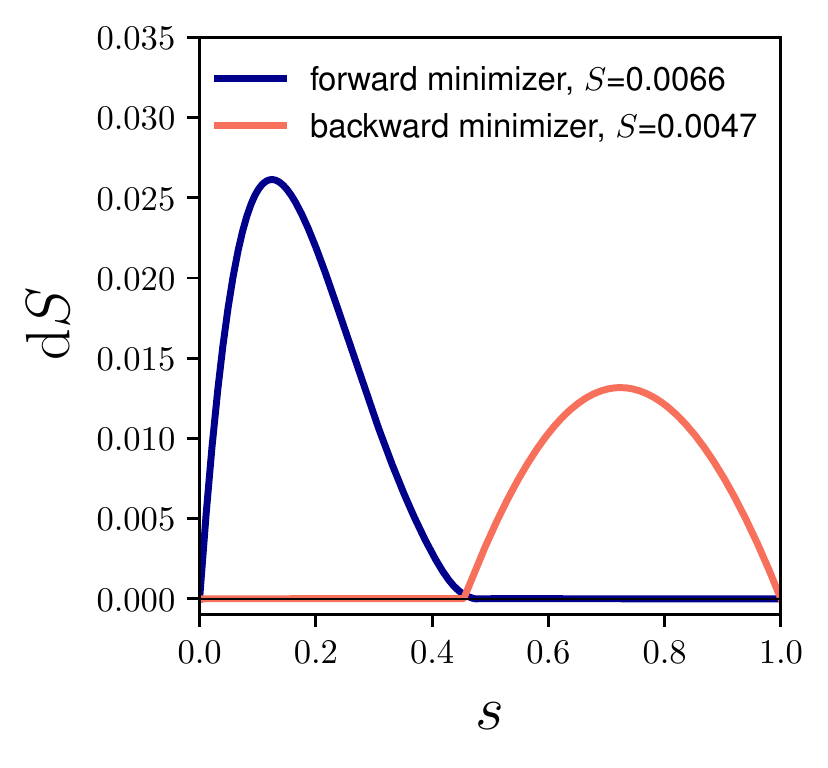}
    };
    \draw (-3.2,2.5) node[anchor=south west] {\bf b)};
  \end{tikzpicture}
    \caption{Optimal reactions paths in the bistable Schl\"ogl model presented in Sec.~\ref{sec:schlogl}, with two reactive compartments and where the particles are allowed to jump from one compartment to another at rate $\gamma$. This system belongs to the class of reaction networks. Here the Hamiltonian is given by $H=H^R+H^D$ with $H^R= \sum_{i=1}^2  w^+(x_i)(e^{\theta_i}-1)+w^-(x_i)(e^{-\theta_i}-1)$ and $H^D=\sum_{i=1}^2  \gamma x_i (e^{\theta_{i-1}-\theta_i} + e^{\theta_{i+1}-\theta_i}-2)$. The flow lines of the mass-action law are shown in panel a), along with its stable fixed points (black dots) and unstable fixed points (black crosses). When the number of agents is large but finite, these stable fixed points become metastable states, and the most likely paths between them are shown as full lines in blue and orange. Also shown in dashed red line is the heteroclinic orbit. All three paths differ, indicative of a a nonequilibrium transition. Panel b) shows the increment along the action of the two minimum action paths, indicating that the bottom left state is most stable under random fluctuations.}
    \label{fig:schlogl_2d}
\end{figure}

\paragraph{Diffusion out-of-equilibrium:} The picture above can be generalized to systems whose evolution is described by the stochastic differential equation
\begin{equation}
\label{eq:sde}
    \dot x = b(x) + \sqrt{\epsilon} \sigma(x) \eta(t),
\end{equation}
even if this equation is not in detailed balance, i.e. it is not possible to write  the drift $b(x)$ as $- D(x) \nabla U(x) + kT \nabla \cdot D(x)$ for $D(x) = (\sigma\sigma^T)(x)$ and some potential $U(x)$. Metastability is  observed with~\eqref{eq:sde} in situations where the noiseless deterministic system $\dot x = b(x)$ has multiple stable fixed points and the noise amplitude is small but finite: the system hovers for long times in the basin around one of these fixed points, but  a noise-driven  transition to another basin eventually occurs. These transitions can be described by the minimum action principle using the Hamiltonian which can again be obtained via WKB analysis of the Fokker Planck equation associated with~\eqref{eq:sde}. It is given by
\begin{equation}
    \label{eq:HWF}
    H(x,\theta) = \< b(x),\theta\>+ \tfrac12 |\sigma(x) \theta|^2.
\end{equation}
Given two stable fixed point $x_a$ and $x_b$ of $\dot x = b(x)$ with adjacent basins of attraction, it is no longer possible in general to solve~\eqref{eq:QP} analytically and calculate $V_{x_a}(x_b)$---to do so requires numerical tools of the type developed below. Still, we know that the rates of transition from $x_a$ to $x_b$ and $x_b$ to $x_a$ satisfy respectively
\begin{equation}
    \label{eq:arrh:non:grad}
    k_{a,b} \asymp e^{-V_{x_a} (x_b)/\epsilon}, \qquad k_{b,a} \asymp e^{-V_{x_b} (x_a)/\epsilon},
\end{equation}
and, with probability $1$ as $\epsilon\to0$, the system performs the transition by following the optimal path  minimizing~\eqref{eq:QP}--for an illustration in the context of Maier-Stein model~\cite{maier1996}, see Fig.~\ref{fig:maier_stein}. The results in SDE with small noise of this type can be made rigorous using Freidlin-Wentzell theory of large deviations (LDT)~\cite{freidlinWentzell1998}. 

\begin{figure}
    \centering
    \begin{tikzpicture}
    \path (0,0) node {               \includegraphics[width=1\columnwidth]{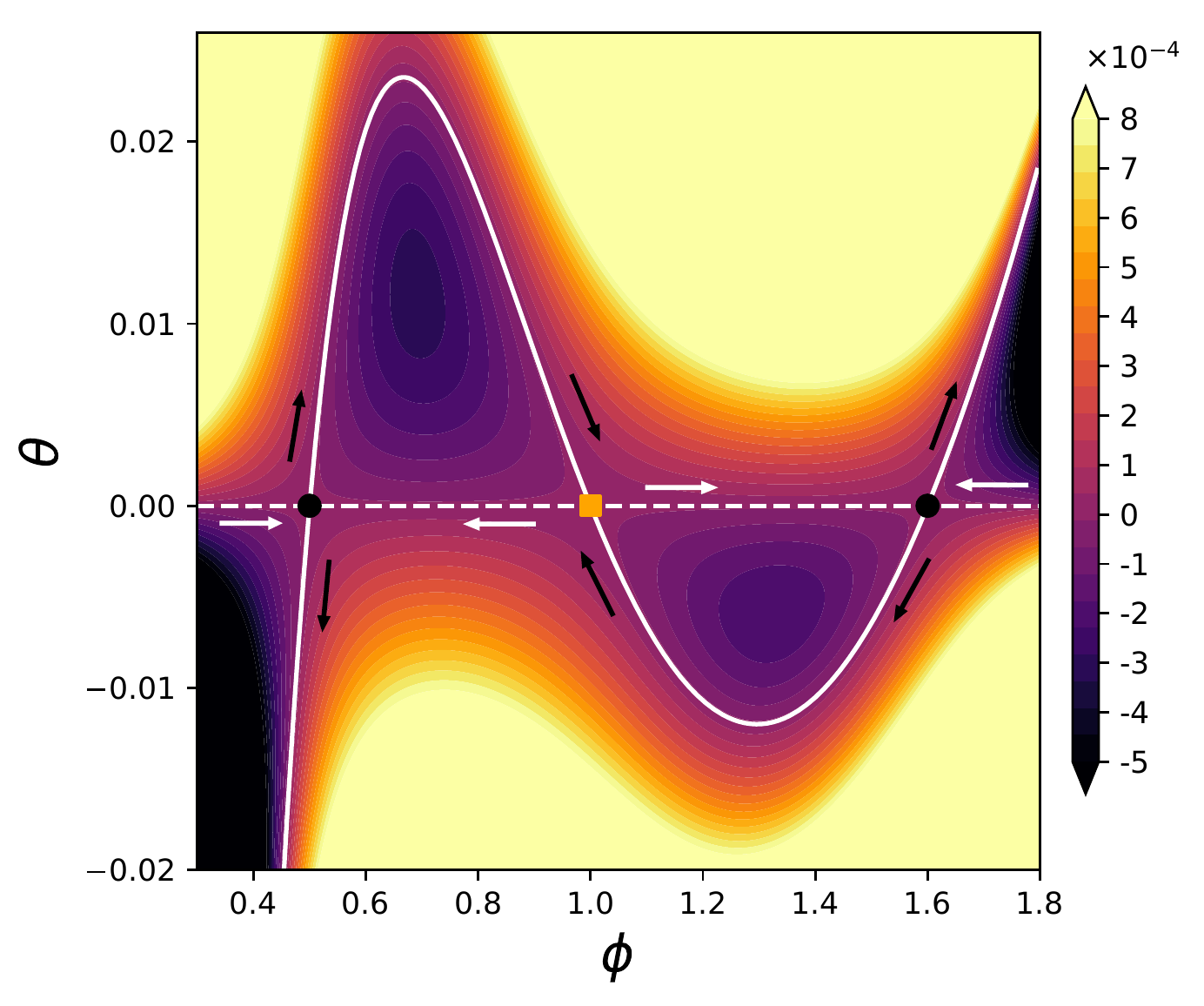}
    };
    \draw (-4,3) node[anchor=south west] {\bf a)};
  \end{tikzpicture}
    \begin{tikzpicture}
    \path (0,0) node {           \includegraphics[width=0.9\columnwidth]{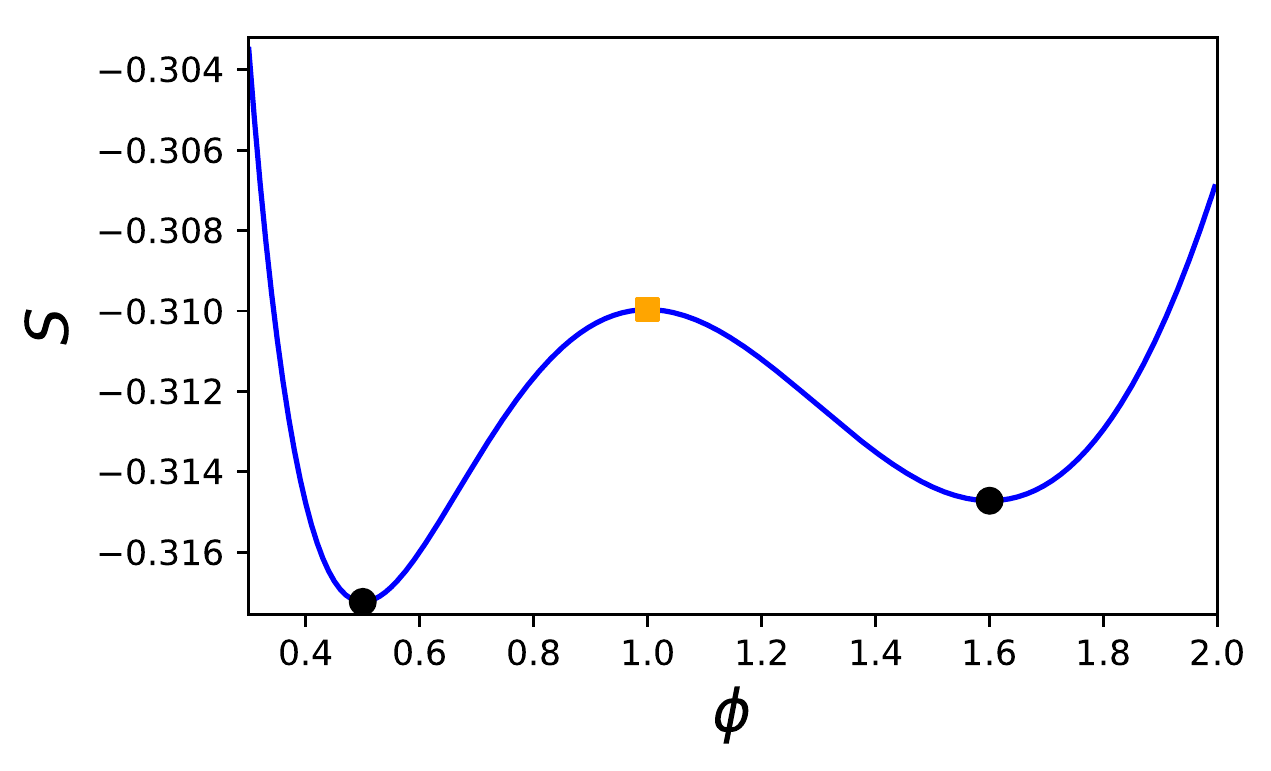}
    };
    \draw (-3,2.2) node[anchor=south west] {\bf b)};
  \end{tikzpicture}
    \caption{a) Contour plot of the Hamiltonian $H(\phi,\theta)$ of the bistable Schl\"ogl model introduced in Sec.~\ref{sec:schlogl}, Eq.~\eqref{eq:chemical_reactions}, with a single reactive compartment containing a large number of particles $\Omega$. 
    The two black disks show the stable fixed points $\phi_a$ (left disk) and $\phi_b$ (right disk) of the law of mass action valid when $\Omega\to\infty$, and the orange square shows its unstable fixed point at $\phi_s=1$. The solid and dashed white line shows the level set $H(\phi,\theta)=0$ on which the solution to Hamilton's equations~\eqref{eq:H1} evolve in the limit when $T\to\infty$.  The horizontal dashed line at $\theta=0$ corresponds to the noiseless dynamics of the mass-action law by which the system relaxes to one of the two stable fixed points $\phi_a$ or $\phi_b$ by moving in the direction of the white arrows. The solid white line, to follow in the direction of the black arrows, shows the solutions of Hamilton's equations on $H(\phi,\theta)=0$ at $\theta\neq0$: these trajectories indicate the most probable path by which the fluctuations can drive the system away from  $\phi_a$ and $\phi_b$, and induce transitions when reaching $\phi_s$. In this example, the action required to drive the system away from $\phi_a$ to some $\phi\le\phi_s$ in its basin of attraction is the area between the white and the dashed solid lines, going from $\phi_a$ to $\phi$: this area is also the quasipotential $V_{\phi_a}(\phi)$ for $\phi\le \phi_s$, and $V_{\phi_b}(\phi)$ for $\phi\ge \phi_s$ can be defined similarly.  b) Global quasipotential $V(\phi)$ obtained by gluing the two quasipotentials $V_{\phi_a}(\phi)$ and $V_{\phi_b}(\phi)$ at $\phi=\phi_s$ after vertical shifting; $V(\phi)$ is also the viscosity solution of the Hamilton-Jacobi equation~\eqref{eq:HJ}.   }
    \label{fig:contour_initial_schlogl}
\end{figure}

\paragraph{Reaction networks:} Consider a well-stirred chemical network between $M$ chemical species, where the quantity of species $i$ is denoted $X_i$. Define the population vector $ X=(X_1,\ldots,X_M)^T$, and assume that there are $R$ reaction channels with rates $w_j(x)$ and change (stoichiometric) vectors $\nu_j\in\mathbb{Z}^M$, i.e.
\begin{equation}
    \label{eq:reactions}
    X \xrightarrow[]{w_j(X)} X+\nu_j, \qquad j=1,\ldots,R.
\end{equation}
When the typical number of agents, $\Omega$, tends to infinity, the dynamics of $x = X/\Omega$ is captured by the mass-action law
\begin{equation}
    \label{eq:mass:action}
    \dot x = \sum_{j=1}^R w_j(x) \nu_j,
\end{equation}
and the minimum action principle is useful to quantify the effects of fluctuations when the number of agents is large but finite. In particular, metastability arises if \eqref{eq:mass:action} has multiple stable fixed points (see Ref.~\cite{dykman1994, tanase2012}), and it can be analyzed using the Hamiltonian
\begin{equation}
    \label{eq:Hreact}
    H(x,\theta) = \sum_{j=1}^R w_j(x) \left(e^{\< \nu_j , \theta\> } - 1 \right).
\end{equation}
Here too this Hamiltonian can be obtained rigorously via Freidlin-Wentzell LDT, or formally via WKB analysis of the system's master equation~\cite{dykman1994} (see also the appendix of~\cite{grafke_cates2017} for a pedagogical derivation), or via a Doi-Peliti field theory computation~\cite{doi1976,*peliti1985, cardy_epidemic_1985, vanWijland1998, lefevre2007}. If $x_a$ and $x_b$ denote two stable fixed points of~\eqref{eq:mass:action} with adjacent basins of attraction, the rates of transition from $x_a$ to $x_b$  and $x_b$ to $x_a$ are respectively given by
\begin{equation}
    \label{eq:arrh:network}
    k_{a,b} \asymp e^{-\Omega V_{x_a} (x_b)},\qquad k_{b,a} \asymp e^{-\Omega V_{x_b} (x_a)},
\end{equation}
and with probability $1$ as $\Omega\to\infty$, when the networks performs the transition, $X/\Omega$ follows the optimal path minimizing~\eqref{eq:QP}. This minimization needs again to be performed numerically in general. As an illustrating example, Fig.~\ref{fig:schlogl_2d} displays reaction paths in the bistable Schl\"ogl model with two reactive compartments: this model belongs to the class of reaction-diffusion networks that will be properly introduced in Sec.~\ref{sec:schlogl_A}. Note that if we reduce the number of compartments to one, the quasipotential of this model can be explicitly obtained from the Hamiltonian, see Fig.~\ref{fig:contour_initial_schlogl}.

\paragraph{Interacting particle systems:} Consider  $N$ particles $x_i\in \Lambda \subset \RR^d$, $i=1,\ldots, N$, that evolve according to
\begin{equation}
\label{eq:sde_manyParticles}
    \dot x_i = b(x_i) + \frac1N \sum_{j=1}^N k(x_i,x_j)+ \sigma(x_i) \, \eta_i(t),
\end{equation}
where $b(x)$ is a drift as in~\eqref{eq:sde}, $k(x,y)$ is some interaction kernel, and $\eta_i(t)$ are independent white-noises. To analyze such interacting particle systems, it is convenient to introduce the empirical density of the particles, $\rho_N(t,x) = N^{-1}\sum_{i=1}^N\delta(x-x_i(t))$. As $N\to\infty$, this empirical density converges towards the density $\rho(t,x)$ that satisfies McKean-Vlasov equation
\begin{equation}
    \label{eq:McKV}
    \partial_t \rho = -\nabla \cdot \left( B(x,[\rho]) \rho \right) + \tfrac12 \nabla \nabla :(D(x) \rho)
\end{equation}
where $B(x,[\rho]) = b(x) + \int_{\RR^d} k(x,y) \rho(y) dy$ and $D(x) = (\sigma\sigma^T)(x)$. Large fluctuations away from the mean-field dynamics~\eqref{eq:McKV} can be captured by the minimum action principle, by using the Hamiltonian
\begin{equation}
    \label{eq:Hinterpart}
    \begin{aligned}
        H(\rho,\theta) &= \int_{\RR^d} \nabla \theta(x) \cdot b(x) \rho(x) dx \\
        & + \int_{\RR^d\times \RR^d} \nabla \theta(x) \cdot k(x,y) \rho(x)\rho(y) dx dy\\
        & + \frac12 \int_{\RR^d} |\sigma(x) \nabla \theta(x) |^2\rho(x) dx.
    \end{aligned}
\end{equation}
This Hamiltonian can again be derived rigorously using LDT~\cite{dawson1987}, and it formally follows from a WKB analysis, now performed on the functional master equation for the empirical particle density.

\paragraph{Fast-slow systems} Consider a system made of a pair of variables $(x,y)\in \RR^{d\times D}$ whose evolution is governed by
\begin{equation}
    \label{eq:fast:slow}
    \left\{ 
    \begin{aligned}
        \dot x &= f(x,y)\\
        \dot y & = \alpha^{-1}b(x,y) + \alpha^{-1/2} \sigma(x,y) \eta(t)
    \end{aligned}\right.
\end{equation}
where $\alpha>0$ measure the separation of time-scale between $x$ and $y$. For small $\alpha$, this separation is large and $y$ evolves much faster than $x$. In particular, when $\alpha\to0$, the dynamics of $x$ is effectively captured by the deterministic limiting equation
\begin{equation}
    \label{eq:lim:x}
    \dot x = F(x)
\end{equation}
where $F(x)$ is obtained by averaging $f(x,y)$ over the stationary distribution of the equation for $y$ at $x$ fixed, assuming that it exists. This equation defines the so-called virtual fast process, which on the fast time scale $\tau = \alpha t$ reads
\begin{equation}
    \label{eq:fast:virtual}
    dy_x/d\tau = b(x,y_x) + \sigma(x,y_x) \eta(\tau) \quad \text{($x$ frozen)}
\end{equation}
Denoting by $\EE^x$ the expectation over the stationary distribution of $y_x(\tau)$, the function $F$ entering~\eqref{eq:lim:x} is given by
\begin{equation}
    \label{eq:F}
    F(x) = \EE^x f(x,y_x).
\end{equation}
If we want to analyze the effect of fluctuations on the dynamics of $x$ when $\alpha$ is small but finite, we can use the minimum action principle with a Hamiltonian that can be derived from LDT~\cite{freidlin_averaging_1978, veretennikov_averaging_1991, kifer_averaging_2004, bouchet_fast_slow_2016}
\begin{equation}
    \label{eq:H:fast:slow}
    H(x,\theta) = \log \EE^x \exp\left( \<\theta,f(x,y_x)\>\right)
\end{equation}
In general, this Hamiltonian will need to be calculated numerically, which makes slow-fast systems of the type above more difficult to treat than the models previously discussed in this section. Still, provided that we can design some numerical routine to estimate $H$ as well as its derivatives $\partial_x H$ and $\partial_\theta H$, the numerical methods presented below are applicable in the context of slow-fast systems too. In models of this type $\epsilon =\alpha$.

\subsection{General setup}
\label{sec:general:setup}

The problems listed in Sec.~\ref{sec:collect} all share a Hamiltonian with the following features:
\begin{itemize}
    \item[A1.] $H(\phi,0) = 0$ for all $\phi\in \mathcal{M}$;
    \item[A2.] $H(\phi,\theta)$ is strictly convex in $\theta$ for all $\phi\in\mathcal{M}$.
    \item[A3.]$H(\phi,\theta)$ is twice differentiable in both its arguments;
\end{itemize}

\noindent
Assumptions A1 and A2 follow from the fact, generically, the Hamiltonian can be expressed as a cumulant generating function, i.e. in the form of an expectation generalizing~\eqref{eq:H:fast:slow}:
\begin{equation}
    \label{eq:H:generic}
    H(\phi,\theta) = \log \EE^\phi \exp\left( \<\theta,F(\phi,y_\phi)\>\right)
\end{equation}
where $F$ is problem-dependent and the expectation is taken over the statistics of some underlying process $y_\phi$ conditional on $\phi$ being fixed. Assumption A3 is added for simplicity, as it guarantees that Hamilton's equations~\eqref{eq:H1} are well posed. 

The aim of this section is to discuss at generic level the meaning we can give to the min-max problem~\eqref{eq:min:max} assuming that the Hamiltonian satisfies these assumptions. In particular, we will show that the conjugate field~$\theta$ appearing in  the min-max problem~\eqref{eq:min:max} can be generically interpreted as  measuring the mean effects of the fluctuations in the system dynamics needed to achieve some rare event, and the minimum of the action as the total cost of these fluctuations from which the probability of the event can be estimated as well as its mechanism and rate (if we add the minimization over $T>0$) . 

\subsubsection{Mean behavior} To begin, notice that for the interpretation that $\theta$ measures the effects of the fluctuations to be consistent, the stochastic system under consideration should, in some appropriate limit in which the fluctuations disappear, satisfy the deterministic evolution equation obtained by setting $\theta=0$ in Hamilton's equations~\eqref{eq:H1}:
\begin{equation}
    \label{eq:nonoise}
    \dot \phi = \partial_\theta H(\phi,\theta=0).
\end{equation}
For example, returning to the problems mentioned in Sec.~\ref{sec:collect}, \eqref{eq:nonoise} reduces to the ODE $\dot x= b(x)$ as $\epsilon\to0$ in the SDE~\eqref{eq:sde} with Hamiltonian~\eqref{eq:HWF}; to the law of mass action~\eqref{eq:mass:action} as $\Omega\to\infty$ for the reaction network~\eqref{eq:reactions} with Hamiltonian~\eqref{eq:Hreact}; to the McKean-Vlasov equation as $N\to\infty$ for the interacting particle system~\eqref{eq:sde_manyParticles} with Hamiltonian~\eqref{eq:Hinterpart}; and to~\eqref{eq:lim:x} as $\alpha\to0$ in the slow-fast system~\eqref{eq:fast:slow} with Hamiltonian~\eqref{eq:H:fast:slow}. More generally, under our assumptions, the solution of \eqref{eq:nonoise} is indeed a special solution with $\theta(t)=0$ of  Hamilton's equations~\eqref{eq:H1}  since Assumptions~A1 and A3 imply that  $\partial_\phi H(\phi,0) = 0$ for all $\phi$. 

\subsubsection{Impact of the fluctuations} At the same time, the solution to the deterministic evolution~\eqref{eq:nonoise} with $\phi(0)=\phi_a$ does not satisfies $\phi(T)=\phi_b$ in general. Therefore, for general boundary conditions $\phi(0)=\phi_a$ and $\phi(T)=\phi_b$, the solution of \eqref{eq:H1} must have $\theta(t)\not=0$---in the minimum action  framework this is a reflection that in the original system fluctuations are needed to drive the system's trajectory away from the solution of~\eqref{eq:nonoise}, and the value of $\theta(t)\not=0$  allows us to quantify the cost/probability of observing the event $\phi(T)=\phi_b$ given that $\phi(0)=\phi_a$. Specifically, under the strict convexity Assumption~A2, we have
\begin{equation}
    \label{eqconvex:cons}
    \<\theta, \partial_\theta H(\theta,\phi)\> - H(\phi,\theta)\ge -H(\phi,0)=0,
\end{equation}
with equality if and only if $\theta=0$.
Since $\dot \phi = \partial_\theta H(\phi,\theta)$ along the solution to Hamilton's equations~\eqref{eq:H1}, we deduce that along this solution
\begin{equation}
    \label{eq:minmax:reform}
    \inf_\phi S_T[\phi]=\inf_\phi \int_0^T \left(  \< \theta, \partial_\theta H(\phi,\theta)\> - H(\phi,\theta)\right) dt \ge0
\end{equation}
Thus the action $S_T[\phi]$ can be indeed interpreted as a cost, which is zero only if $\theta(t)=0$ (i.e. when the event can occur without fluctuations), and is strictly positive otherwise (i.e. when the event requires fluctuations). For the problems listed in Sec.~\ref{sec:collect} the form of the integrand $\< \theta, \partial_\theta H\> - H$ is  $|\sigma(x)\theta|^2\ge0 $ for the SDE~\eqref{eq:sde}; $\sum_{j=1}^R a_j(x)\left( \<\nu_j,\theta\> e^{\<\nu_j,\theta\>}- e^{\<\nu_j,\theta\>} + 1\right) \ge0$ for the reaction network~\eqref{eq:reactions}; $\frac12 \int_{\RR^d} |\sigma(x) \nabla \theta(x) |^2\rho(x) dx\ge0$ for the interacting particle system~\eqref{eq:sde_manyParticles}; and $\EE^x f(x,y_x) e^{\<\theta,f(x,y_x)\>}/\EE^x  e^{\<\theta,f(x,y_x)\>}-H$ for the slow-fast system~\eqref{eq:fast:slow}.

In this interpretation, minimizing $S_T[\phi]$ amounts to minimizing the cost of the fluctuations or, equivalently, finding the most likely fluctuation that drives the event $\phi(T)=\phi_b$ given that $\phi(0)=\phi_a$, leading to the asymptotic estimate~\eqref{eq:prob} for the probability of the event.

 \subsubsection{Long-time limit.} Turning now our attention to the problem $\inf_{T\ge0} \inf_\phi S_T[\phi]$, its interpretation is easiest if we assume that the noiseless equation~\eqref{eq:nonoise} has  $N$ stable fixed points $\phi_1$, $\phi_2$, ..., $\phi_N$, and the basins of attraction of these fixed points under~\eqref{eq:nonoise}, denoted respectively as $B_1,\ldots B_N$,  partition $\mathcal{M} = \cup_{i=1}^N \bar B_i$ (note that $B_i\cap B_j = \emptyset$ if $i\not= j$ by definition). In this case, we can calculate the quasipotentials $V_{\phi_i}(\phi_j)$ of every $\phi_i$ and $\phi_j$ with $i\not=j$ such that these points have adjacent basins: these are defined as $V_{\phi_i}(\phi_j) = \inf_{T>0} \inf_{\phi} S_T[\phi]$ for all paths $\{\phi(t)\}_{T\in[0,T]}$  such that $\phi(0)=\phi_i$, $\phi(T)=\phi_j$, and $\phi(t)\in\overline{B_i \cup B_j}$ for all $t\in[0,T]$. Consistently  we also set $V_{\phi_i}(\phi_j)=+\infty$ if $B_i$ and $B_j$ have no common boundary.
These quasipotentials quantify the cost of the fluctuations needed to escape $\phi_i$ conditional on entering $\phi_j$ next, and these costs can be used to deduce the asymptotic rate of these transition events. More precisely,  as the parameter $\epsilon$ measuring the amplitude of the macroscopic fluctuations tends to zero, the system dynamics can be approximated by a Markov jump process (MJP) between the metastable states $\phi_1$, ..., $\phi_N$, with rates given asymptotically by
\begin{equation}
    \label{e:pair:QP} 
    k_{i,j} = e^{-V_{\phi_i}(\phi_j)/\epsilon} \qquad i,j=1,\ldots,N \quad (i\not=j).
\end{equation}
Questions about the asymptotic behavior of the system's dynamics can be answered by analyzing this MJP. Since the rates are all vanishing exponentially at different rates as $\epsilon\to0$, this process is quite singular and can be analyzed by the method of decomposition into cycles developed by Freidlin and Wentzell~\cite{freidlinWentzell1998} (see also~\cite{graham1986}). This method is however rather intricate, and in practice it is often simpler to solve specific questions in the MJP directly (e.g. what is its invariant distribution or what is the mean first passage time from state $i$ to state $j$), and take the limit as $\epsilon\to0$ afterward. Note also that the quasipotentials $V_{\phi_i}(\phi)$ for $i=1,\ldots,N$ can be used to construct a global nonequilibrium potential $\phi$, solution of the Hamilton-Jacobi equation~\eqref{eq:HJ}~\cite{freidlinWentzell1998, graham1986}; since this construction is not doable in practice for the high-dimensional examples we are interested in, we will not dwell upon it here.  We will however look at the minimizers of $\inf_{T\ge0} \inf S_T[\phi]$, which can exist if the path $\{\phi(t)\}_{t\in[0,T]}$ is reparametrized using arclength rather than physical on any minimizing sequence: these geometric paths give the mechanism of the transition between $\phi_i$ and $\phi_j$.

\medskip

The statements made in this section summarize what can be deduced from the minimum action principle when it applies. As stated early, these results can be proven rigorously in specific cases using e.g. tools for large deviation theory (LDT), in the small noise~\cite{freidlinWentzell1998} and weak interaction limit~\cite{dawson1987}, or in more complicated setups involving  hydrodynamic limits~\cite{spohn_large_1991, kipnis1999}. We can however envision situations beyond the realm of LDT where min-max problem like~\eqref{eq:min:max} with a Hamiltonian $H$ satisfying Assumptions A1, A2, and A3 arise (e.g. from the Martin-Siggia-Rose-Jensen-De Dominicis~\cite{martin1973,*janssen1976,*de_dominicis1976} or the Doi-Peliti formalism~\cite{doi1976, *peliti1985}, or from macroscopic fluctuation theory~\cite{bertini2015,baek2018}), and can be interpreted as above. The methods that we propose below are of general purpose and were designed with such general situations in mind.

\section{Computational Aspects}
\label{sec:theo:comp}

\subsection{Min-max on the action as a wave propagation problem}
\label{sec:minmax:T}

\begin{figure*}
\begin{minipage}{\linewidth}
\begin{algorithm}[H]
\label{alg:1}
\caption{: Action Minimization by Gradient Descent Ascent}
\begin{algorithmic}[1]
\State \textbf{Inputs:} $M\in \NN$; a path $\{\phi^0_i\}_{i\in I}$ with $\phi_0^0=\phi_a$ and $\phi_M^0=\phi_b$; the functions $f(u,v)$ and $g(u,v)$; $T>0$, $\Delta\tau>0$, $\alpha>0$.
\State \textbf{Initialization:} For every $i\in I$, take $\theta^0_i=0$, and set $u_i^0 = \phi_i^0+\alpha \theta_i^0$ and $v_i^0 = \phi_i^0-\alpha \theta_i^0$; set $\Delta t=T/M$.
\For{$n\ge0$}
\State 
Update $u$ with  an implicit upwind scheme, namely,  solve $ \{u^{n+1}_i\}_{i\in I}$ sequentially from $i=M$ to $i=0$ using: 
\begin{align*}
\begin{cases}
    u^{n+1}_M=- v_M^{n}+2\phi_b\\[4pt]
    \dfrac{u_i^{n+1}-u_i^{n}}{\Delta \tau} = \dfrac{u_{i+1}^{n+1}-u_i^{n+1}}{\Delta t} +f(u^n_{i+1},v^{n}_{i+1}), \qquad i=M-1,\dots,0
   \end{cases}
\end{align*}
\State Update $v$ with an implicit upwind scheme, namely, solve $\{v_i^{n+1}\}_{i\in I}$ sequentially from $i=0$ to $i=M$ using: 
 \begin{align*}
      \begin{cases}
        v^{n+1}_0=-u_0^{n+1}+2\phi_a\\[4pt]
        \dfrac{  v_i^{n+1}-v_i^n}{\Delta \tau} = -\dfrac{ v_{i}^{n+1}- v_{i-1}^{n+1}}{\Delta t} +g(u_{i-1}^{n+1},v_{i-1}^{n}),\qquad i=1, \dots, M
       \end{cases}
   \end{align*}
\State Compute $\{\phi_i^{n+1} =\frac12(u_i^{n+1}+v_i^{n+1})\}_{i\in I}$ and $\{\theta_i^{n+1}=\frac12\alpha^{-1} (u_i^{n+1}-v_i^{n+1})\}_{i\in I}$ (if needed).
\EndFor
\end{algorithmic}
\end{algorithm}
\end{minipage}
\end{figure*}

In this section, we discuss how to solve the min-max problem stated in~\eqref{eq:min:max}, i.e. 
\begin{equation}
    \label{eq:minmax:I}
    \inf_\phi S_T[\phi] = \min_{\phi} \max_{\theta} I_T(\phi,\theta)
\end{equation}
where we defined the functional
\begin{equation}
    \label{eq:I:def}
    I_T(\phi,\theta)=\int_0^T  \left( \langle \dot\phi,\theta \rangle - H(\phi,\theta)\right) dt 
\end{equation}
and the optimization is to be performed over trajectories $\{\phi(t),\theta(t)\}_{t\in[0,T]}$ subject to the boundary conditions $\phi(0)=\phi_a$, $\phi(T)=\phi_b$.  
If the functional $I_T$ is convex with respect to $\phi$ and concave with respect to $\theta$, then it is well-known~\cite{kose1956,cherukuri_saddle_point_2017} that this min-max problem can be solved by  amortizing the minimization and maximization over small alternating steps of steepest descent in $\phi$ and steepest ascent in $\theta$. If these steps are infinitesimal in some artificial optimization time~$\tau$, this gradient descent ascent (GDA) method leads to the  evolution equation
\begin{align}
   \partial_\tau \phi =-\alpha\delta I_T/\delta \phi,\qquad
    \alpha\partial_\tau \theta = \delta I_T/\delta \theta,
    \label{eq:system_gradient_descent-ascent}
\end{align}
where for convenience we have introduced a parameter~$\alpha>0$  that sets the relative time scales over which $\phi$ and $\theta$ evolve -- in the jargon of GDA this is referred to two time scale GDA~\cite{lin_tin_jordan2020}. Calculating the functional derivatives, the system~\eqref{eq:system_gradient_descent-ascent} is explicitly given by
\begin{align}
   \displaystyle \partial_\tau \phi  = \alpha\partial_t\theta + \alpha\partial_\phi H,\qquad
    \displaystyle \alpha\partial_\tau \theta = \partial_t\phi - \partial_\theta H.
    \label{eq:system_gradient_descent-ascent_hyperbolic}
\end{align}
These equations for $\{\phi(\tau,t),\theta(\tau,t)\}$ are to be solved with the boundary conditions (in physical time $t$)
\begin{align}
    \phi(\tau,t=0)=\phi_a,\qquad
    \phi(\tau,t=T)=\phi_b.
    \label{eq:boundary_conditions_phi}
\end{align}
for some initial conditions (in optimization time $\tau$)
\begin{align}
    \phi(\tau=0,t)=\phi^0(t),\qquad
    \theta(\tau=0,t)=\theta^0(t),
    \label{eq:initial_conditions}
\end{align}
with $\phi^0(t)$ such that $\phi^0(0) = \phi_a$ and $\phi^0(T) = \phi_b$.

It is easy to see that the fixed points (in $\tau$) of~\eqref{eq:system_gradient_descent-ascent_hyperbolic} are solution to Hamilton's  equations~\eqref{eq:H1} that satisfy $\phi(0)=\phi_a$ and $\phi(T) = \phi_b$.  
In Appendix~\ref{app:convergence_small_alpha} we show that: (i) there is a one-to-one correspondence between the fixed points of~\eqref{eq:system_gradient_descent-ascent_hyperbolic} and the critical points of the  action $S_T[\phi]$, and (ii) if $\alpha$ is small enough these fixed points are stable if and only if  they are local minimizers of the action. Thus solving~\eqref{eq:system_gradient_descent-ascent_hyperbolic} is indeed a way to perform $\inf_\phi S_T[\phi]$.
For illustrative purposes, we derive in Appendix~\ref{app:analytical_solution_GDA} how the GDA converges to the instanton for an Ornstein-Uhlenbeck process.

Let us now show how to put \eqref{eq:system_gradient_descent-ascent_hyperbolic} in a form that is convenient for numerical integration.
Since \eqref{eq:system_gradient_descent-ascent_hyperbolic} is an hyperbolic system of partial differential equations (PDEs), it is useful to diagonalize the problem and introduce the fields $u=\phi+\alpha\theta$ and $v=\phi-\alpha\theta$ that propagate along characteristics and verify
\begin{align}
    \partial_\tau u &=\partial_t u +f(u,v)\label{eq:u}\\
    \partial_\tau v &= -\partial_t v +g(u,v),\label{eq:v}
\end{align}
where we have defined 
\begin{align}
    f(u,v)&= \alpha \partial_\phi H - \partial_\theta H\\
    g(u,v)&= \alpha \partial_\phi H + \partial_\theta H.
\end{align}
The boundary conditions now only involve the propagating fields 
\begin{align}
    &v(\tau,t=0) = -u(\tau,t=0)+2\phi_a\label{eq:vBC}\\
    &u(\tau,t=T) = -v(\tau,t=T)+2\phi_b.\label{eq:uBC}
\end{align}
This formulation shows that the system made of~\eqref{eq:u} and \eqref{eq:v} is well-posed under these boundary conditions (see \cite{siam_hyperbolic04}) since the fields  $v$ and $u$ propagate respectively  forward  and backward in physical time $t$ as the optimization time $\tau$ increases. It also immediately suggests an algorithm to solve Eqs.~\eqref{eq:u} and \eqref{eq:v} based on Strang splitting~\cite{glowinski_strang2016}: to update the fields at every iteration step in~$\tau$, first update $u$ at $v$ fixed by propagating the final condition  at $t=T$ for $u$ in~\eqref{eq:uBC} towards $t=0$ using~\eqref{eq:u} with forward differentiation in $t$, then update $v$ at $u$ fixed by propagating the initial condition at $t=0$ for $v$ in~\eqref{eq:vBC} towards $t=T$ using~\eqref{eq:v} with backward differentiation in $t$. Of course, the convergence does not change if the algorithm starts by updating $v$ before updating $u$.

In practice, the  continuous paths $\phi(\tau,t)$ and $\theta(\tau,t)$ are discretized in physical time on $M+1$ points with index $i\in I=\{0,\cdots,M\}$ such that $T=M\Delta t$,  and we use index $n\in\NN_0$ to encode the evolution of the paths in optimization time~$\tau$ using steps of size $\Delta \tau$, so that  for any field $\psi(\tau,t)$, $\psi^n_i\equiv \psi^n(i\Delta t)$. The details are given in Algorithm~1. The stability of the code relies on two important features. First, advection of the fields is treated with an implicit upwind scheme for both $v$ and $u$. Second, reaction terms $g$ and $f$ are also evaluated on an upwind grid point with respect to the direction of advection. Reaction terms could also be evaluated on site $i$ but we empirically found that the upwind implementation strongly stabilizes the code when dealing with spatially extended diffusive fields. The stability analysis of the numerical scheme is detailed in Appendix~\ref{app:stability_numerical_scheme}.

We should also emphasize that our interest resides in the fixed point of the dynamics that solves Hamilton's equation in physical time $t$, rather than in the details of the dynamics in algorithmic time $\tau$, which has no physical relevance. This consideration enjoins us to look for the largest time-step $\Delta \tau $ that still provides a converging algorithm. The time-step $\Delta t$, however, crucially needs to remains smaller than some characteristic time $t_c$ needed to correctly resolve the dynamics of the instanton. This issue is discussed in more details in Appendix~\ref{app:stability_numerical_scheme}.

Finally, it is important to mention that a higher-order finite difference stencil for the advection of the fields can be implemented while keeping the same algorithmic complexity. Such a scheme can significantly improve computation time since we need a smaller number of grid points to get the same accuracy as the first-order scheme, introduced in the text for purpose of simplicity. The second-order scheme is presented in Appendix~\ref{app:higher_order_scheme}.

\subsection{Geometric formulation on unbounded time intervals}
\label{sec:minmax:G}

\begin{figure*}
\begin{minipage}{\linewidth}
\begin{algorithm}[H]
\label{alg:2}
\caption{: Geometric Action Minimization by Gradient Descent Ascent}
\begin{algorithmic}[1]
\State \textbf{Inputs:} $M\in \NN$; two stable fixed points $\phi_a$ and $\phi_b$ of the noiseless dynamics where $H(\phi_{a,b},0)=\partial_\theta H(\phi_{a,b},0) =0$; a path $\{\hat \phi^0_i\}_{i\in I}$ with $\hat \phi_0^0=\phi_a$ and $\hat\phi_M^0=\phi_b$, such  that $|\hat \phi^{0}_{i+1}-\hat \phi^{0}_{i}|$ is constant in $i$; the functions $f(u,v)$,   $g(u,v)$, and $\lambda(u,v)$; $\Delta\tau>0$, $\alpha>0$.
\State \textbf{Initialization:} For every $i\in I$, take $\hat\theta^0_i=0$, and set $u_i^0 = \hat\phi_i^0+\alpha \hat\theta_i^0$ and $v_i^0 = \hat\phi_i^0-\alpha \hat\theta_i^0$; set $\Delta s=1/M$. 
\For{$n\ge0$}
   \State Update $u$ with  an implicit upwind scheme, namely,  solve $ \{u^{n+1}_i\}_{i\in I}$ sequentially from $i=M$ to $i=0$ using: 
   \begin{align*}
    \begin{cases}
    u^{n+1}_M=- v_M^{n}+2\phi_b\\[4pt]
        \dfrac{u_i^{n+1}-u_i^{n}}{\Delta \tau} = \lambda_i(u^n,v^{n}) \dfrac{u_{i+1}^{n+1}-u_i^{n+1}}{\Delta s} +f(u_i^n, v_i^{n}), \qquad i=M-1,\dots,0
       \end{cases}
   \end{align*}
   \State Update $v$ with  an implicit upwind scheme, namely,  solve $ \{v^{n+1}_i\}_{i\in I}$ sequentially from $i=0$ to $i=M$ using: 
   \begin{align*}
       \begin{cases} v^{n+1}_0=-u_0^{n+1}+2\phi_a\\[4pt]
     \dfrac{v_i^{n+1}-v_i^n}{\Delta \tau} = -\lambda_i(u^{n+1},v^n) \dfrac{ v_{i}^{n+1}- v_{i-1}^{n+1}}{\Delta s} +g(u_i^{n+1},v_i^n), \qquad i=1,\ldots,M\\
       \end{cases}
   \end{align*}

    \State Compute  $\{\bar \phi^{n+1}=\frac12(u^{n+1}+v^{n+1})\}_{i\in I}$ and $\{\bar \theta^{n+1}=\frac12\alpha^{-1}(u^{n+1}-v^{n+1})\}_{i\in I}$.  
    \State Interpolate $\{(\bar\phi_i^{n+1},\bar \theta_i^{n+1})\}_{i\in I}$ onto a path $\{(\hat\phi_i^{n+1}, \hat\theta_i^{n+1})\}_{i\in I}$ such that $|\hat \phi^{n+1}_{i+1}-\hat \phi^{n+1}_{i}|$ is constant in $i$, as in the string method. 
   \State Set $\{u^{n+1}= \hat\phi^{n+1}+\alpha \hat\theta^{n+1}\}_{i\in I}$ and $\{v_i^{n+1}= \hat\phi_i^{n+1}-\alpha \hat\theta_i^{n+1}\}_{i\in I}$.
  \EndFor
\end{algorithmic}
\end{algorithm}
\end{minipage}
\end{figure*}

Let us now turn to the min-max problem 
\begin{equation}
    \label{eq:minmax:II}
    V_{\phi_a}(\phi_b)=\inf_{T>0} \inf_\phi S_T[\phi]=\inf_{T>0}\inf_{\phi} \sup_{\theta} I_T(\phi,\theta)
\end{equation}
where $I_T(\phi,\theta)$ is the functional defined in~\eqref{eq:I:def} and the inner min-max is again to be performed over trajectories $\{\phi(t),\theta(t)\}_{t\in[0,T]}$ subject to the boundary conditions $\phi(0)=\phi_a$, $\phi(T)=\phi_b$.

In general a trajectory $\{\phi(t),\theta(t)\}_{t\in[0,T]}$ of finite time duration $T$ cannot solve~\eqref{eq:minmax:II}, i.e. we can always reduce the value of the inner min-max by increasing $T$. This lack of optimizer complicates the solution of~\eqref{eq:minmax:II}. To proceed, it is useful to follow the strategy of the geometric minimum action method (GMAM) in~\cite{VE2008gMAM}, and parametrize the physical time as $t(s)$ for $s\in [0,1]$. Writing $dt/ds=\lambda^{-1}(s)$, and introducing $(\hat \phi(s),\hat \theta(s))= (\phi(t(s)),\theta(t(s))$, the minimization over $T$ in~\eqref{eq:minmax:II} can be turned into a minimization over $\lambda$:
\begin{equation}
    \label{eq:min:max:2}
    V_{\phi_a}(\phi_b)=\min_{\lambda\ge0}\min_{\hat\phi}\max_{\hat \theta} \int_0^1 \left(\langle\hat\phi',\hat\theta\rangle -\lambda^{-1}H(\hat\phi,\hat\theta)\right) ds.
\end{equation}
where $\hat \phi' = d\hat \phi/ds$ and the  min-max  is  to be performed over the triple $\{\hat\phi(s),\hat\theta(s),\lambda(s)\}_{s\in[0,1]}$ subject to the boundary conditions $\hat\phi(0)=\phi_a$, $\hat\phi(1)=\phi_b$. Since we have added degrees of freedom by representing $T$ by the function $\lambda(s)$, we can add a constraint on the parametrization of $\hat \phi(s)$, e.g. by imposing that $|\hat \phi'|= cst$ (in $s$), in which case $s$ is normalized arclength along the path~$\hat\phi$. This choice is convenient numerically as it guarantees that discretization points in $s$ will be uniformly distributed along the path.

The min-max problem~\eqref{eq:min:max:2} is now in a form that can be solved using GDA, similar to what we did for~\eqref{eq:minmax:I}. As shown in Appendix~\ref{app:geometric_GDA}, it is however convenient to treat $\lambda$ separately, and this leads to the following GDA equations (compare~\eqref{eq:system_gradient_descent-ascent_hyperbolic})
\begin{align}
    \partial_\tau \hat\phi = \alpha\lambda \partial_s \hat \theta + \alpha\partial_{\hat\phi} H, \qquad 
    \alpha\partial_\tau \hat\theta =\lambda \partial_s\hat\phi - \partial_{\hat\theta} H,
    \label{eq:gMAM_xprime}
\end{align}
with $\lambda$ given by
\begin{align}
    \label{eq:lambda:def}
    \lambda = \frac{|\langle \partial_{\theta} H,  \partial_s\hat\phi \rangle| + \sqrt{\max(0,\langle \partial_{\theta} H, \partial_s\hat\phi \rangle^2 - 4 H |\partial_s\hat\phi|^2)}}{2|\partial_s\hat\phi|^2}.
\end{align}
Eqs.~\eqref{eq:gMAM_xprime} are to be solved under the constraint that $|\partial_s \hat \phi | = cst$ (in $s$, not $\tau$), with the boundary conditions (in $s$) 
\begin{align}
    \hat\phi(\tau,s=0)=\phi_a,\qquad
    \hat\phi(\tau,s=1)=\phi_b,
    \label{eq:boundary_conditions_hatphi}
\end{align}
for some initial conditions (in optimization time $\tau$)
\begin{align}
    \hat\phi(\tau=0,s)=\hat\phi^0(s),\qquad
    \hat \theta(\tau=0,s)=\hat\theta^0(s),
    \label{eq:initial_conditions_ggda}
\end{align}
with $\hat\phi^0(s)$ such that $\hat\phi^0(0) = \phi_a$ and $\hat\phi^0(1) = \phi_b$.  

Similar to what we did with~\eqref{eq:system_gradient_descent-ascent_hyperbolic}, the system of equations~\eqref{eq:gMAM_xprime} as well as the expression~\eqref{eq:lambda:def} for $\lambda$ can be put in a form suitable for numerical solution by rewriting them in terms of the fields $u = \hat \phi+\alpha \hat \theta $ and $ v = \hat \phi-\alpha \hat \theta $. For brevity we will not write these equations explicitly and refer the reader to Algorithm~2 for their discretized version.  At convergence, $\hat\phi$ and $\hat \theta$ satisfy
\begin{align}
    \lambda \hat\phi' = \partial_{\theta} H,\qquad \lambda \hat\theta' = -\partial_{\phi} H.
    \label{eq:gMAM_xprime_lambda}
\end{align}
If we insert the first of these equation in~\eqref{eq:lambda:def}, we can reorganize this equation into
\begin{align}
    \label{eq:lambda:conv}
    \lambda = \frac{\lambda |\hat \phi'|^2 + \sqrt{\lambda^2 |\hat \phi'|^4 - 4 H |\hat\phi'|^2 }}{2|\hat\phi'|^2}.
\end{align}
This equation shows that at convergence we must also have 
\begin{equation}
    \label{eq:H:zero}
    \forall s\in[0,1] \ : \ H(\hat\phi(s),\hat\theta(s)) = 0.
\end{equation}
This is consistent with the fact that the term involving $\lambda^{-1} H$ in~\eqref{eq:min:max:2} can also be interpreted as a Lagrangian multiplier term added to the objective function to enforce the constraint that $H=0$, see Ref.~\cite{VE2008gMAM} for details. 

In practice the  continuous paths $\hat \phi(\tau,s)$ and $\hat \theta(\tau,s)$ are discretized in $s$ on a grid of $M+1$ points with index $i\in I=\{0,\cdots,M\}$ such that $1=M\Delta s$, and we use the index $n\in \NN_0$ to encode their evolution with step size $\Delta \tau$ in artificial time $\tau$ so that for any field $\psi(\tau,s)$, $\psi^n_i\equiv \psi^n(i\Delta s)$. The details are given in Algorithm~2, which is a modified version of Algorithm~1 that includes the step of  reparametrization of the path. Note that we have kept the implicit upwind discretization for the advection term. Also, the reaction terms $f$ and $g$, and the coefficient $\lambda$ should be evaluated at the same grid point, as originating from the same term $\lambda^{-1}H(\hat \phi,\hat \theta)$ in \eqref{eq:min:max:2}. We found that evaluating these terms at the target grid point $i$ is better for stability in general.

Algorithms~1 and~2 are simple to implement and only require evaluating the first derivative of the Hamiltonian $H(\phi,\theta)$ with respect to its arguments, just as we would have to do to solve Hamilton's equations~\eqref{eq:H1}. It was used to calculate the paths shown in Figs.~\ref{fig:simple_example}, \ref{fig:maier_stein}, and \ref{fig:schlogl_2d}.  In the more complicated examples we treat below, because the state variable is a field that depends on space as well as time, and as a result the PDEs~\eqref{eq:u} and \eqref{eq:v} involve spatial derivative too, some additional consideration must be given to the way we discretize space and evaluate these derivatives to ensure that the resulting scheme is numerically stable. As usual with PDEs, this issue needs to be addressed on a case by case basis, depending of the nature of the PDE.

\section{First-Order Phase Transitions in a Nonequilibrium Ginzburg-Landau (GL) system}
\label{sec:modified_GL_mainSection}

A possible starting point of the method borrows from Landau's theory of phase transitions where, relying on the symmetries of the system, we postulate the free energy of the macroscopic variable of interest $\phi$ rather than deriving it from a microscopic distribution~\cite{landau1937theory,landau_phaseTransitions,ginzburg2009, de_gennes1971}. When the system is in equilibrium, this procedure is well understood~\cite{hohenberg1977,chaikin1995}.  For nonequilibrium systems, macroscopic fluctuation theory (MFT)~\cite{bertini_current_fluctuations_2006,bertini2015,baek2018} offers a generalization of Landau's approach where we directly starts from the action.  In what follows, we apply a similar approach to study phase transitions in a modified Ginzburg-Landau system.

\subsection{Nonequilibrium GL dynamics}
\label{subsec:modified_GL}

Following the general framework introduced in \cite{nardini2017prx,catesHouches2019}, the stochastic evolution of a non-conservative field $\rho(t,x)$ can formally be described by the Langevin equation 
\begin{align}
    \partial_t\rho = -\mu([\rho],x) + \sqrt{2\epsilon}\,\eta,
    \label{eq:SPDE_GL}
\end{align}
where the drift $\mu([\rho],x)$ can be interpreted as a chemical potential, $\eta(t,x)$ is a standard Gaussian white noise in space and time whose amplitude is measured by $\epsilon>0$, and where for simplicity we have set the mobility to 1. For a field $\rho$ defined on a one-dimensional domain, say $[0,1]$, the action corresponding to this equation is 
\begin{align}
    S_T[\rho]= \frac12 \int_0^T \int_0^1 |\partial_t\rho  + \mu([\rho],x)|^2 dx dt,
    \label{eq:action_GL}
\end{align}
and it will be the subject of our investigations.

The dynamics \eqref{eq:SPDE_GL} is in detailed balance when $\mu=\mu_E$, with $\mu_E$ such that it can be written as a derivative of a free energy functional $\mathcal F[\rho]$,
\begin{equation}
    \label{eq:mu:eq}
    \mu_E([\rho],x)=\frac{\delta \mathcal F[\rho]}{\delta \rho(x)}.
\end{equation}
In this case, at equilibrium, the field configurations are distributed according to the Gibbs measure associated with the free energy $\mathcal F$, at temperature $\epsilon$.

Here we will be mostly interested in the active version of dynamics \eqref{eq:SPDE_GL}, when $\mu=\mu_E+\mu_A$ with  $\mu_A$ that cannot be cast into the form~\eqref{eq:mu:eq}, or, equivalently, does not satisfy the functional Schwarz relation~\cite{grafke_cates2017,obyrne2020}:
\begin{align}
    \frac{\delta \mu_A([\rho],x)}{\delta \rho(y)}
    -\frac{\delta \mu_A([\rho],y)}{\delta \rho(x)}\not=0.
\end{align}
When $\mu_A\not=0$, the stationary distribution of the field configurations, if it exists, is a nonequilibrium distribution which is not available in closed form. 

For concreteness, we will focus on the following example that displays a nonequilibrium first-order phase transition: we assume that the field $\rho(t,x)$ is one-dimensional, with $x\in [0,1]$ and periodic boundary conditions, and we use chemical potentials given by
\begin{align}
    &\mu_E([\rho],x) = -(D\partial_x^2\rho+\rho-\rho^3 +h)
     \label{eq:mu_E}\\
    &\mu_A([\rho]) = -\kappa \int_0^1 \rho^2(y) dy,
     \label{eq:mu_A}
\end{align}
such that~\eqref{eq:SPDE_GL} becomes
\begin{align}
    \partial_t \rho=D\partial^2_x \rho +\rho-\rho^3 +h+\kappa \int_0^1 \rho^2(y)dy+  \sqrt{2\epsilon}\,\eta.
    \label{eq:spde_GL_extField}
\end{align}
Here $D>0$ is a diffusion constant which effectively depends on the system size in the dimensionless variables we use (reducing $D$ is equivalent to enlarging the domain size), $h$ the strength of an externally applied field, and $\kappa$ the strength of the nonequilibrium coupling. The chemical potential $\mu_E$ is borrowed from the Ginzburg-Landau $\phi^4$ theory, often referred to as Model A~\cite{hohenberg1977}.
A direct functional integration shows that $\mu_E = \delta \mathcal F/\delta \rho$ with
\begin{equation}
    \label{eq:muE:der}
    \mathcal F[\rho] = \int_0^1  \left(\tfrac12 D |\partial_x\rho|^2 + \tfrac14 (1-\rho(x))^2 - h \rho(x) \right) dx.
\end{equation}
The chemical potential $\mu_A$ was chosen because the phenomenology that it brings remains straightforward without being trivial. Specifically, $\mu_A$ acts  as an additional uniform applied field, that is active and depends nonlinearly on the value of $\rho$, rather than being externally imposed.  While $\mu_E$ drives the system towards the minimizers of the energy~\eqref{eq:muE:der}, which are homogeneous state solutions of $\rho-\rho^3+h$, $\mu_A$ homogeneously pushes the field upward when $\kappa>0$ and downward when $\kappa<0$. Therefore, the applied field $h$ and $\mu_A$ have competing effects when $h$ and $\kappa$ have opposite signs. Since there is a region in the $(\kappa,h)$ space where the noiseless dynamics has two stable fixed points (see Sec.~\ref{sec:pb}), this means that the system can undergo a nonequilibrium first-order phase transition for critical values of $h$ and $\kappa$ which we will determine in Sec.~\ref{sec:GL:mam}.

\begin{figure}
    \centering
    \includegraphics[width=0.9\columnwidth]{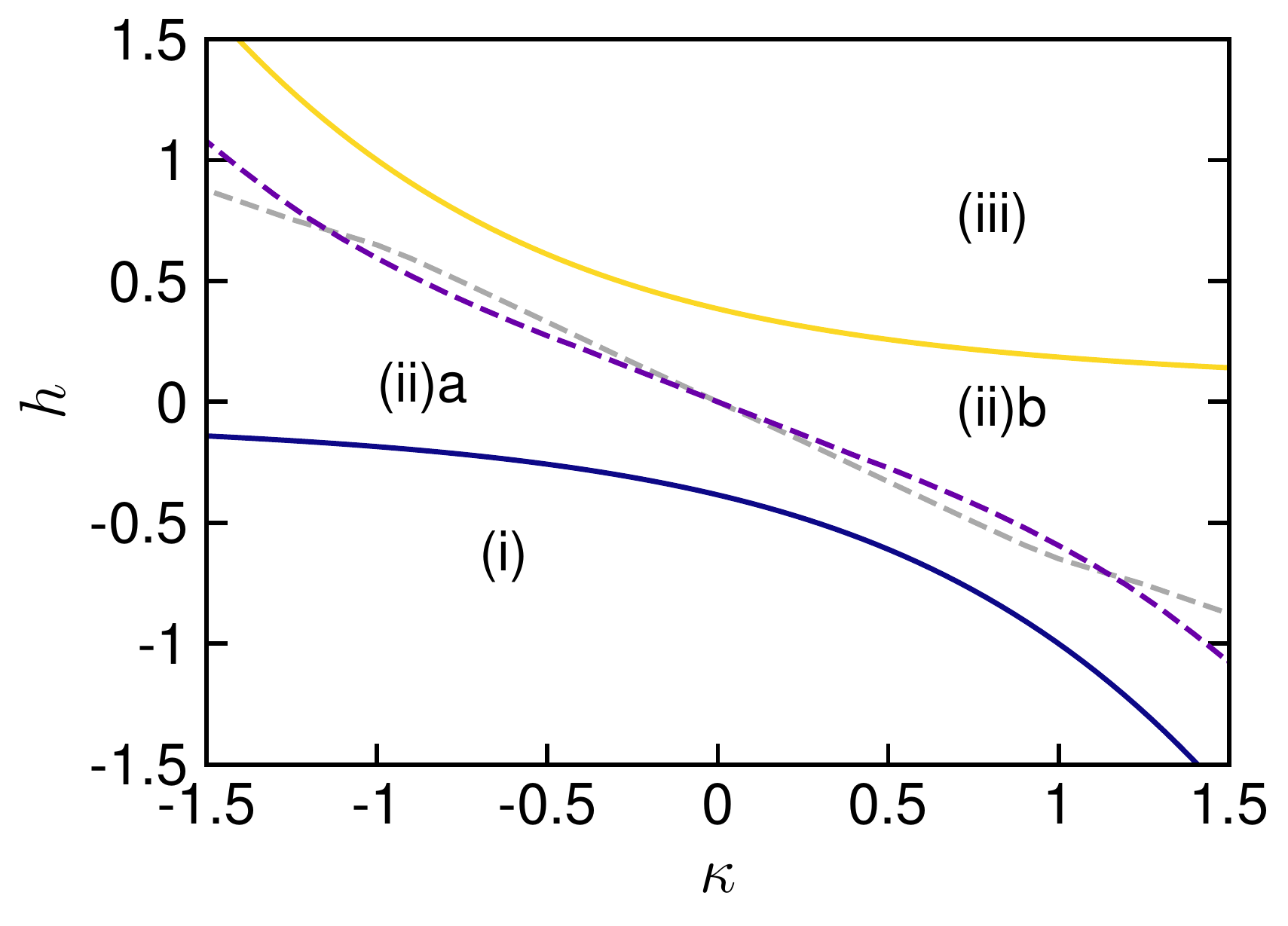}
    \caption{Phase diagram of the modified Ginzburg-Landau system. The phase diagram is divided in 4 regions. Region (i): $\rho_-$ is the only phase. Region (ii): coexistence region where the noiseless dynamics has two stable homogeneous fixed points $\rho_\pm$, and one unstable homogeneous fixed point $\rho_c$. Region (ii)a: $\rho_-$ is the stable phase. Region (ii)b: $\rho_+$ is the stable phase.  Region (iii): $\rho_+$ is the only phase. Yellow solid line: $h_c^+(\kappa)$. Purple solid line: $h_c^-(\kappa)$. Purple dashed line: phase transition curve between region (ii)a and (ii)b obtained by the minimum action method. Grey dashed line: phase transition curve obtained by treating the transition as if the system were in equilibrium (wrong prediction). }
    \label{fig:phaseDiagram_noneq_GL_gentle}
\end{figure}

\begin{figure*}
  \begin{tikzpicture}
    \path (0,0) node {    \includegraphics[width=0.68\columnwidth]{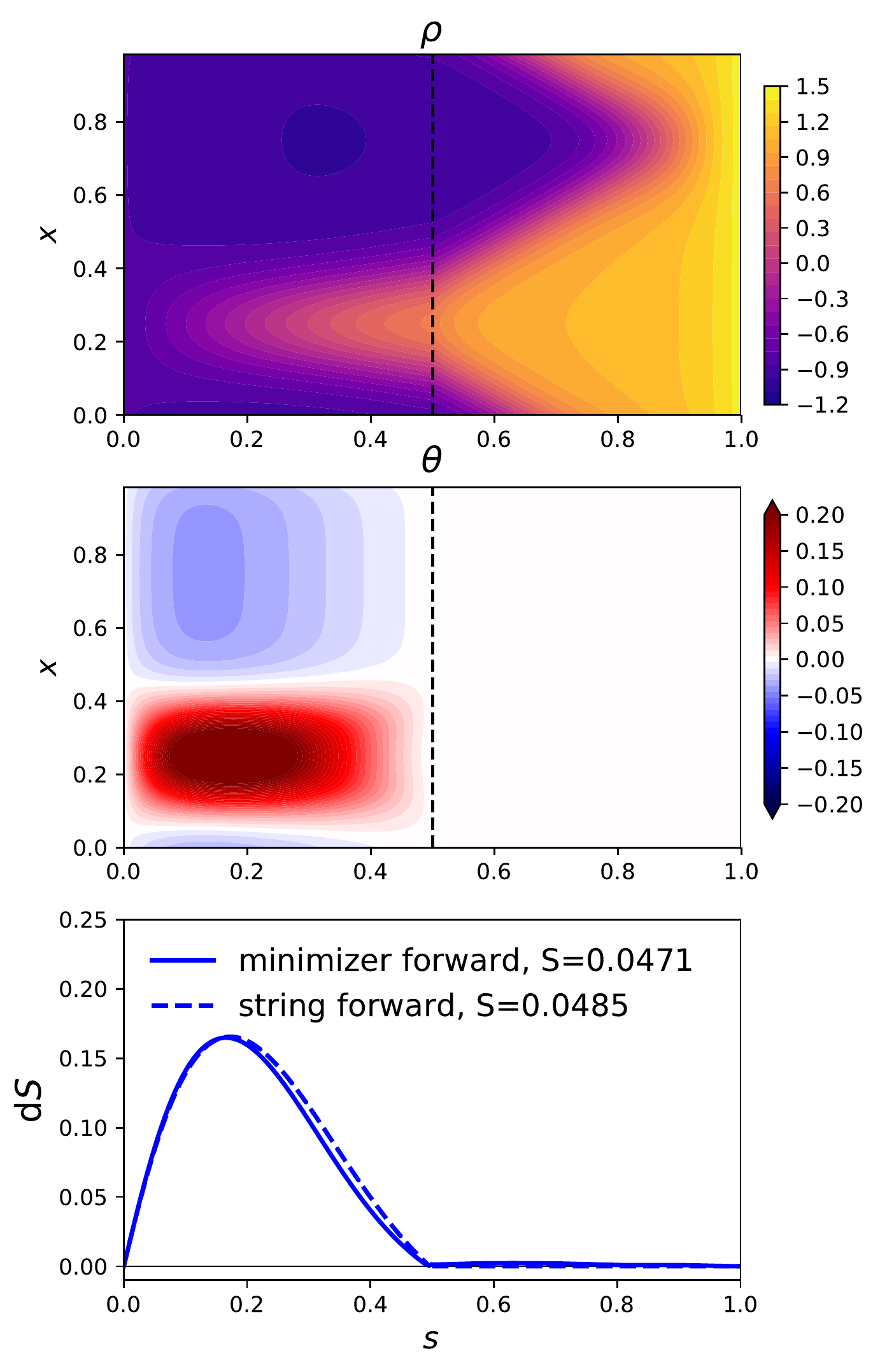}
    };
    \draw (-2.6,4.4) node[anchor=south west] {\bf a)};
  \end{tikzpicture}
  \hspace{-12pt}
  \begin{tikzpicture}
    \path (0,0) node {\includegraphics[width=0.68\columnwidth]{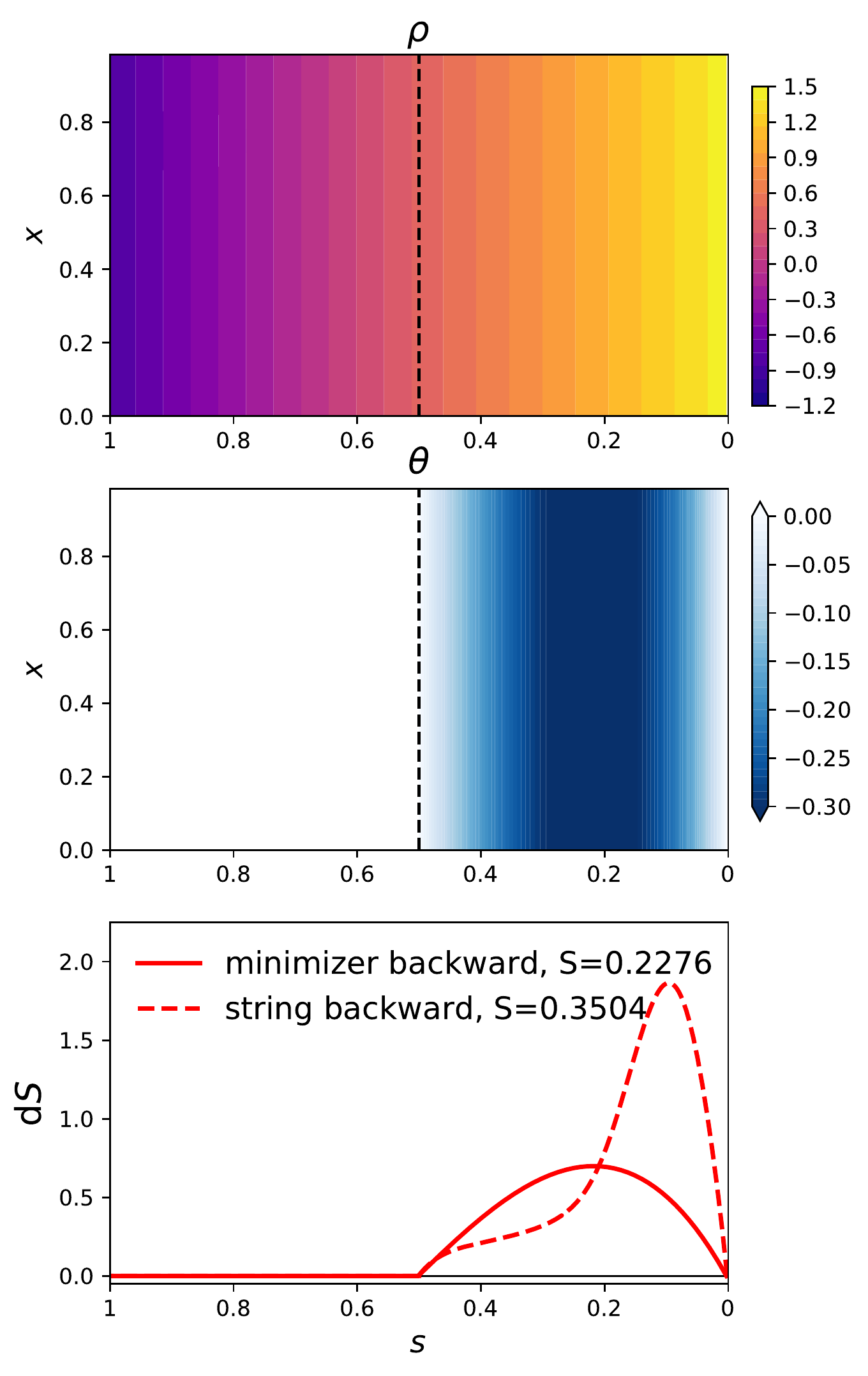}};
    \draw (-2.6,4.4) node[anchor=south west] {\bf b)};
  \end{tikzpicture}
  \hspace{-12pt}
  \begin{tikzpicture}
    \path (0,0) node {    \includegraphics[width=0.68\columnwidth]{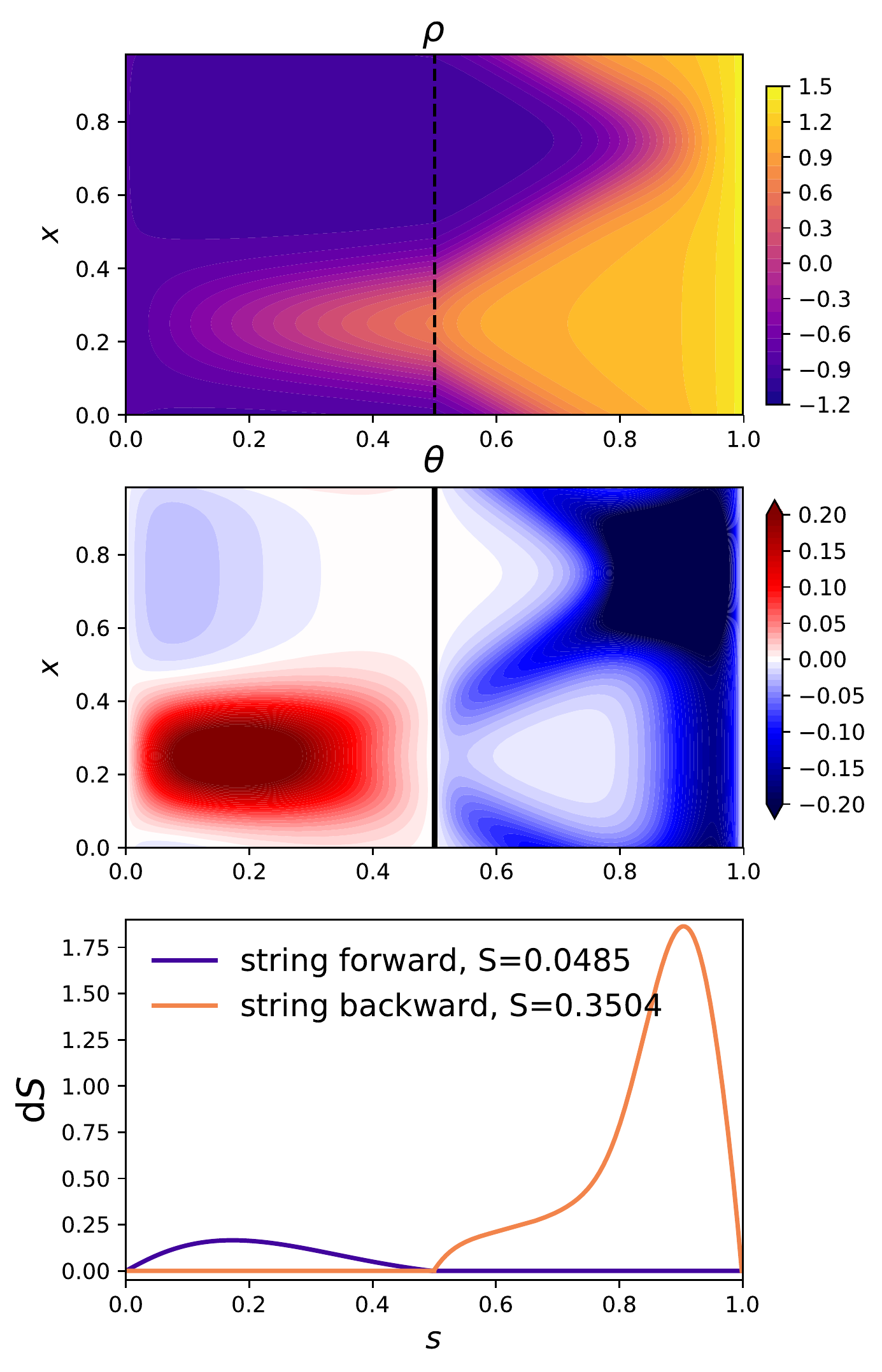}
    };
    \draw (-2.6,4.4) node[anchor=south west] {\bf c)};
  \end{tikzpicture}
    \caption{Nonequilibrium GL system for $D=5\times 10^{-3}$, $\kappa=1$, $h=-0.5$, hence $\rho_+=1.45161$ and $\rho_-=-0.854638$. For these values, we have $V_{\rho_-}(\rho_+)<V_{\rho_+}(\rho_-)$, indicating that $\rho_+$ is the stable phase. Upper panels: contourplots of the paths $\hat\rho(s,x)$; middle panels: contourplots  of the conjugate momentum $\hat \theta(s,x)$; lower panels: Lagrangian along the paths.  Panels a): minimum action path from $\rho_-$ to $\rho_+$; panels b): minimum action path from $\rho_+$ to $\rho_-$. Notice the strong difference between the forward and backward paths. The black dashed line at $s=0.5$ marks the critical nucleus, past which $\theta$ and $L$ are both zero as they should.   Panel c): heteroclinic orbit joining $\rho_-$ and $\rho_+$ calculated by the string method. Since the ascent is set as the reverse descent here, we plot on one graph the conjugate momentum and the Lagrangian in both the forward and reversed directions.}
    \label{fig:contour_modifiedGL}
\end{figure*}

\subsection{Phase boundaries for coexisting homogeneous fixed points}
\label{sec:pb}

Numerical evidence indicates that the stable fixed points of the noiseless dynamics (i.e. Eq.~\eqref{eq:spde_GL_extField} with $\epsilon=0$) are homogeneous states. As a result, they are solutions to $\rho-\rho^3 +h +\kappa \rho^2=0$. In the domain where this equation has three real roots, $\rho_-$, $\rho_c$  and $\rho_+$ with $\rho_-<\rho_c<\rho_+$, $\rho_-$ and $\rho_+$ are stable fixed points of the noiseless dynamics whereas $\rho_c$ is an unstable point.  The coexistence region in the parameter space $(\kappa, h)$ where both $\rho_-$ and $\rho_+$ are present is marked as region (ii) in Fig.~\ref{fig:phaseDiagram_noneq_GL_gentle}; it is where $h\in[h_c^-,h_c^+]$, with $h_c^-$ and $h_c^+$ given by 
\begin{align}
    h_c^-=-\frac{1}{27} \left(\sqrt{\kappa ^2+3}+\kappa \right)^2 \left(2 \sqrt{\kappa ^2+3}-\kappa \right),
\end{align}
shown as a blue line in Fig.~\ref{fig:phaseDiagram_noneq_GL_gentle}, and
\begin{align}
    h_c^+=\frac{1}{27} \left(\kappa -\sqrt{\kappa ^2+3}\right)^2 \left(2 \sqrt{\kappa ^2+3}+\kappa \right),
\end{align}
shown as a yellow line in Fig.~\ref{fig:phaseDiagram_noneq_GL_gentle}. 
Exactly on these boundaries, only two real roots coexist, and one state is thus marginally stable. In regions (i) and (iii) only one stable state exists.

Our next goal will be to analyze the relative stability of $\rho_-$ and $\rho_+$ under the effect of the noise, i.e. derive the phase diagram of the system. Even though we lack a  free energy that yields the stationary measure, we expect $\rho_+$ to be the stable phase for $h,\kappa>0$ and $\rho_-$ for $h,\kappa<0$. However, when $h$ and $\kappa$ have opposite signs (and thus opposite effects on $\rho$), determining the most likely phase becomes nontrivial.

\subsection{Extracting the minimum action paths}
\label{sec:GL:mam}

To assess whether $\rho_+$ or $\rho_-$ is the stable phase in the coexistence region, we will compute the difference of the minimal actions $\Delta S\equiv V_{\rho_-}(\rho_+)-V_{\rho_+}(\rho_-)$, obtained from the minimum action paths from $\rho_-$ to $\rho_+$ and vice-versa. The Hamiltonian entering the action is the one associated with Eq.~\eqref{eq:SPDE_GL}:
\begin{align}
    H(\rho,\theta) = \langle -\mu[\rho],\theta\rangle_{L^2} + \langle \theta,\theta\rangle_{L^2},
    \label{eq:hamiltonian_modified_GL}
\end{align}
where the scalar product of two functions $f$ and $g$ is given by $\langle f,g\rangle_{L^2}=\int_0^1 f(x) g(x) dx$. To obtain the phase diagram, these calculations must be repeated for a set of values $(\kappa,h)$ in the coexistence region to compute the minimal action difference as a function of these parameters, $\Delta S(\kappa,h)$: From Eq.~\eqref{eq:P:01}, the line of phase transition is then the curve where $\Delta S(\kappa,h)=0$. 

In practice we use Algorithm~2 with $M=400$ copies along the path, $N_x=64$ points of space discretization, $\Delta \tau=10^{-3}$, and $\alpha = 0.33$, and we monitor convergence by looking at the decay of the action.
The result of these computations is shown in Fig.~\ref{fig:phaseDiagram_noneq_GL_gentle}, where  the purple dashed line frontier in the phase diagram corresponding to $\Delta S(\kappa,h)=0$. The stable phase is $\rho_-$ below the line (region (ii)a) , and  $\rho_+$ above it (region(ii)b). 

For comparison, we also compute the phase diagram under the (wrong) assumption that the escape paths were given by the heteroclinic orbits followed in a time-reversed way. This would have to be the case in equilibrium by time-reversal symmetry. While these escape paths are incorrect in general in nonequilibrium systems, their respective cost in the action gives an upper bound on the actual minima $V_{\rho_-}(\rho_+)$ and $V_{\rho_+}(\rho_-)$. Denoting by $S^\mathrm{het}_{\rho_-}(\rho_+)$ and $S^\mathrm{het}_{\rho_+}(\rho_-)$ the  actions along the heteroclinic orbit, we compute for instance $S^\mathrm{het}_{\rho_-}(\rho_+)$ as
\begin{align}
    S^\mathrm{het}_{\rho_-}(\rho_+)= \int_0^1 \left( \< \partial_s \hat \rho,\hat \theta\> - \lambda^{-1} H(\hat \rho,\hat \theta)\right) ds,
\end{align}
where the path $\hat \rho(s)$ and the parametrization $\lambda(s)$ ($s\in[0,1]$) have been obtained by the string method~\cite{ERenVE2002,string2007} that identifies the heteroclinic orbit between $\hat\rho(0)=\rho_-$ and $\hat\rho(1)=\rho_+$, and $\theta$ solves
\begin{align}
    \lambda \partial_s \hat\rho=\partial_\theta H(\hat\rho,\hat\theta).
\end{align}
As a sanity check, we did verify that one always has $V_{\rho_-}(\rho_+)\leq S^\mathrm{het}_{\rho_-}(\rho_+)$ and $V_{\rho_+}(\rho_-)\leq S^\mathrm{het}_{\rho_+}(\rho_-)$, namely, that the minimizer of the action is always smaller than the action along the heteroclinic orbit. It is also worth noticing that these bonds offer no information about the location of the phase transition line:  The line where $\Delta S^\mathrm{het}(\kappa,h)=0$ is plotted as the grey dashed line in Fig.~\ref{fig:phaseDiagram_noneq_GL_gentle}, and it is different from the actual transition line $\Delta S(\kappa,h) =0$

The minimum action paths also give physical insights about the mechanism of the transition: the contour plot in $(s,x)$ space of these paths are shown in Fig.~\ref{fig:contour_modifiedGL} for the specific value $(\kappa,h)=(1,-0.5)$ (which is in region (ii)a): panel (a) shows the forward path from $\rho_-$ to $\rho_+$, and panel (b) the reversed path from $\rho_+$ to $\rho_-$. Also shown in panel (c) of the figure is the heteroclinic orbit. The actual path $\{\hat \rho(s)\}_{s\in[0,1]}$ are shown in the first row, while the second row displays the conjugate momentum $\{\hat \theta(s)\}_{s\in[0,1]}$, and the third the action increment (i.e. the Lagrangian) $\< \hat \theta(s),\hat \rho'(s)\>$ along the path.

As can be seen in Fig.~\ref{fig:contour_modifiedGL} the forward and the backward minimum action paths are different, and they cross the separatrix (marked as a dashed black line in the figure) at different places: that is, the critical nucleii for the forward and backward transitions are different (for the backward path this `critical nucleus' is actually flat).  This is a signature of time-symmetry breaking that can be intuitively explained as follows: When $\kappa>0$, as in Fig.~\ref{fig:contour_modifiedGL}, the nonequilibrium term $\kappa \int \rho^2dx $ favors the movement from $\rho_- $ to $\rho_+$ but opposes the one from $\rho_- $ to $\rho_+$. In the forward path from $\rho_-$ to $\rho_+$, it is  better to have  $\int \rho^2dx $ large, which favors nucleation;  conversely, in the backward path from $\rho_+$ to $\rho_-$, it is better to have  $\int \rho^2dx $ small, and the way to minimize this quantity given the value of its changing mean is to have $\rho$ spatially uniform. Notice however that this is a finite size effect: If $D$ were decreased to even smaller values, we would observe nucleation events in both directions (albeit different ones in each).

Note also that the interesting part of these minimum action paths is their escape half, where the noise is needed and the action increment is therefore positive: it is the first half for the forward path in panel (a) and the second half for the backward path in panel (b); for the heteroclinic orbit shown in panel (c), we display both halves at once since the forward and reversed paths are symmetric. In all situations, past the critical nucleus, the paths simply follow the noiseless dynamics, and this half of the path may not be unique if the critical nucleus has more than one unstable direction. This non-uniqueness has no impact on the action however, since the Lagrangian is zero along the solution of the noiseless dynamics.

The effective nonequilibrium Ginzburg-Landau-like dynamics we have considered in this section can be modified to include different nonequilibrium terms, like for example  $\mu_A=\kappa |\partial_x\rho|^2$ which naturally appears in the coarse-grained field description of interface growth phenomena~\cite{kardar1986} and active matter systems~\cite{cates2015, nardini2017prx, wittkowski2014, solon2018binodal}. We could also use our approach to compute the phase diagram of modified Cahn-Hilliard systems, which naturally emerge in active matter field theories~\cite{tailleur2008, nardini2017prx,catesHouches2019,grafke2017, solon2018binodal,obyrne2021}.


\section{Phase transitions in a bistable reaction-diffusion system}
\label{sec:schlogl}

\subsection{The Schl\"ogl model}
\label{sec:schlogl_A}
In 1972, Schl\"ogl introduced the following chemical reaction network~\cite{schlogl1972}
\begin{align}
\label{eq:chemical_reactions}
    A\xrightleftharpoons[k_1]{k_0}X, \qquad 2X+B \xrightleftharpoons[6k_3]{2k_2}3X,
\end{align}
with microscopic rates $k_i>0$, and where the concentration of $A$ and $B$ are held constant. In a certain regime of the reaction rates, this system displays metastability between a low density and a high density phase. Here we will consider a spatially extended variant of this model introduced by T\u{a}nase-Nicola and Lubensky~\cite{tanase2012}, in which a one dimensional domain is split into $L\in \NN$ well-stirred compartments: the molecules only react within their compartment, and randomly jump to neighboring ones with rate $\gamma>0$. We also impose periodic boundary conditions. 

When the number $n_i$ of molecules in compartment $i\in\{1,\dots,L\}$ is large, it is convenient to introduce the rescaled $\rho_i=n_i/\om$, where $\om\gg1$ is the typical number of molecules per compartments. In the limit as $\om\to\infty$, the law of mass action for $\rho_i$ is a (discrete) reaction-diffusion equation
\begin{equation}
    \label{eq:Schlogl:MF}
    \dot \rho_i = \gamma (\rho_{i+1}+\rho_{i-1}-2\rho_i) + w_+(\rho_i) - w_-(\rho_i)
\end{equation}
where $w_+(\rho_i)=\lambda_0+\lambda_2\rho_i^2$ and $w_-(\rho_i)=\lambda_1\rho_i+\lambda_3\rho_i^3$ are the rescaled reaction rates with $\lambda_i=k_i\om^{i-1}$ (see App.~\ref{app:detail_largeDeviation_schlogl}). \eqref{eq:Schlogl:MF} can be written as a gradient flow: 
\begin{equation}
    \label{eq:Schlogl:MF:2}
    \dot \rho_i = -\partial_{\rho_i} E(\rho), 
\end{equation}
where $\rho=(\rho_1,\dots,\rho_L)$ and we introduced 
\begin{equation}
    \label{eq:scholgl:E}
    E(\rho) = \sum_{i=1}^L \left(\tfrac12\gamma(\rho_{i+1}-\rho_i)^2 +U(\rho_i)\right)
\end{equation}
with
\begin{equation}
    U(\rho_i) = -\lambda_0\rho_i -\tfrac{1}{3}\lambda_2\rho_i^3 + \tfrac{1}{2}\lambda_1\rho_i^2 +\tfrac{1}{4}\lambda_3\rho_i^4.
\end{equation}
We recognize a (discrete) Ginzburg-Landau free energy, from which we conclude that the stable fixed points of \eqref{eq:Schlogl:MF} are homogeneous states as long as $\gamma$ is not too small. For the value of $(\lambda_0,\lambda_1,\lambda_2,\lambda_3)$  that we will consider here, there are two such fixed points,  $\rho_i=\rho_\pm$ for all $i\in\{1,\dots,L\}$, where $\rho_-<\rho_+$ are the largest and smallest roots of $w_+(z)=w_-(z)$.

The gradient structure of~\eqref{eq:Schlogl:MF:2} may suggest that $\rho_-$ is the stable phase if $E(\rho_-)<E(\rho_+)$ whereas $\rho_+$ is  if $E(\rho_-)>E(\rho_+)$. This conclusion is however incorrect, as already observed by T\u{a}nase-Nicola and Lubensky~\cite{tanase2012} who showed that the system undergoes a nonequilibrium  first-order phase transition when $\gamma$ changes. Since $E(\rho_\pm)=U(\rho_\pm)$ and, therefore, is independent of $\gamma$, the phase transition cannot be predicted by analyzing $E(\rho)$ only: the system is not in detailed balance with respect to the Gibbs measure associated to $E(\rho)$. What was not determined in~\cite{tanase2012} is the critical value $\gamma_c$ at which the phase transition occurs. This is the question we address next.

\begin{figure}
    \centering
    \includegraphics[width=0.8\columnwidth]{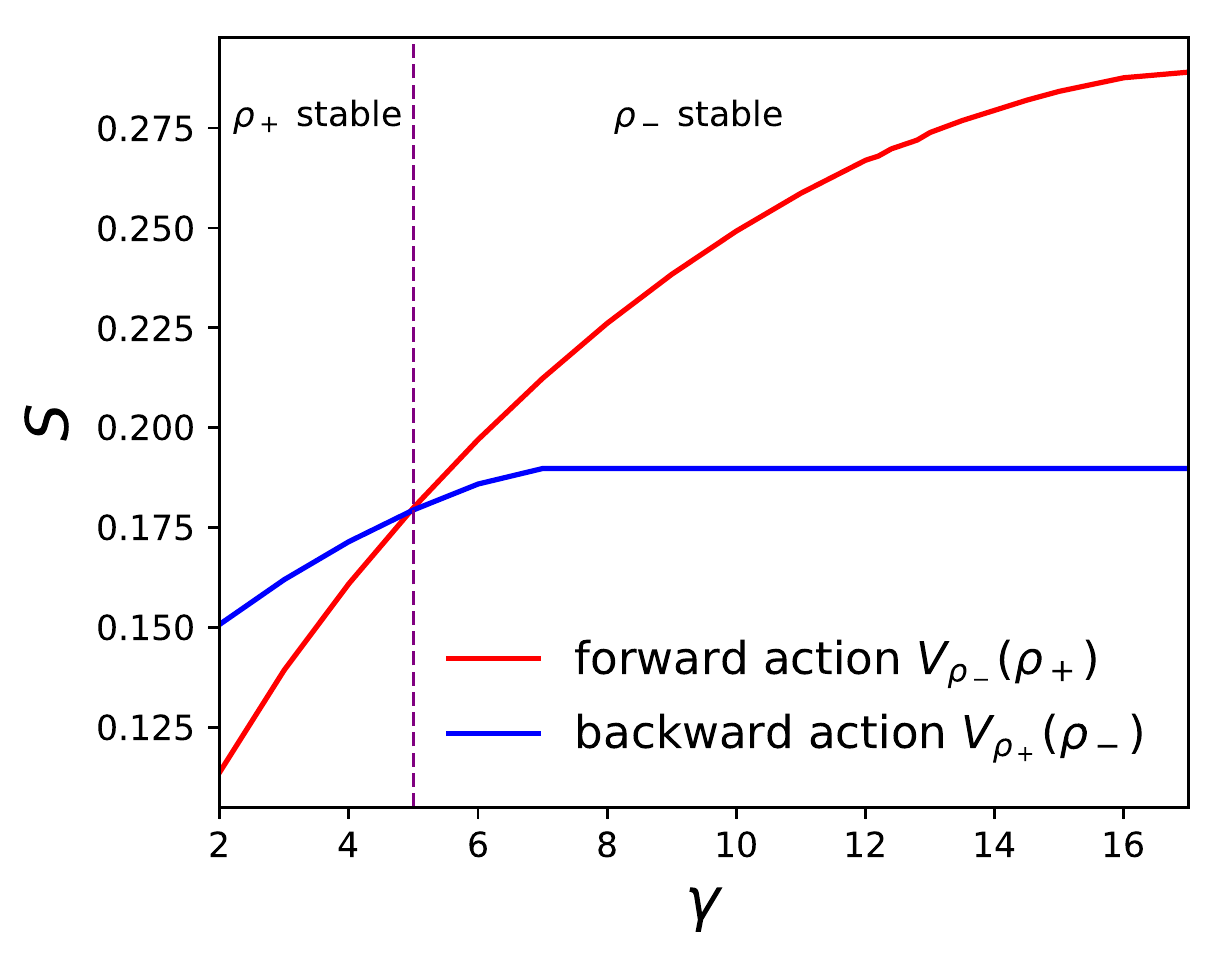}
    \caption{Phase diagram of the Schl\"ogl model in the limit of a large number of particles jumping between well-stirred reactive boxes. The parameters $\lambda_i$ are fixed and we vary the jump rate $\gamma$. For $\gamma<\gamma_c=5\pm0.1$, we have $V_{\rho_+}(\rho_-)>V_{\rho_-}(\rho_+)$, which indicates that $\rho_+$ is  stable state phase, whereas $\rho_-$ is for $\gamma>\gamma_c$. We use the parameters from~\cite{grafke2017}: $\lambda_0=0.8$, $\lambda_1=2.9$, $\lambda_2=3.1$, $\lambda_3=1$, and took $L=40$.}
    \label{fig:phase_diagram_schlogl}
\end{figure}

\begin{figure*}
  \begin{tikzpicture}
    \path (0,0) node {    \includegraphics[width=0.68\columnwidth]{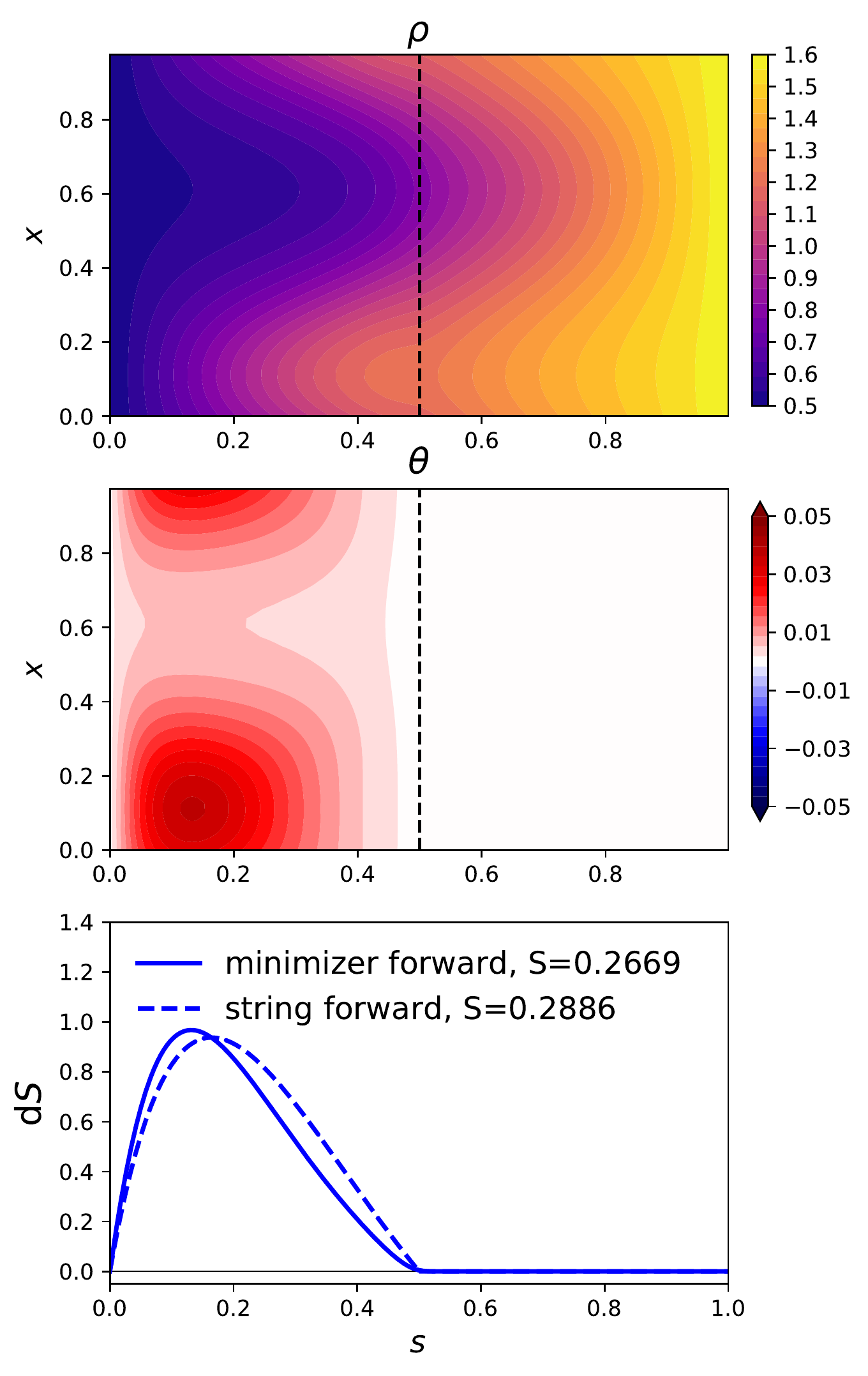}
    };
    \draw (-2.6,4.4) node[anchor=south west] {\bf a)};
  \end{tikzpicture}
  \hspace{-12pt}
  \begin{tikzpicture}
    \path (0,0) node {\includegraphics[width=0.68\columnwidth]{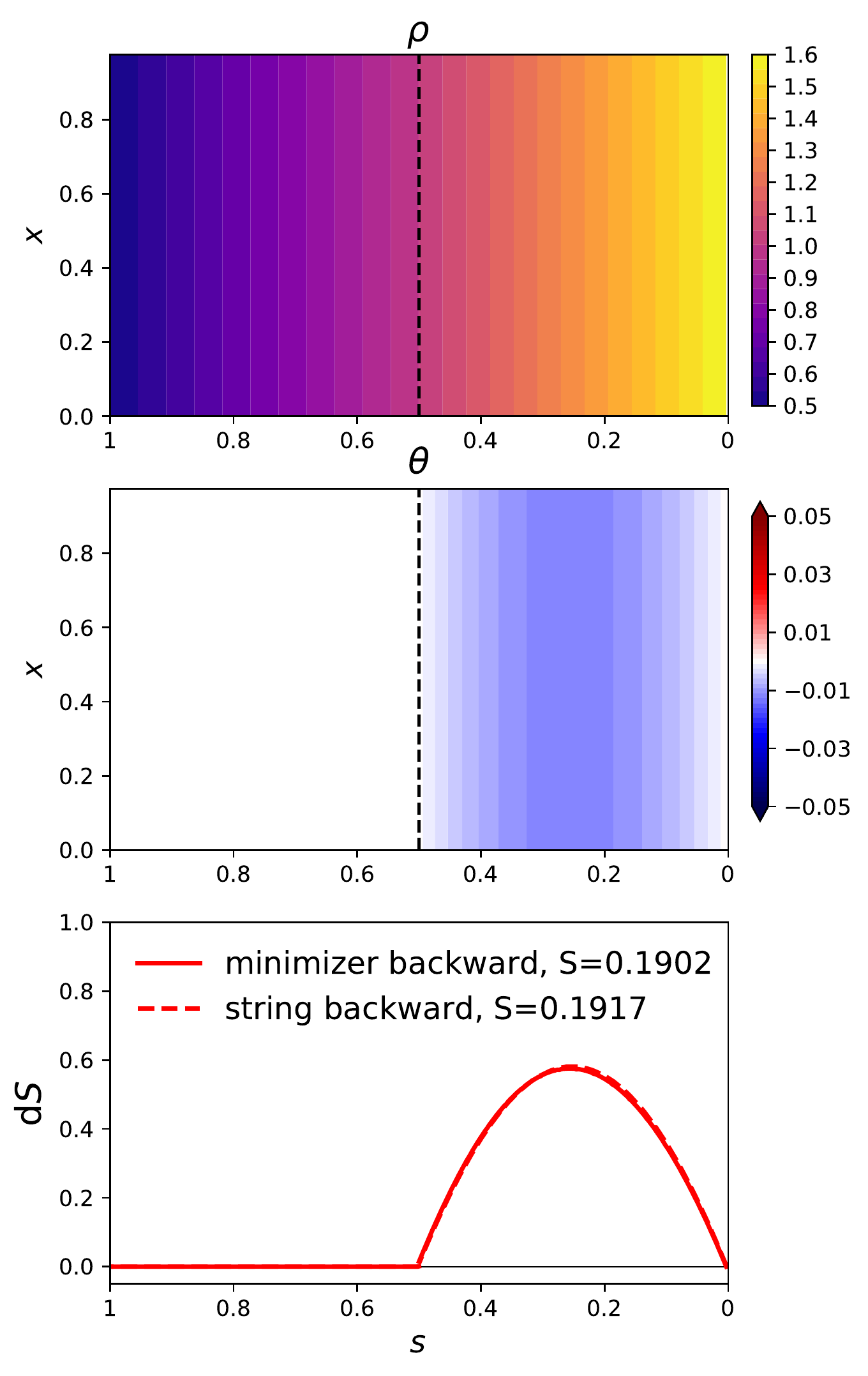}};
    \draw (-2.6,4.4) node[anchor=south west] {\bf b)};
  \end{tikzpicture}
  \hspace{-12pt}
  \begin{tikzpicture}
    \path (0,0) node {    \includegraphics[width=0.68\columnwidth]{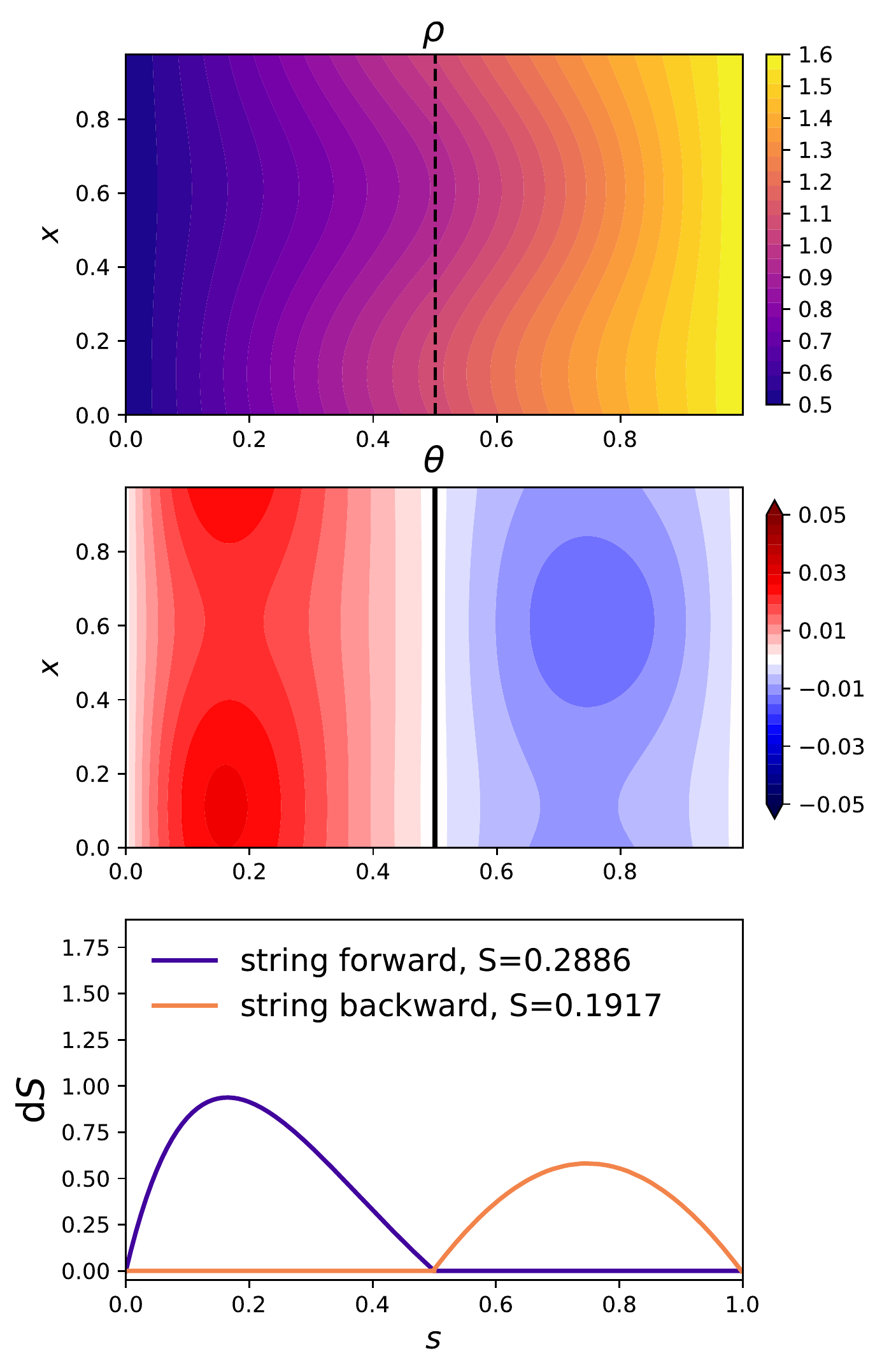}
    };
    \draw (-2.6,4.4) node[anchor=south west] {\bf c)};
  \end{tikzpicture}
  \caption{ Spatially extended Schl\"ogl model for $L=40$ reactive boxes, $(\lambda_0,\lambda_1,\lambda_2,\lambda_3)=(0.8,2.9,3.1,1)$, and $\gamma=12$. For these values we have $\rho_s=1$ and $\rho_+=1.6$. The figure is organized as Fig.~\ref{fig:contour_modifiedGL}. As  spatial coordinate, we use $x=i/L$ and plot the transitions paths and associated momenta as if they were continuous in space.
  }
    \label{fig:contour_schlogl_gamma12}
\end{figure*}

\subsection{Change of relative stability with increasing jumping rate}
\label{sec:change:stab:Schol}

The spatially extended Schl\"ogl model is a reaction network of the type considered in Sec.~\ref{sec:collect}, and its phase diagram can be analyzed by minimizing the action associated with a Hamiltonian similar to~\eqref{eq:Hreact}. Using the structure of the model,  it is natural to decompose $H=H^R+H^D$, with $H^R$ accounting for the reaction and  $H^D$ for the jumps:
\begin{align}
    H^R(\rho,\theta)&= \sum_{i=1}^L w_+(\rho_i)(e^{\theta_i}-1)+w_-(\rho_i)(e^{-\theta_i}-1), \label{eq:hamiltonian_reaction}\\
    H^D(\rho,\theta)&=\gamma\sum_{i=1}^L \rho_i (e^{\theta_{i-1}-\theta_i} + e^{\theta_{i+1}-\theta_i}-2),
    \label{eq:hamiltonian_diffusion}
\end{align}
We use this Hamiltonian in the action that we minimize using Algorithm~2 to calculate the quasipotentials $V_{\rho_-}(\rho_+)$ and $V_{\rho_-}(\rho_+)$. We repeat these calculations for different values of the jump rate $\gamma$ while keeping the rates fixed at $(\lambda_0,\lambda_1,\lambda_2,\lambda_3)=(0.8,2.9,3.1,1)$, for which $\rho_-=0.5$ and $\rho_+=1.6$. We use $L=40$ compartments and in Algorithm~2 we set  $M=400$ ($\Delta s = 2.5\times 10^{-3}$), $\alpha=1$, and $\Delta \tau=0.01$. The graphs of $V_{\rho_-}(\rho_+)$ and $V_{\rho_-}(\rho_+)$ versus $\gamma$ are shown in Fig.~\ref{fig:phase_diagram_schlogl}. These results indicate that the nonequilibrium first-order phase transition occurs at $\gamma_c\simeq5$: $\rho_+$ is the stable phase for $\gamma>\gamma_c$, while $\rho_-$ is the stable one for $\gamma<\gamma_c$. We stress again that this result cannot be deduced by looking at $E(\rho)$ even though the law of mass action can be written as the gradient flow~\eqref{eq:Schlogl:MF:2}. This is of course not a contradiction:  the Schl\"ogl model is not in detailed balance and lacks time reversal symmetry because this property also depends on the nature of the noise.

The non-equilibrium nature of the phase transition can be confirmed by looking at the transition paths from $\rho_-$ and $\rho_+$ and vice-versa. They are shown in Fig.~\ref{fig:contour_schlogl_gamma12}, in which we use the same plotting conventions as in Fig.~\ref{fig:contour_modifiedGL}. These results are for $\gamma=12$, when $\rho_-$ is the stable phase. We can see that the forward (panel a)  and the backward (panel b) paths are different and go through different critical nucleii (marked as a dashed vertical black line on the graphs). These path are also different from the heteroclinic orbit (panel c). If we increase the value of $\gamma$, the forward path eventually become homogeneous too (results not show): this is consistent with the fact that at high $\gamma$, the system behaves essentially as one single well-stirred compartment. In that limit we can calculate the nonequilibrium steady distribution of the system and use it to calculate $V_{\rho_-}(\rho_+)$ and  $V_{\rho_-}(\rho_+)$: this is done in App.~\ref{app:detail_largeDeviation_schlogl} and gives the same values as the ones obtained by Algorithm~2 when $\gamma$ is large and the transition paths are both homogeneous.  Conversely, if we decrease the value of $\gamma$, the backward path becomes inhomogeneous too (results not shown). This transition from homogeneous to inhomogeneous backward path occur around $\gamma=7$. How to gain intuition about these changes of behavior is harder on this model than in the GL model of Sec.~\ref{sec:modified_GL_mainSection} because the noise is non-Gaussian, and ultimately the shape of the minimum action paths depend on a complex interplay between many effects in the dynamics.

\subsection{Comparison with microscopic simulations}
\label{sec:micro}

\begin{figure}
    \centering
    \includegraphics[width=0.95\columnwidth]{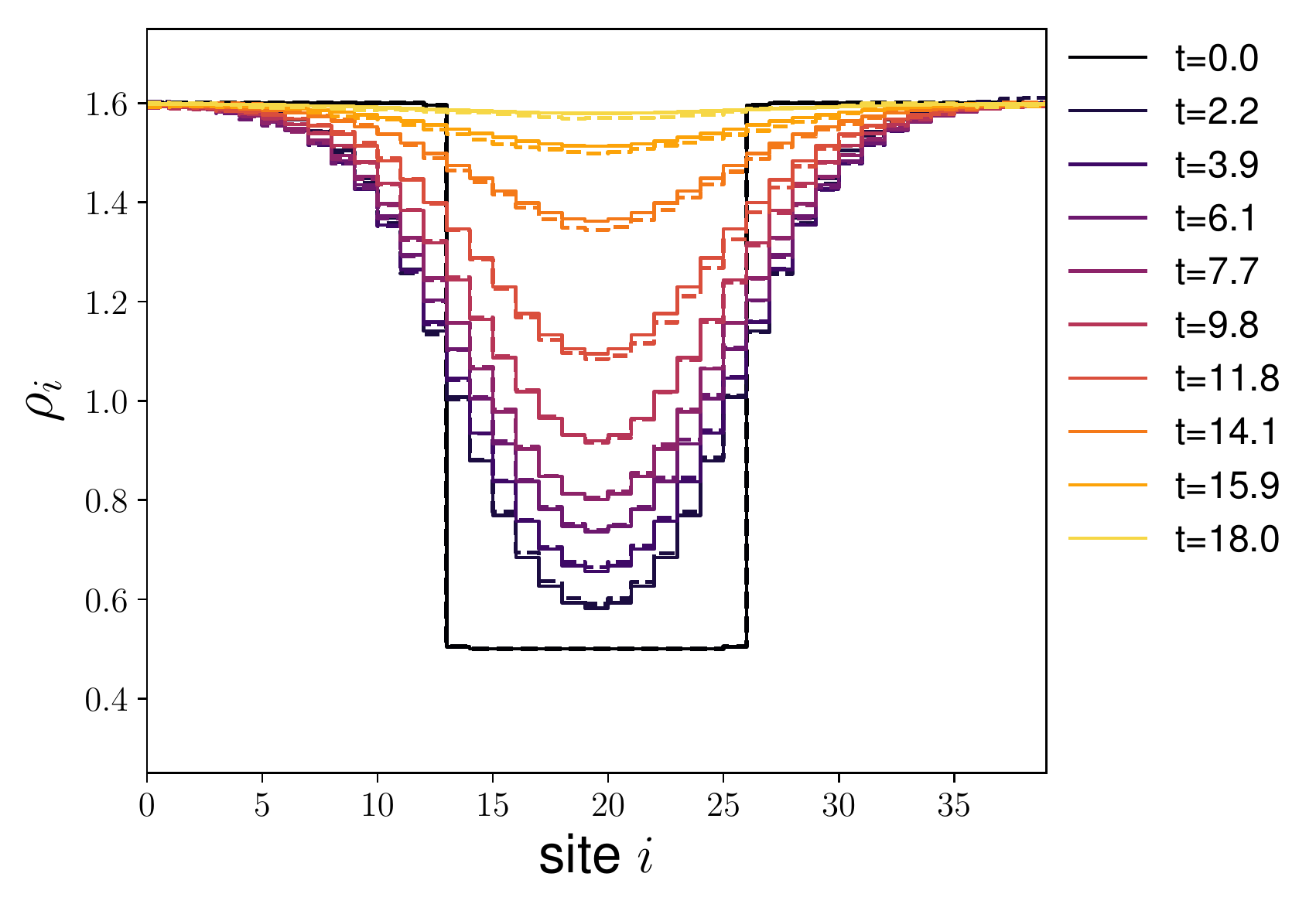}
    \caption{Spatially extended Schl\"ogl model: Relaxation of the microscopic system with a large number of particles per site starting from a step profile at $t=0$. The microscopic dynamics (dashed lines) closely follows the solution of~\eqref{eq:Schlogl:MF} (solid lines) with same initial condition, and converges to the stationary fixed point $\rho=\rho^+$ within the same physical time $t\simeq 18$. Parameters: $\lambda_0=0.8$, $\lambda_1=2.9$, $\lambda_2=3.1$, $\lambda_3=1$, $\gamma=4$, $L=40$, $\Omega=16\times 10^4$. }
    \label{fig:compare_downhill_simus_PDE}
\end{figure}
\begin{figure}
    \centering
      \begin{tikzpicture}
    \path (0,0) node {       \includegraphics[width=0.8\columnwidth]{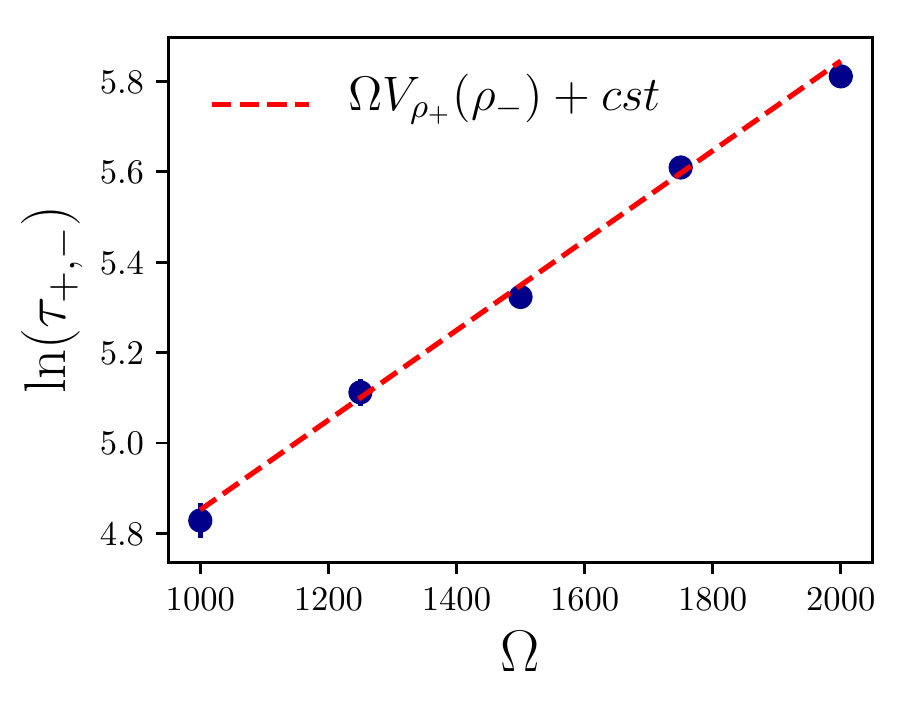}
    };
  \end{tikzpicture}
    \caption{ 
   Logarithm of the average nucleation time $\tau_{+,-}$ needed to reach the separatrix starting from a state metastable state $\rho_+$ close to linear instability, as a function of the typical number of particles $\om$ per site. In this regime, large deviation theory can be tested with Monte Carlo simulations in reasonable time. LDT does not predict the prefactor in front of the exponential scaling.
    Blue dots: results of MC simulations. Dashed line: slope predicted by theory $V_{\rho_+}(\rho_-)=9.931\times 10^{-4}$. Parameters: $\lambda_0=0.728$, $\lambda_1=2.9$, $\lambda_2=3.1$, $\lambda_3=1$, $\gamma=20$, $L=40$. }
    \label{fig:compare_logTau_N_deltaS}
\end{figure}

In this section we corroborate the conclusions of Sec.~\ref{sec:change:stab:Schol} by performing Markov Chain Monte Carlo (MCMC) simulations of the microscopic system. 

For the parameters value used in Sec.~\ref{sec:change:stab:Schol} it is not possible to calculate this way the phase diagram shown in Fig.~\ref{fig:phase_diagram_schlogl}: this is because $\om$ needs to be large in order for the phases $\rho_-$ and $\rho_+$ to be (meta)stable under the noise, and the timescales of the forward or backward transition between the phases $\rho_-$ and $\rho_+$ is of the order of the nucleation times $e^{\om V_{\rho_-}(\rho_+)}$ and $e^{\om V_{\rho_+}(\rho_-)}$. These timescales are too big to be accessible with MCMC. We can however perform two types of experiments to test the results of the minimum action principle:

First we can check that MCMC simulations of the microscopic system initiated with some nonuniform profile in the compartments behave as predicted by Eq.~\eqref{eq:Schlogl:MF}. The results of these simulations are shown in Fig.~\ref{fig:compare_downhill_simus_PDE} where we compare the evolution of a step profile: the microscopic system follows the deterministic dynamics~\eqref{eq:Schlogl:MF}, as expected.  To do these calculation, we fix $\om=1.6 \times 10^5$, set the number of molecules in each compartment $i\in\{1,\dots,L\}$ to be $n_i=\lfloor \rho_i\om \rfloor$ for all $i$, and simulate the microscopic dynamics exactly with  Gillespie algorithm~\cite{gillespie1976,elf_spontaneous_2004} using the microscopic rates $k_i=\lambda_i\om^{1-i}$. 

Second, to check the prediction of the minimum action principle in terms of nucleation times, we can change the rates $\lambda_i$ to make one of two phases (say $\rho_+$) only very weakly metastable, i.e. such that  $V_{\rho_+}(\rho_-) \ll 1$.  We can then calculate the mean escape time  $ \tau_{+,-}$ from this state towards $\rho_-$, repeat this calculation for different values of $\om$ and check that $\tau_{+,-} \asymp e^{\om V_{\rho_+}(\rho_-)}$. The result is shown in Fig.~\ref{fig:compare_logTau_N_deltaS}, which confirms that the prediction from the minimum action framework  explains  the microscopic simulations.

\subsection{Continuous limit}
\label{sec:cont:Schol}

Finally, let us consider the continuous space limit of the model by sending the number of compartments $L\to\infty$. 
To this end, let us set $\rho_i=\tilde\rho(x_i)/L$ and $\theta_i=\tilde\theta(x_i)$, with $x_i=i/L$ and where $\tilde\rho(x)$ and $\tilde\theta(x)$ are fields on $x\in[0,1]$. Let us also  set $\gamma =DL^2$ for some diffusion coefficient $D>0$, and $\lambda_i = \tilde\lambda_i L^{i-1}$ for some rescaled rates $\tilde \lambda_i$.  Assuming that $D$ and $\tilde\lambda_i$ are $O(1)$ in $L$, in the limit as $L\to\infty$, it is easy to verify that the Hamiltonians $H^R$ and $H^D$ in~\eqref{eq:hamiltonian_reaction} and ~\eqref{eq:hamiltonian_diffusion} now become
\begin{align}
\begin{split}
    \tilde H^R(\tilde\rho,\tilde\theta) &= \int_0^1 \left( \tilde w_+(\tilde\rho) (e^{\tilde\theta}-1)+\tilde w_-(\tilde\rho)(e^{-{\tilde\theta}}-1) \right) dx\\
    \tilde H^D(\tilde\rho,\tilde \theta) &= D \int_0^1 \left(\tilde\rho \partial^2_x \tilde\theta + \tilde \rho(\partial_x\tilde\theta)^2 \right) dx,
    \label{eq:continuous_hamiltonian_dean_kawasaki}
\end{split}
\end{align}
where we defined $\tilde w_+(\tilde\rho)=\tilde \lambda_0+\tilde \lambda_2\tilde\rho^2$, $\tilde w_-(\tilde\rho)=\tilde \lambda_1\tilde\rho+\tilde \lambda_3\tilde\rho^3$.
Note that in this limit the discrete Poisson jumps of the molecules are approximated as a Gaussian noise on the density: we have recovered the multiplicative Gaussian noise that appears in the Dean-Kawasaki equation~\cite{dean1996, lefevre2007}. The structure of the Poisson noise of the reaction is however left unaffected by the limit. 

We used Algorithm~2 to calculate transition pathways at continuous level by minimizing the action associated with $\tilde H = \tilde H^R + \tilde H^D$. The pathways (not shown) are not significantly different from those shown in Fig.~\ref{fig:contour_schlogl_gamma12} when $L=1/40$: this indicates that the system was already close to its continuous limit at that value of $L$.

\section{Conclusion}
\label{sec:conclusion}

In summary, the analysis of first-order phase transitions and other activated processes in nonequilibrium systems can be reduced to the minimization of an action in situations where these processes are rare and occur via reproducible pathways. The approach can be justified rigorously within the framework of large deviation theory (LDT), and minimum action principles can also be derived formally in other instances using e.g. the  Martin-Siggia-Rose-Janssen-De Dominicis~\cite{martin1973,*janssen1976,*de_dominicis1976} or the Doi-Peliti~\cite{doi1976, *peliti1985} formalisms. The minimum of the action can be used to generalize Arrhenius law, and its minimizer to explain the mechanism of the transitions, including the shape of the critical nucleus that serves as transition state. Concrete predictions however rest on our ability to solve this minimization problem, which often needs to be done numerically. 

Here we developed algorithms to perform these calculations, both in finite and infinite times. These algorithms are designed to be used directly within the Hamiltonian formulation of the action, which leads to a min-max problem, and do not require the user to calculate the Lagrangian beforehand. In particular it can be used for systems where the fluctuations are non-Gaussian and the Lagrangian is typically unavailable in closed form. The applicability of our method was tested on two nontrivial examples involving spatially extended systems undergoing nonequilibrium phase transitions: a modified Ginzburg-Landau equation perturbed by noise and a reaction-diffusion system based on the Schl\"ogl model. In both cases the method allowed us to calculate the phase diagram of the systems, compute the paths of the transitions, and identify the critical nucleus. 

We hope that our work will pave the way for a systematic approach to study activated processes in other interesting examples, like active matter systems undergoing a motility-induced phase separation or a flocking transition, or other systems which display phase transitions but are not in detailed balance. These applications will depend on the possibility to derive an appropriate minimum action principle, potentially through coarse-graining, which is a non-trivial question on its own.

\begin{acknowledgments}
  {We thank Jasna Brujic,  Tobias Grafke, Tobias K\"uhn, and Fr\'ed\'eric van Wijland for useful comments. This work was supported by the Materials Research Science and Engineering Center (MRSEC) program of the National Science Foundation under Grants No. NSF DMR-1420073 and in part by Grant No. NSF DMR-1710163.
  RZ would like to thank Laboratoire MSC Paris for hospitality. RZ and EVE would also like to thank Center for Data Science ENS Paris for hospitality.}
\end{acknowledgments}

\newpage
\appendix

\section{Convergence of the min-max for $\alpha$ small but finite}
\label{app:convergence_small_alpha}

We would like to show that if a path is stable with respect to the Lagrangian minimization, then the path is also stable with respect to the Hamiltonian min-max algorithm, for $\alpha$ small enough.
We start from~\eqref{eq:system_gradient_descent-ascent}, but in this section, we conveniently rescale artificial time $\tau= \tilde \tau/\alpha$, we set $\tilde \alpha=\alpha^2$ and drop the tilde such that the evolution equations read 
\begin{align}
\begin{cases}
    \partial_\tau x = \partial_t \theta +\partial_xH\\
    \alpha\partial_\tau\theta =\partial_t x -\partial_\theta H.
    \end{cases}
\end{align}

We focus on the dynamical system subjected to an additive Gaussian white noise, as presented in Eqs.~\eqref{eq:sde} and \eqref{eq:HWF}, i.e the Hamiltonian takes the form
\begin{align}
    H(x,\theta) = \langle b(x),\theta\rangle +\frac{1}{2}|\theta|^2.
\end{align}
We assume that a minimum action path $\{(x^*, \theta^*)\}_{t\in[0,T]}$ has been obtained. 
We look at a perturbed path $(x^*+X, \theta^*+\Theta)$ with $X=(x_1,\cdots, x_P)^T$ and $\Theta=(\theta_1,\cdots, \theta_P)^T$ the perturbations, and we assess the conditions for path relaxation to the minimum action path. The evolution of the perturbation reads
\begin{align}
    \begin{cases}
    \partial_\tau x_p =  \dot \theta_p
+\theta_j^* \dfrac{\partial^2 b^*_j}{\partial x_k \partial x_p}x_k + \theta_j \dfrac{\partial b^*_j}{\partial x_p}\\
\alpha \partial_\tau \theta_p =  \dot x_p - \dfrac{\partial b^*_p}{\partial x_k}x_k - \theta_p,
\end{cases}
\end{align}
where we have used the Einstein convention for the sum on repeated indices, and the shorthand notations $\frac{\partial b^*_p}{\partial x_k}=\frac{\partial b_p}{\partial x_k}|_{x^*}$ and $\frac{\partial^2 b^*_j}{\partial x_k\partial x_p}=\frac{\partial^2 b_j}{\partial x_k\partial x_p}|_{x^*}$, for any indices $j$, $k$, $p$.

The evolution of the fields can now be cast into the following form:
\begin{align}
\begin{cases}
    \partial_\tau X= \mathcal L_1 X+\mathcal L_2 \Theta\\
    \partial_\tau\Theta = \alpha^{-1}(\mathcal L_3 X -\Theta)
    \end{cases}
    \label{eq:operator_hamiltonian_evolution}
\end{align}
where 
\begin{align}
    (\mathcal L_1)^{(pk)} &= \theta_j^* \dfrac{\partial^2 b^*_j}{\partial x_k \partial x_p} \\
    (\mathcal L_2)^{(pk)} &= (\delta_{pk}\partial_t + \frac{\partial b_k^*}{\partial x_p})\\
     (\mathcal L_3)^{(pk)} &= (\delta_{pk}\partial_t - \frac{\partial b_p^*}{\partial x_k})
\end{align}
In the Lagrangian algorithm, the equation $\Theta=\mathcal L_3 X$ is always verified, so the evolution of the perturbation $X$ is simply given by 
\begin{align}
    \partial_\tau X=(\mathcal L_1 +\mathcal L_2\mathcal L_3)X.
\end{align}
Any perturbation $X$ vanishes  if the eigenvalues of the operator $\mL=\mathcal L_1 +\mathcal L_2\mathcal L_3$ are all of negative real part.  The operator $\mL$ is self-adjoint (since $\mL$ is the Hessian of the action $S$), and one can thus extract a basis of normalized orthogonal eigenvectors $X_n^\mL$. Let us denote by $\mu_n$ the n-th eigenvalue and $X_n^\mL$ the corresponding eigenvector. We have $\mL X_n^\mL = \mu_n X_n^\mL$, with $\mu_n<0$.

Now, in the Hamiltonian algorithm, we would like to find the conditions under which any perturbation $(X,\Theta)$ close to a path of minimum action vanishes when evolving in artificial time $\tau$.
The perturbation will vanish if and only if the eigenvalues $\lambda_n$ of the linear operator given in Eq.~\eqref{eq:operator_hamiltonian_evolution} have a negative real part.
The eigenvalue $\lambda_n$ associated to the eigenvector $(\Theta_n,X_n)^T$ should verify
\begin{align}
\begin{cases}
     \mathcal L_1 X_n+\mathcal L_2 \Theta_n =\lambda_n X_n \\
     \alpha^{-1}(\mathcal L_3 X_n -\Theta_n) =\lambda_n \Theta_n.
     \label{eq:eigensystem_0}
\end{cases}
\end{align}
The second equation in \eqref{eq:eigensystem_0} yields $\Theta_n= (1+\lambda_n\alpha)^{-1}\mathcal L_3 X_n$, assuming that there exists $\alpha>0$ such that $(1+\lambda_n\alpha)\neq0$ for every $n$.
Injecting this result into the first equation of \eqref{eq:eigensystem_0} yields a closed equation for $X_n$ and $\lambda_n$
\begin{align}
    (1+\lambda_n\alpha)(\mathcal L_1 X_n-\lambda_n X_n) +\mathcal L_2 \mathcal L_3 X_n =0.
    \label{eq:Xn_operator_before_expansion}
\end{align}
For $\alpha\ll 1$, the system~\eqref{eq:operator_hamiltonian_evolution} can be seen as a perturbation of the Lagrangian problem, where an additional degree of freedom $\Theta$ relaxes to $\mathcal L_3 X$ on a fast time scale $1/\alpha$.  
This suggests to look for eigenvalues with a specific form: (i) a first set of eigenvalues $\lambda_n^{(1)}$ should be the perturbed eigenvalues $\mu_n$ with perturbed $X_n^\mL$ as corresponding eigenvectors, (ii) a second set of eigenvalues $\lambda_n^{(2)}$ is expected to scale as $O(\alpha^{-1})$ and encodes the fast relaxation of the variable $\Theta$ to $\mathcal L_3 X$.

Therefore we look for eigenvalues of a general form 
\begin{equation}
\label{eq:eig:expan}
    \lambda_n=\frac{\eta_n}{\alpha}+\nu_n + O(\alpha)
\end{equation}
with $\eta_n$ and $\nu_n$ of $O(1)$. Expanding at leading order $O(\alpha^{-1})$ in \eqref{eq:Xn_operator_before_expansion} yields
\begin{align}
    \eta_n(1+\eta_n)X_n=0,
\end{align}
implying that $\eta_n=0$ or $\eta_n=-1$. The case $\eta_n=0$ leads us to consider the first case (i) mentioned above, in which we look for eigenvalues $\lambda_n^{(1)}=\mu_n+\alpha \xi_n$, with $\xi_n=O(1)$ and we expand \eqref{eq:Xn_operator_before_expansion} to leading order. At order $0$ in $\alpha$, we find that $X_n$ must solve
\begin{align}
    (\mL_1+\mL_2\mL_3) X_n -\mu_n X_n=0,
\end{align}
where we recognize the operator $\mL=\mL_1+\mL_2\mL_3$, which confirms that we expand around the eigenvalues $X_n^\mL$ of the Lagrangian system. We write $X_n=X_n^\mL+\alpha Y_n$ with $\|Y_n\|=O(1)$.
Now expanding \eqref{eq:Xn_operator_before_expansion} at order 1 in $\alpha$, and using the relation $\mL X_n^\mL=\mu_n X_n^\mL$, we get
\begin{align}
    \mL Y_n-\mu_n Y_n = \xi_n X_n^{\mL}+\mu_n \mL_2\mL_3 X_n^{\mL}.
    \label{eq:expand_first_order}
\end{align}
By taking the scalar product with the eigenvector $X_n^\mL$ on both sides of \eqref{eq:expand_first_order}, we obtain
\begin{align}
    \xi_n=- \mu_n \langle X_n^{\mL},\mL_2 \mL_3 X_n^{\mL}\rangle.
\end{align}
The brackets $\<\cdot,\cdot\>$ stand for the scalar product in $L^2([0,T];{\mathbb R}^P)$ which is given by 
\begin{align}
    \langle\phi,\psi\rangle = \int_0^T  \phi_p(t)\psi_p(t) dt.
\end{align}
Using the fact that $\mL_2^*=-\mL_3$, we have
\begin{align}
    \xi_n= \mu_n \| \mL_3 X_n^{\mL}\|^2,
\end{align}
where $\|\cdot\|$ is the norm associated with $\<\cdot,\cdot\>$.
Inserting this result as well as $\eta_n=0$ in \eqref{eq:eig:expan}, we deduce that 
\begin{align}
\label{eq:set2}
    \lambda^{(1)}_n = \mu_n(1+\alpha \| \mL_3 X_n^{\mL}\|^2 +O(\alpha^2)).
\end{align}
This shows that for $\alpha$ small enough, the sign of the real part of $\lambda_n$ and $\mu_n$ are the same.

Now,  in case (ii) where $\eta_n=-1$, at leading order 1 in $\alpha$  \eqref{eq:Xn_operator_before_expansion} becomes
\begin{align}
    -\nu_nX_n=\mL_2\mL_3X_n.
\end{align}
Hence using again the fact that $\mL_2^*=-\mL_3$, $\nu_n$ verifies
\begin{align}
    \nu_n=\frac{\|\mL_3 X_n\|^2}{\|X_n\|^2},
\end{align}
where $X_n$ is an eigenvector of $\mL_2\mL_3$. Therefore we have another set of eigenvalues given by
\begin{align}
\label{eq:set1}
    \lambda^{(2)}_n = -\alpha^{-1} +\| \mL_3 X_n^{\mL_2\mL_3}\|^2 +O(\alpha),
\end{align}
with $X_n^{\mL_2\mL_3}$ an eigenvector of $\mL_2\mL_3$. Since $\lambda^{(2)}_n <0$ if $\alpha$ is small enough, the associated eigenvectors are always stable. Therefore the stability of the fixed points of the Hamiltonian system is determined by the sign of $\lambda^{(1)}_n$, which is the same as the sign of the eigenvalues $\mu_n$ of the Lagrangian system for $\alpha$ small enough.

\section{Convergence of the GDA in an analytically solvable case }
\label{app:analytical_solution_GDA}

To gain insight about the convergence of the GDA algorithm, we consider an Ornstein-Uhlenbeck (OU) process in one dimension for which the evolution equations of the GDA are amenable to analytic solution. The example is also relevant since it is the dynamics verified by each Fourier mode of a freely diffusive field in a one-dimensional periodic box. The stability of the numerical scheme associated with this example will be analyzed in Appendix~\ref{app:stability_numerical_scheme}.

For an OU process, the Hamiltonian is given by
\begin{align}
    H(x,\theta) = -\zeta x\theta+\frac{1}{2}\theta^2,
\end{align}
where $\zeta>0$ is the stiffness of the confining harmonic potential. We look for the instanton joining  $x_a$ to $x_b\not=x_a$, with $x_a$ possibly nonzero (i.e. not the stable point of the deterministic dynamics).
The GDA equations read
\begin{align}
\begin{cases}
    \partial_\tau x = \alpha(\partial_t \theta -\zeta\theta)\\
    \alpha\partial_\tau \theta = \partial_t x +\zeta x -\theta,
    \label{eq:app_OU_GDA}
\end{cases}
\end{align}
subject to the boundary conditions
\begin{align}
    x(\tau,0)=x_a,\quad x(\tau,T)=x_b,
    \label{eq:app_BC_x}
\end{align}
and to the initial conditions 
\begin{align}
    x(0,t)=x_0(t),\quad \theta(0,t)=\theta_0(t).
    \label{eq:app_IC_x_theta}
\end{align}
Again, as in the body of text, we have introduced a scale $\alpha>0$ that controls the relative evolution of $x$ and $\theta$ in time $\tau$. 
By taking the derivative with respect to $\tau$ of the first equation in~\eqref{eq:app_OU_GDA} and applying the operator $\partial_t - \zeta$ to the second equation, a closed PDE for the function $x(\tau,t)$ can be obtained:
\begin{align}
    \partial_\tau^2 x = -\frac{1}{\alpha}\partial_\tau x +\left( \partial_t^2 x -\zeta^2 x\right).
    \label{eq:app_OU_1}
\end{align}
For $\alpha>0$, we set $\gamma(t)=x_a(1-t/T)+x_b t/T$ and define
\begin{align}
    \varphi(\tau,t)\equiv x(\tau,t) - \gamma(t)
    \label{eq:app_OU_split_x}
\end{align}
such that $\varphi(\tau,0)=\varphi(\tau,T)=0$.  To proceed, let us use the Fourier decomposition of $\varphi$:
\begin{align}
    \varphi(\tau,t) =\sum_{q=1}^\infty  \varphi_q(\tau)\sin(q\pi t/T),
    \label{eq:app_OU_fourier_phi}
\end{align}
where we have taken the convention
\begin{align}
    \varphi_q(\tau) = \frac{2}{T}\int_0^T  \varphi(\tau,t) \sin(q\pi t/T) dt.
\end{align}
Inserting \eqref{eq:app_OU_split_x} and \eqref{eq:app_OU_fourier_phi} into \eqref{eq:app_OU_1} and projecting, we arrive at the equation verified by the modes $\varphi_q$ ($q\in \mathbb{N}$):
\begin{align}
    \alpha \partial_\tau^2\vphiq + \partial_\tau\vphiq + \alpha \omega_q^2 \vphiq = \alpha \zeta^2\frac{2(x_a-x_b(-1)^q)}{\pi q},
\end{align}
with $\omega_q^2=\zeta^2+(\pi q/T)^2$.
We recognize the equation for the damped harmonic oscillator, implying that the modes will eventually decay exponentially in $\tau$ to their stationary value. Setting 
\begin{equation}
    \Delta_q\equiv 1-4\alpha^2\omega_q^2
\end{equation}
and defining
\begin{align}
    \bar\varphi_q = \frac{\zeta^2(x_a-x_b(-1)^q)}{\pi q \omega_q^2}.
\end{align}
we have: for $\Delta_q>0$
\begin{equation}
    \begin{aligned}
    \vphiq(\tau) = \bar\varphi_q &+ C_{1q} e^{- \tau/(2\alpha)}  e^{\sqrt{\Delta_q}\tau/(2\alpha)} \\
    & +C_{2q}e^{- \tau/(2\alpha)}  e^{-\sqrt{\Delta_q}\tau/(2\alpha)};
\end{aligned}
\end{equation}
for $\Delta_q<0$, 
\begin{equation}
\begin{aligned}
    \vphiq(\tau) = \bar\varphi_q &+ C_{1q} e^{- \tau/(2\alpha)}   \cos\big(\sqrt{|\Delta_q|}\tau/(2\alpha)\big) \\ &+C_{2q}e^{- \tau/(2\alpha)}  \sin\big(\sqrt{|\Delta_q|}\tau/(2\alpha)\big);
\end{aligned}
\end{equation}
 and for $\Delta_q=0$,
\begin{align}
    \vphiq(\tau) =  \bar\varphi_q + (C_{1q} + C_{2q} \tau) e^{-\tau/(2\alpha)}.
\end{align}
In each cases the constants $C_{1q}$ and $C_{2q}$ are then determined by the initial conditions $x_0(t)$ and $\theta_0(t)$.

This computation addresses the question of the convergence rate, which is different for each mode $q$. For a given $q\geq 1$, the best decay rate $\lambda_q$ is obtained when choosing $\alpha=1/(2\omega_q)$, which corresponds to the damped critical regime. In practice, we need to use the same $\alpha$ for  all modes $q$. The convergence to the final path is then determined by the smallest rate, which must be chosen as large as possible. Since the decay rates $\lambda_q$ are ordered according to $\lambda_1\leq \lambda_2\leq...\leq \lambda_q$, this prescribes the choice of $\alpha=1/(2\omega_1)$ that maximises the convergence rate of the mode $q=1$, for which we have $\lambda_1=\omega_1$. This also yields identical rate $\lambda_q=\omega_1$ for all modes $q$ and the modes $q>1$ will display damped oscillations.

\section{Stability of the numerical scheme}
\label{app:stability_numerical_scheme}

In this section, we analyze the stability of Algorithm 1 for the Ornstein-Uhlenbeck system introduced in Appendix~\ref{app:analytical_solution_GDA}.
Starting from~\eqref{eq:app_OU_GDA}, where $x$ is subject to the boundary conditions~\eqref{eq:app_BC_x}, we set $u=x+\alpha\theta$ and $v=x-\alpha \theta$. The GDA equations in these variables becomes
\begin{align}
    \begin{cases}
    \partial_\tau u = \partial_t u + f(u,v)\\
    \partial_\tau v = -\partial_t v + g(u,v)
    \end{cases}
    \label{eq:app_u_v_hyperbolic}
\end{align}
with 
\begin{align}
    f(u,v) &= \left(\zeta +\frac{1}{\alpha}\right) v -\frac{1}{\alpha}u\\
    g(u,v) &=  \left(-\zeta  + \frac{1}{\alpha} \right) u - \frac{1}{\alpha} v,
\end{align}
and with the boundary conditions
\begin{align}
    v(\tau,t=0)&= -u(\tau,t=0)+2x_a\\
    u(\tau,t=0)&= -v(\tau,t=0)+2x_b.
\end{align}

Since \eqref{eq:app_u_v_hyperbolic} are linear equations, Algorithm 1 can be  expressed in terms of matrix multiplications.
To this end, let us  define $r=\dtau/\Delta t$, the ratio between the algorithm evolution time step and the physical time step, and  set $U^n=(u_{0}^n,\cdots,u_{M}^n)^T$ and $V^n=(v_{0}^n,\cdots,v_{M}^n)^T$, such that $(U^n,V^n)^T=(u_{0}^n,\cdots,u_{M}^n,v_{0}^n,\cdots,v_{M}^n)^T$.
The scheme we prescribe in Algorithm 1 thus writes in a block-matrix form, where each block $K$, $L$, $\{G_i\}_{1\leq i\leq 4}$ is a $(M+1)\times(M+1)$ matrix, and where $\mathbf{1}$ is the identity matrix:
we solve $(U^{n+1},V^{n+1})^T$ such that
\begin{align}
    \begin{pmatrix}
        K & 0\\
        0 & \mathbf{1}
    \end{pmatrix}
    \begin{pmatrix}
        U^{n+1}\\
        V^{n}
    \end{pmatrix}
    &=\begin{pmatrix}
        G_1 & G_2\\
         0 & \mathbf 1
    \end{pmatrix}     
    \begin{pmatrix}
        U^{n}\\
        V^{n}
    \end{pmatrix}
    +R_1,\\
    \begin{pmatrix}
        \mathbf 1 & 0\\
        0 & L
    \end{pmatrix}
    \begin{pmatrix}
        U^{n+1}\\
        V^{n+1}
    \end{pmatrix}
    &=\begin{pmatrix}
        \mathbf 1 &  0\\
        G_3 & G_4
    \end{pmatrix}     
    \begin{pmatrix}
        U^{n+1}\\
        V^{n}
    \end{pmatrix}
    +R_2,
\end{align}
with
\begin{align}
    K &=     \begin{pmatrix}
    1+r    & -r  & 0  & \dots& 0\\
    0      & 1+r & -r &\ddots  &  \vdots\\
    \vdots & \ddots & \ddots &\ddots  &  0\\
    \vdots& & 0&1+r & -r\\
    0& & &0 & 1
    \end{pmatrix};\\
     L  &=     \begin{pmatrix}
    1    & 0  & 0  & \dots& 0\\
    -r      & 1+r & 0 &\ddots  &  0\\
    0 & \ddots & \ddots &\ddots  & \vdots\\
    \vdots& & -r&1+r & 0\\
    0& & 0&-r & 1+r
    \end{pmatrix}
    \end{align}
and, setting $z\equiv \dtau(\zeta+\frac{1}{\alpha})$, and $\bar z\equiv \dtau(-\zeta+\frac{1}{\alpha})$,
    \begin{align}
    G_1 &= \begin{pmatrix}
    1    & -\frac{\dtau}{\alpha}   & 0  & \dots& 0\\
    0      &  1 & -\frac{\dtau}{\alpha}  &\ddots  &  \vdots\\
    \vdots & \ddots & \ddots &\ddots  &  0\\
    \vdots& & 0&1 & -\frac{\dtau}{\alpha} \\
    0& & 0&0 & 0
    \end{pmatrix};
    \end{align}
    \begin{align}
     G_2 &=\begin{pmatrix}
    0   & z   & 0  & \dots& 0\\
    0      & 0 & z  &\ddots  &  \vdots\\
    \vdots & \ddots & \ddots &\ddots  &  0\\
    \vdots& & 0&0 & z \\
    0& & 0&0 &  -1
    \end{pmatrix};
    \end{align}
    \begin{align}
    G_3 &=     \begin{pmatrix}
    -1    &  &   & & \\
     \bar z &  0 &  & \mathbf{0}  &  \\
    0 & \ddots & \ddots &  & \\
    \vdots& &  \bar z& 0 & \\
    0& & 0&  \bar z & 0
    \end{pmatrix}\\
    G_4 &=     \begin{pmatrix}
    0  &  &   & & \\
    -\frac{\dtau}{\alpha}  & 1 &  & \mathbf{0}  &  \\
    0 & \ddots & \ddots &  & \\
    \vdots& & -\frac{\dtau}{\alpha} & 1 & \\
    0& & 0&-\frac{\dtau}{\alpha} & 1
    \end{pmatrix},
\end{align}
and with $R_1=(\mathbf{0}_M,2x_{b},\mathbf{0}_{M+1})^T$ and $R_2=(\mathbf{0}_{M+1}, 2x_{a},\mathbf{0}_M)^T$, and where $\mathbf 0_M$ indicates a list with $M$ zeros.

The scheme is stable if the eigenvalues of the matrix $Q$
\begin{align}
    Q\equiv 
    \begin{pmatrix}
        \mathbf 1 & 0\\
        0 & L^{-1}
    \end{pmatrix}
    \begin{pmatrix}
        \mathbf 1 &  0\\
        G_3 & G_4
    \end{pmatrix}
    \begin{pmatrix}
        K^{-1} & 0\\
        0 & \mathbf{1}
    \end{pmatrix}
    \begin{pmatrix}
        G_1 & G_2\\
         0 & \mathbf 1
    \end{pmatrix}
\end{align}
are of module $\leq1$. To check the stability for different values of  $\dtau$ and $\Delta t$, we prescribe a value of $\zeta$, which defines a typical relaxation time scale $t_\zeta=1/\zeta$ of the noiseless dynamics.
We then either prescribe the final time $T$ and vary the number of points $M$ along the path, or we fix the number of points and vary $\Delta t$, which changes the final time $T$. In particular, it is important to  check the stability for a final time $T\gg t_\zeta$, such that a transition from $x_a$ or $x_b$ to the critical point $x=0$ can follow the deterministic flow on most of the path (such transitions only need noise close to the critical point that must be reached exactly at time $t=T$). To correctly resolve the instanton, we should then typically take $\Delta t < t_\zeta $.

\begin{figure}
\begin{tikzpicture}
    \path (0,0) node {    \includegraphics[width=0.49\columnwidth]{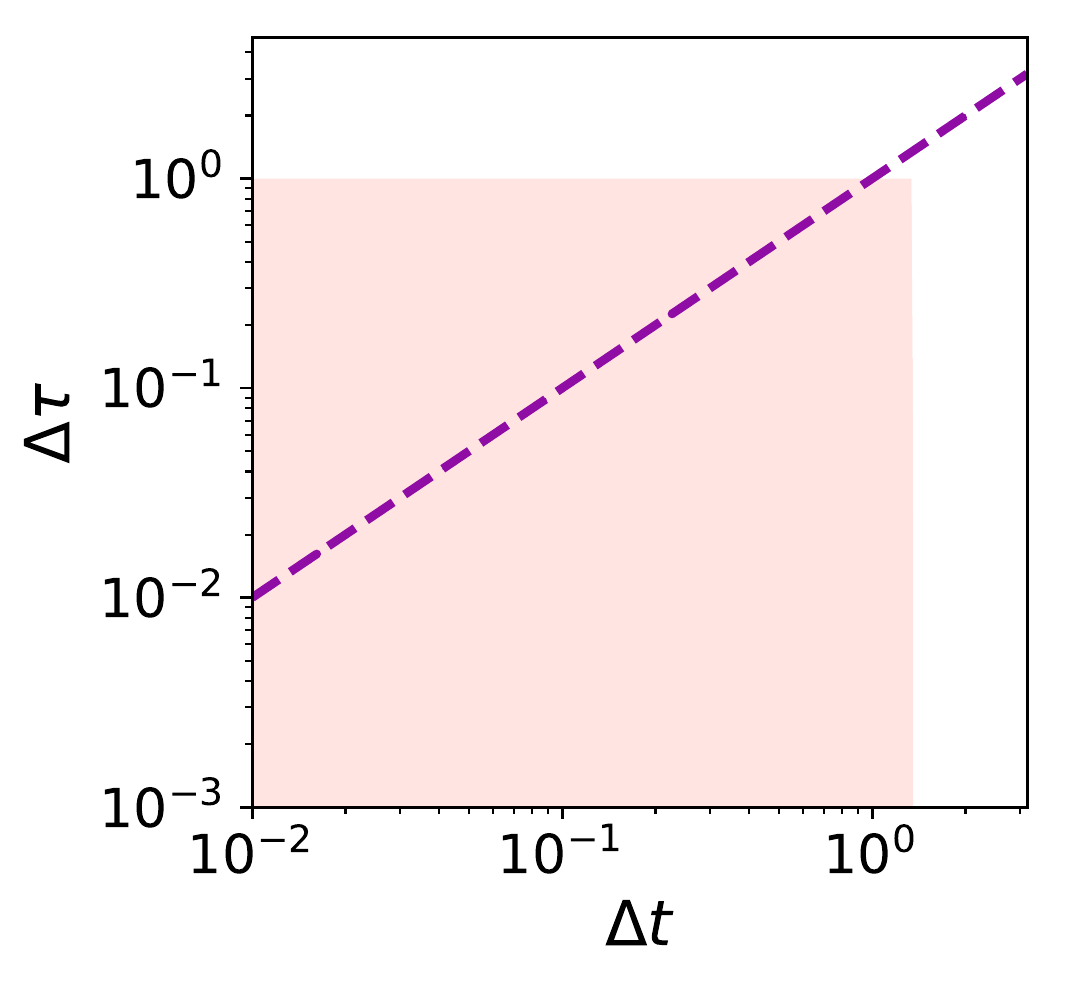}
    };
    \draw (-1.8,1.5) node[anchor=south west] {\bf a)};
    \draw (-1,1.3) node[anchor=south west] {$\zeta=1$, $\alpha=1$};
  \end{tikzpicture}
    \hspace{-13pt}
  \begin{tikzpicture}
    \path (0,0) node {    \includegraphics[width=0.49\columnwidth]{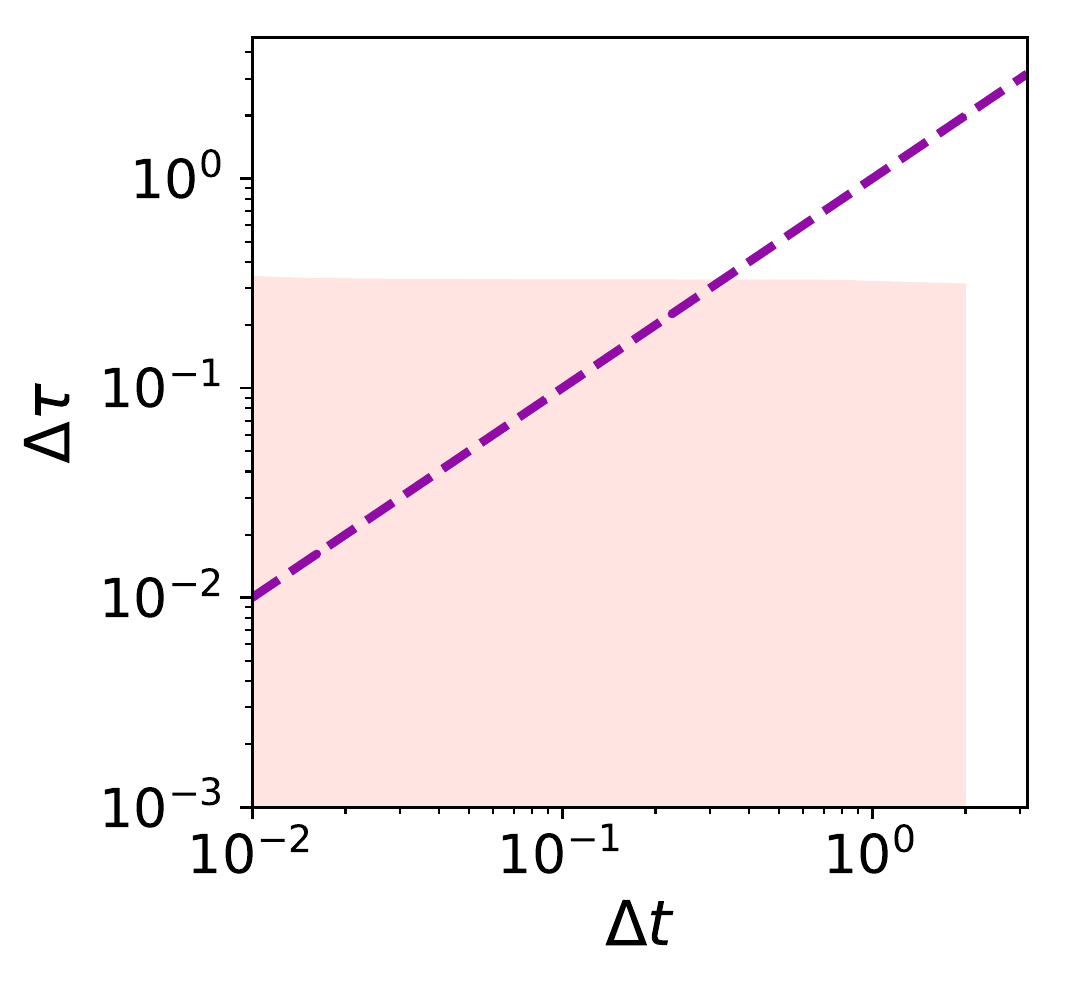}
    };
    \draw (-1.8,1.5) node[anchor=south west] {\bf b)};
    \draw (-1,1.3) node[anchor=south west] {$\zeta=5$, $\alpha=1$};
  \end{tikzpicture}
  \begin{tikzpicture}
    \path (0,0) node {    \includegraphics[width=0.49\columnwidth]{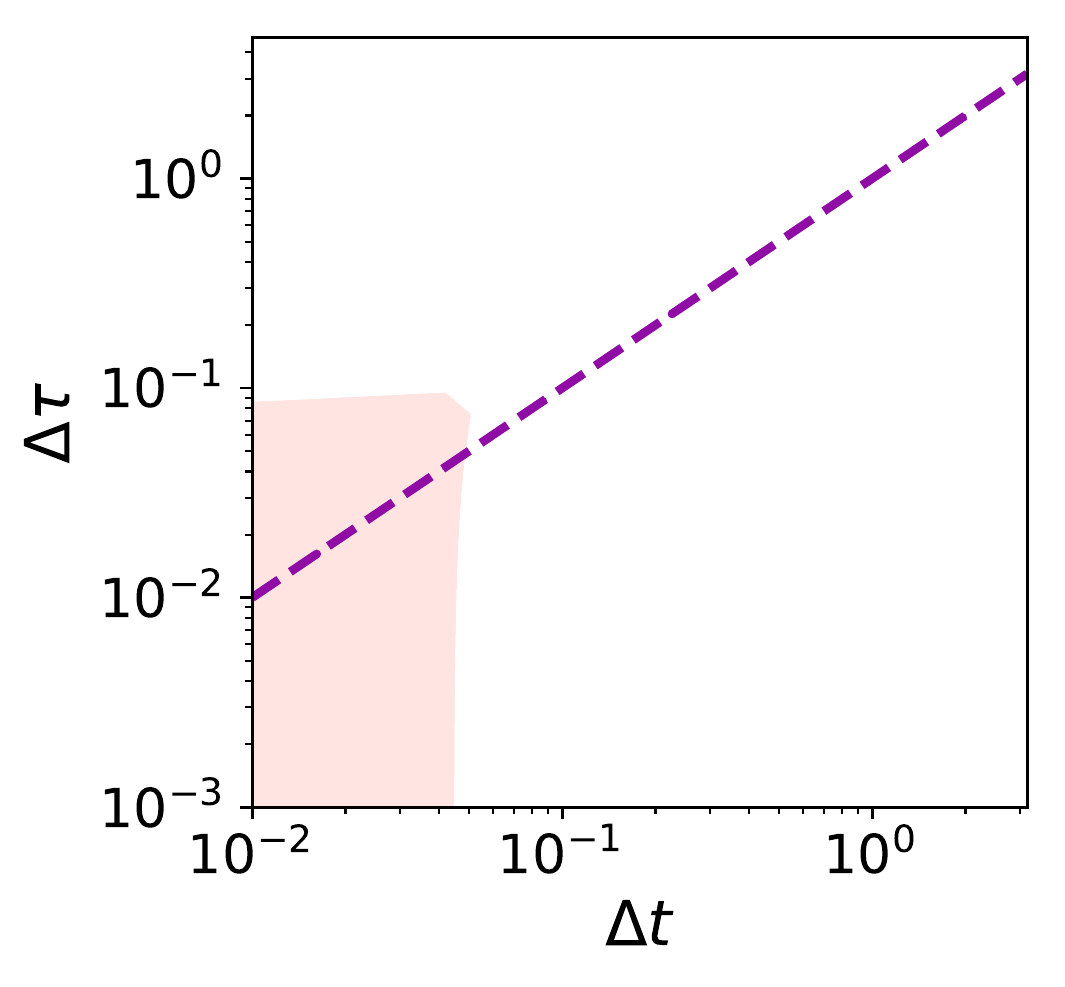}
    };
    \draw (-1.8,1.5) node[anchor=south west] {\bf c)};
    \draw (-1,1.3) node[anchor=south west] {$\zeta=5$, $\alpha=0.05$};
  \end{tikzpicture}
    \hspace{-13pt}
  \begin{tikzpicture}
    \path (0,0) node {    \includegraphics[width=0.49\columnwidth]{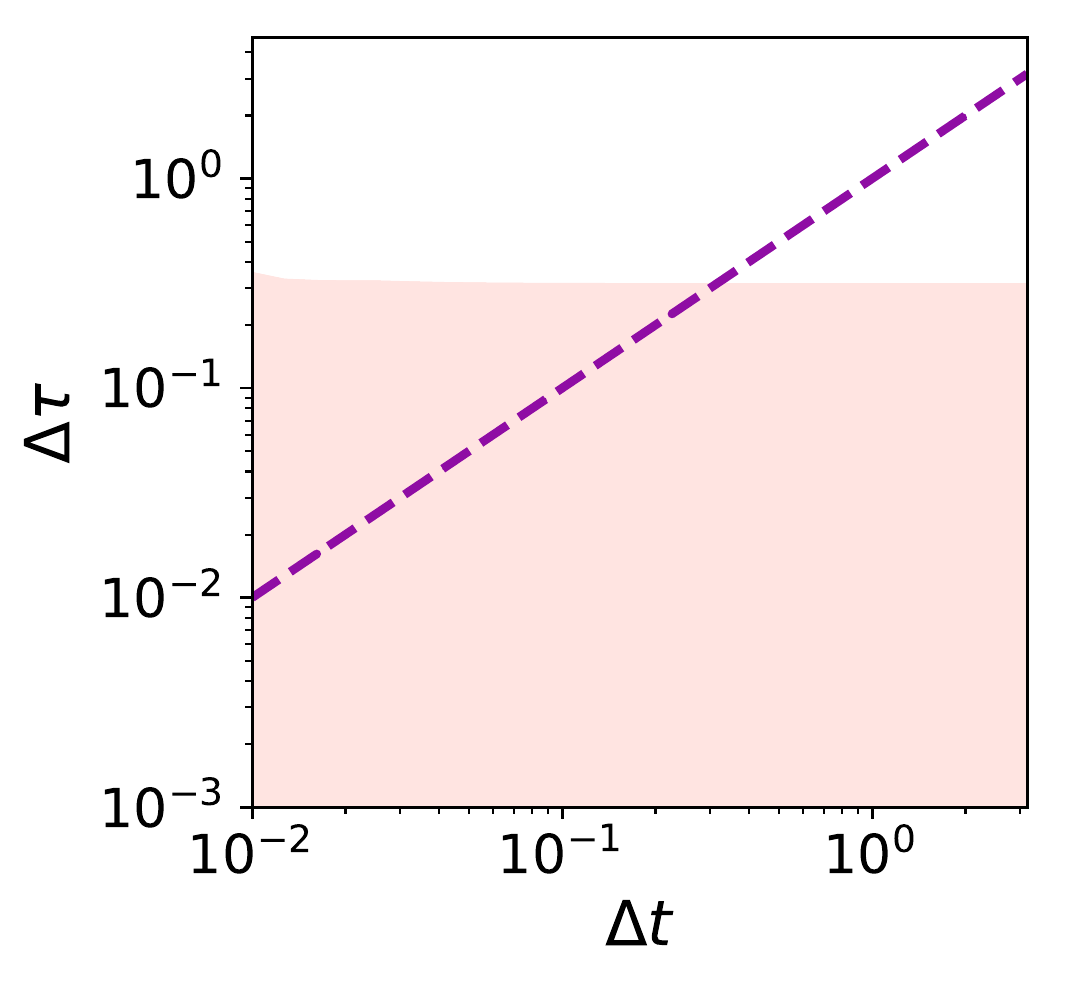}
    };
    \draw (-1.8,1.5) node[anchor=south west] {\bf d)};
    \draw (-1,1.3) node[anchor=south west] {$\zeta=5$, $\alpha=20$};
  \end{tikzpicture}
  \begin{tikzpicture}
    \path (0,0) node {\includegraphics[width=0.49\columnwidth]{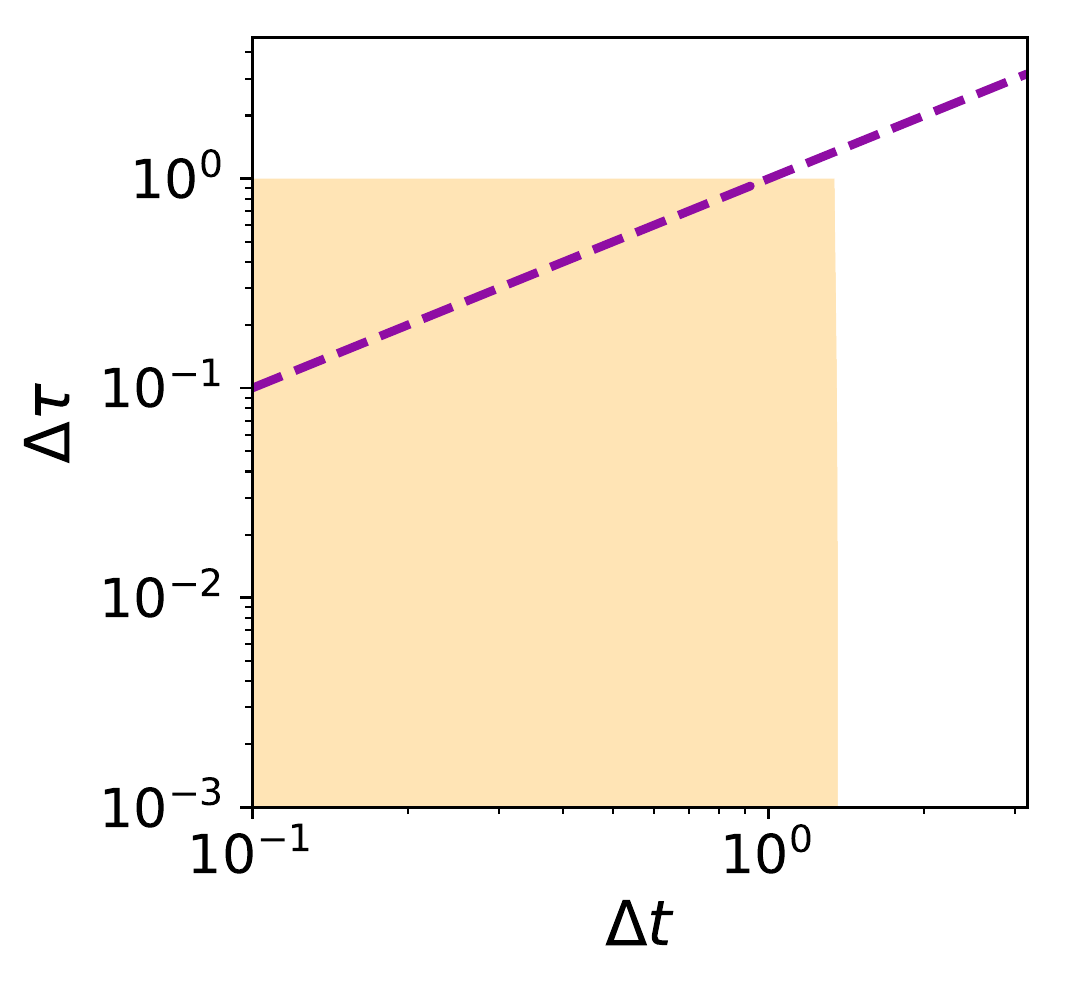}};
    \draw (-1.8,1.5) node[anchor=south west] {\bf e)};
    \draw (-1,1.3) node[anchor=south west] {$\zeta=1$, $\alpha=1$};
  \end{tikzpicture}
  \hspace{-13pt}
  \begin{tikzpicture}
    \path (0,0) node {\includegraphics[width=0.49\columnwidth]{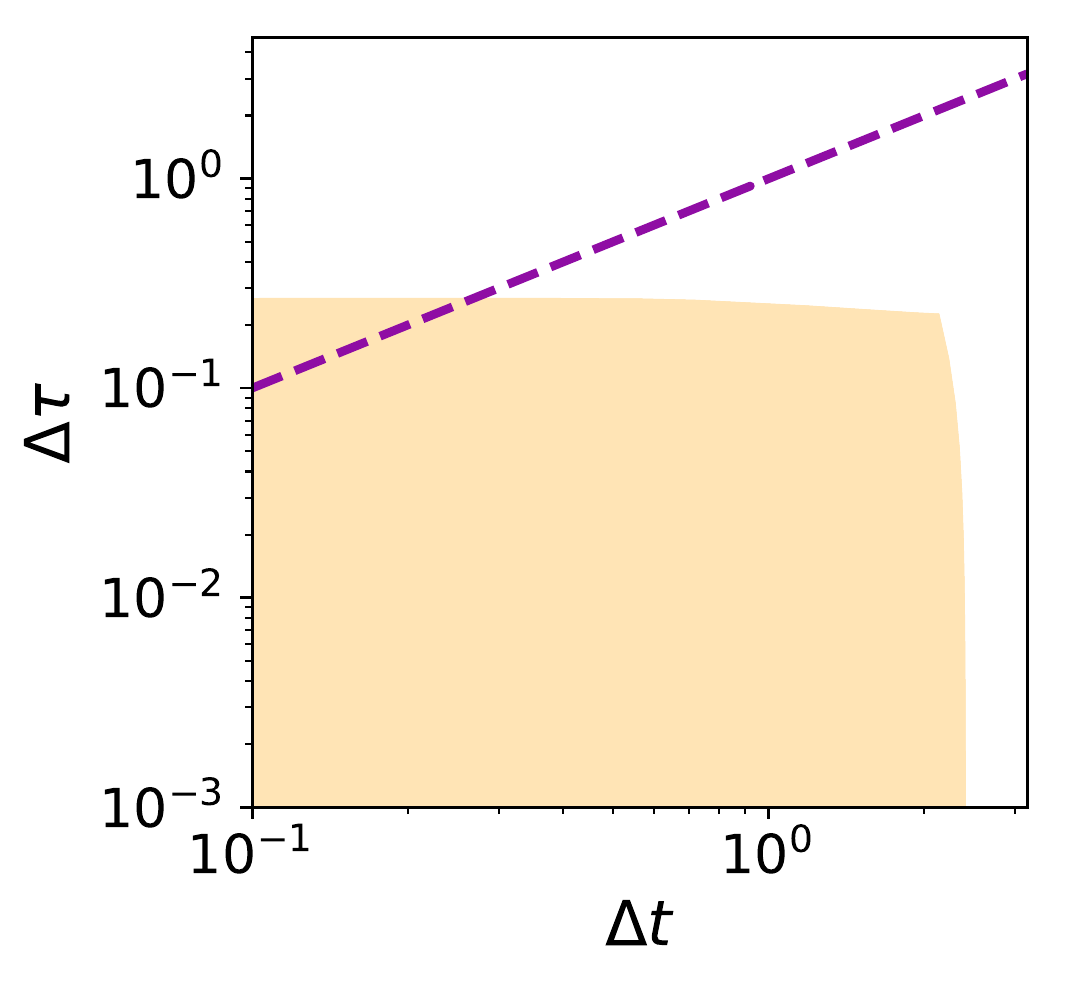}};
    \draw (-1.8,1.5) node[anchor=south west] {\bf f)};
    \draw (-1,1.3) node[anchor=south west] {$\zeta=5$, $\alpha=1$};
  \end{tikzpicture}
    \caption{Stability of the numerical scheme represented as the shaded zone in the space $(\Delta t, \dtau)$. The dashed line materializes the usual CFL conditions for explicit advection scheme, $\dtau = \Delta t$, which is relaxed here since the scheme is semi-implicit.
    For (a), (b), (c) and (d) we fix the number of discretization points $M=100$ and $T$ varies according to $T=M\Delta t$. For (e) and (f) we fix $T=10 t_k$, and $M$ is then given by $M=T/\Delta t$.}
    \label{fig:app_contour_plot_stability}
\end{figure}

In Fig.~\ref{fig:app_contour_plot_stability} we display the stability region of the scheme in space $(\Delta t,\Delta \tau)$ for some values of the parameters. The result shows that the stability region depends on $\zeta$ and $\alpha$. Since the fixed point (the instanton) is independent of the dynamical parameters $\Delta \tau$ and $\alpha$, we should take the best values that bring stability and convergence. Interestingly sending $\alpha\to0$ is not the best choice for stability, even if it looks appealing at first sight since it corresponds to solving the Legendre-Fenchel transform $\argmax_\theta(\langle\dot x,\theta\rangle -H(x,\theta))$ for a given $x$. It is instead preferable to take $\alpha=O(1)$ to update $x$ and $\theta$ on similar time scales.
It is also worth noticing that the use of an implicit scheme for advection has relaxed the Courant-Friedrichs-Lewy (CFL) stability condition, i.e. one can take $\Delta \tau>\Delta t $ and still have a stable scheme.
Unfortunately, it is clear from the graphs that the limit $\Delta \tau\to\infty$ is not stable. This limit is however interesting since it corresponds to an infinitely fast update of $u^{n+1}$ at $v^n$ fixed, i.e. that $u^{n+1}$ solves in the continuous limit
\begin{align}
    0 = \partial_t u^{n+1} +f(u^n,v^n)
\end{align}
at $v^n$ fixed, and, following, $v^{n+1}$ solves
\begin{align}
    0 = -\partial_t v^{n+1} +g(u^{n+1},v^{n+1})
\end{align}
at $u^{n+1}$ fixed. This greedy procedure would bring the convergence to the fixed point in a few steps, when it converged. Here instead, $\Delta \tau$ must remain finite to guarantee convergence. 

Finally, let us discuss the issue of preconditioning the scheme for large values of $\zeta$, which is relevant e.g. if $\zeta$ refers to the different relaxation rates of the different Fourier modes of a diffusive field (we typically have $\zeta(k)=Dk^2$ with $D$ the diffusion coefficient of the field). Specifically, we would like to choose time step $\Delta \tau$ for which the numerical scheme remains stable when evolving separately each Fourier component in a semi-spectral method for solving PDEs. 
Preconditioning the evolution \eqref{eq:app_OU_GDA} aims at keeping the same $\Delta \tau$ for each mode, and allowing higher modes (large $\zeta(k)$) to relax slower than lower modes (small $\zeta(k)$). This procedure does not change the final path, which still solves Hamilton's equations.
Such preconditioning for the GDA reads
\begin{align}
\begin{cases}
    \partial_\tau u =(1+\frac{1}{\alpha}+\zeta)^{-1}[\partial_t u + f(u,v)],\\
    \partial_\tau v = (1+\frac{1}{\alpha}+\zeta)^{-1}[-\partial_t v +g(u,v)].
    \label{eq:app_OU_system_fourier_conditioned}
\end{cases}
\end{align}
The shape of the matrices $K$, $L$ and $\{G_i\}_{1\leq i\leq 4}$ is unchanged but $\Delta \tau$ should be modified following the substitution rule: 
\begin{align}
    \Delta \tau \to \Delta \tau (1+\frac{1}{\alpha}+\zeta)^{-1},
\end{align}
which also modifies $r$, $z$, and $\bar z$.
We find that this procedure ensures the stability of the scheme up to $\Delta \tau=1$, for all $\zeta$, assuming $\Delta t$ is small enough. 

The semi-analytical proof presented here in the Ornstein-Uhlenbeck setup provides insights on the behavior of the system, but the problems we are usually interested in display strong non-linearities, and their stability cannot be analyzed through the spectrum of a linear operator. We keep in mind however that a few ingredients should be reused and that they stabilize the code in general: the advection should be treated with an implicit upwind scheme, the reaction terms can be evaluated upwind, and preconditioning the dynamics allows us to take the same time step $\Delta \tau $ for each Fourier mode in a semi-spectral scheme.

\section{Higher-order scheme}
\label{app:higher_order_scheme}

 In Algorithms~1 and 2 that we have presented in the text, the derivative of $u$ and $v$ with respect to physical time $t$ (or w.r.t parametrization $s$) is approximated with a first-order finite difference upwind derivative. This finite difference scheme is straightforward to implement but it is only first-order accurate, i.e. the error with respect to the analytical solution decreases as $O(M^{-1})$ when $M$ increases, $M$ being the number of points used to discretize the interval $[0,T]$ (or interval $[0,1]$ if working with reparametrized time $s$). Yet, it is possible to implement a higher-order stable finite difference scheme for the advection, while keeping large steps $\Delta \tau$. This can be done by keeping the implicit and upwind features of the scheme, while approximating the derivatives with the second-order difference stencil. The second-order stencil involves two upwind grid points, which means that it can only be used for points $i=2,...,M$ in forward advection for $v_i$, and for $i=M-2,...,0$ for backward advection for $u_i$. The values $v_1$ and $u_{M-1}$ will still be computed with a first-order finite-difference stencil. We have checked that the higher-order scheme indeed significantly improves the accuracy.

The higher-order implementation of Algorithms 1 and 2 are detailed in Algorithms 3 and 4, respectively. The algorithmic complexity of the higher-order schemes is the same as the algorithmic complexity of lower-order ones. Indeed, in both cases the implicit fields $u^{n+1}$ and $v^{n+1}$ simply solve a triangular system.

\begin{figure*}
\begin{minipage}{\linewidth}
\begin{algorithm}[H]
\label{alg:1_2order}
\caption{: Action Minimization by Higher-Order Gradient Descent Ascent}
\begin{algorithmic}[1]
\State Follow Steps 1 and 2 of Algorithm~1.
\For{$n\ge0$}
\State 
Update $u$ with  an implicit upwind scheme, namely,  solve $ \{u^{n+1}_i\}_{i\in I}$ sequentially from $i=M$ to $i=0$ using: 
\begin{align*}
\begin{cases}
    u^{n+1}_M=- v_M^{n}+2\phi_b\\[4pt]
    \dfrac{u_{M-1}^{n+1}-u_{M-1}^{n}}{\Delta \tau} = \dfrac{u_{M}^{n+1}-u_{M-1}^{n+1}}{\Delta t} +f(u^n_{M},v^{n}_{M}),\\[4pt]
    \dfrac{u_i^{n+1}-u_i^{n}}{\Delta \tau} = \dfrac{-u_{i+2}^{n+1}+4u_{i+1}^{n+1}-3u_i^{n+1}}{2\Delta t} +f(u^n_{i+1},v^{n}_{i+1}), \qquad i=M-2,\dots,0
   \end{cases}
\end{align*}
\State Update $v$ with an implicit upwind scheme, namely, solve $\{v_i^{n+1}\}_{i\in I}$ sequentially from $i=0$ to $i=M$ using: 
 \begin{align*}
      \begin{cases}
        v^{n+1}_0=-u_0^{n+1}+2\phi_a\\[4pt]
        \dfrac{  v_1^{n+1}-v_1^n}{\Delta \tau} = -\dfrac{ v_{1}^{n+1}- v_{0}^{n+1}}{\Delta t} +g(u_{0}^{n+1},v_{0}^{n})\\[4pt]
        \dfrac{  v_i^{n+1}-v_i^n}{\Delta \tau} = -\dfrac{ 3v_{i}^{n+1} - 4v_{i-1}^{n+1} +v_{i-2}^{n+1} }{2\Delta t} +g(u_{i-1}^{n+1},v_{i-1}^{n}),\qquad i=2, \dots, M
       \end{cases}
   \end{align*}
\State Compute $\{\phi_i^{n+1} =\frac12(u_i^{n+1}+v_i^{n+1})\}_{i\in I}$ and $\{\theta_i^{n+1}=\frac12\alpha^{-1} (u_i^{n+1}-v_i^{n+1})\}_{i\in I}$ (if needed).
\EndFor
\end{algorithmic}
\end{algorithm}
\end{minipage}
\end{figure*}

\begin{figure*}
\begin{minipage}{\linewidth}
\begin{algorithm}[H]
\label{alg:2_2order}
\caption{: Geometric Action Minimization by Higher-Order Gradient Descent Ascent}
\begin{algorithmic}[1]
\State Follow Steps 1 and 2 of Algorithm 2. 
\For{$n\ge0$}
   \State Update $u$ with  an implicit upwind scheme, namely,  solve $ \{u^{n+1}_i\}_{i\in I}$ sequentially from $i=M$ to $i=0$ using: 
   \begin{align*}
    \begin{cases}
    u^{n+1}_M=- v_M^{n}+2\phi_b\\[4pt]
    \dfrac{u_{M-1}^{n+1}-u_{M-1}^{n}}{\Delta \tau} = \lambda_{M-1}(u^n,v^{n}) \dfrac{u_{M}^{n+1}-u_{M-1}^{n+1}}{\Delta s} +f(u_{M-1}^n, v_{M-1}^{n}),\\[4pt]
    \dfrac{u_i^{n+1}-u_i^{n}}{\Delta \tau} = \lambda_i(u^n,v^{n}) \dfrac{-u_{i+2}^{n+1}+4u_{i+1}^{n+1}-3u_i^{n+1}}{2\Delta s} +f(u_i^n, v_i^{n}), \qquad i=M-2,\dots,0
   \end{cases}
   \end{align*}
   \State Update $v$ with  an implicit upwind scheme, namely,  solve $ \{v^{n+1}_i\}_{i\in I}$ sequentially from $i=0$ to $i=M$ using: 
   \begin{align*}
       \begin{cases} v^{n+1}_0=-u_0^{n+1}+2\phi_a\\[4pt]
       \dfrac{v_1^{n+1}-v_1^n}{\Delta \tau} = -\lambda_1(u^{n+1},v^n) \dfrac{ v_{1}^{n+1}- v_{0}^{n+1}}{\Delta s} +g(u_1^{n+1},v_1^n),\\[4pt]
     \dfrac{v_i^{n+1}-v_i^n}{\Delta \tau} = -\lambda_i(u^{n+1},v^n) \dfrac{ 3v_{i}^{n+1} - 4v_{i-1}^{n+1} +v_{i-2}^{n+1} }{2\Delta s} +g(u_i^{n+1},v_i^n), \qquad i=2,\ldots,M\\
       \end{cases}
   \end{align*}
    \State Follow Steps 5, 6 and 7 of Algorithm 2.
  \EndFor
\end{algorithmic}
\end{algorithm}
\end{minipage}
\end{figure*}

\section{Details on the calculation of $\lambda$ in the geometric formulation}
\label{app:geometric_GDA}

We detail here the steps that led us to consider the geometric formulation of the gradient descent ascent (GDA) scheme with a specific treatment to $\lambda$.
We start from Eq.~\eqref{eq:min:max:2}, and we notice that the minimization of $\lambda$ can be performed before the minimization over $ \hphi$. This reads
\begin{equation}
\label{eq:action:lamb1}
    \min_{\hphi}\min_{\lambda\ge0}\max_{ \htheta} \int_0^1 \left(\langle\hphi',\hat \theta\rangle -\lambda^{-1}H(\hphi,\hat \theta)\right) ds.
\end{equation}
The $\max$ on $\htheta$ can be performed pointwise in $s$ and this gives the following relation that the minimum action path should satisfy:
\begin{align}
    \lambda \hphi' = \partial_\htheta H,
    \label{eq:gMAM_xprime_append}
\end{align}
which implicitly defines $\htheta(s) \equiv\vartheta(\hphi(s),\hphi'(s), \lambda(s))$. Inserting this expression in~\eqref{eq:action:lamb1}, the minimization in $\lambda$ gives the equation
\begin{align}
    \langle \hphi',\partial_\lambda \vartheta\rangle + \lambda^{-2}H-\lambda^{-1}\langle \partial_\htheta H, \partial_\lambda \vartheta \rangle =0,
     \label{eq:gMAM_infimum_lambda}
\end{align}
which can be simplified to obtain $\lambda$ explicitly. Indeed, by taking the derivative of \eqref{eq:gMAM_xprime_append} with respect to $\lambda$, we get $\hphi'=(\partial_\htheta^2 H)\partial_\lambda\vartheta$, or $\partial_\lambda\vartheta=(\partial_\htheta^2 H)^{-1} \hphi'$, since $C^{-1}\equiv(\partial_\htheta^2 H)$ is invertible under Assumption A3 that $H$ strictly convex in $\theta$. Inserting this expression for $\partial_\lambda \vartheta$ in Eq.~\eqref{eq:gMAM_infimum_lambda}, and extracting the root, we get
\begin{align}
\label{eq:lambda_complete}
    \lambda = \frac{\langle \partial_\htheta H, C \hphi' \rangle + \sqrt{\langle \partial_\htheta H, C \hphi' \rangle^2 - 4 H \langle \hphi', C \hphi' \rangle} }{2\langle \hphi',C \hphi'\rangle}.
\end{align}
Eq.~\eqref{eq:lambda:def} follows from this equation if we modify a few terms to guarantee that $\lambda\ge0$ (which may not always be satisfied since~\eqref{eq:gMAM_xprime_append} only holds at convergence and not during the optimization).  
Note that equations \eqref{eq:gMAM_xprime_append} and \eqref{eq:lambda_complete} impose $H=0$ along the trajectory, for any definite positive $C$. This enjoins us to consider replacing $C$ by the identity in our numerical algorithms in order to avoid computing $(\partial_\htheta^2 H)$ and its inverse. Note also that the value $H=0$ is the only possible value since $H(0)=H(1)=0$ (endpoints are critical points) and that $H(s)=H(0)$ for every $s\in[0,1]$ (Hamiltonian system). In this sense, the coefficient 
$\lambda^{-1}$ can also be seen as a Lagrange multiplier enforcing $H=0$.
Finally, the $\min$ on $\hphi$ brings the second Hamilton equation
\begin{align}
    \lambda \htheta' = -\partial_\hphi H.
\end{align}

\section{Modified Ginzburg-Landau dynamics discretized on two sites}
\label{app:low_dim_modifiedGL}

\begin{figure*}
 \begin{tikzpicture}
    \path (0,0) node {    \includegraphics[width=0.68\columnwidth]{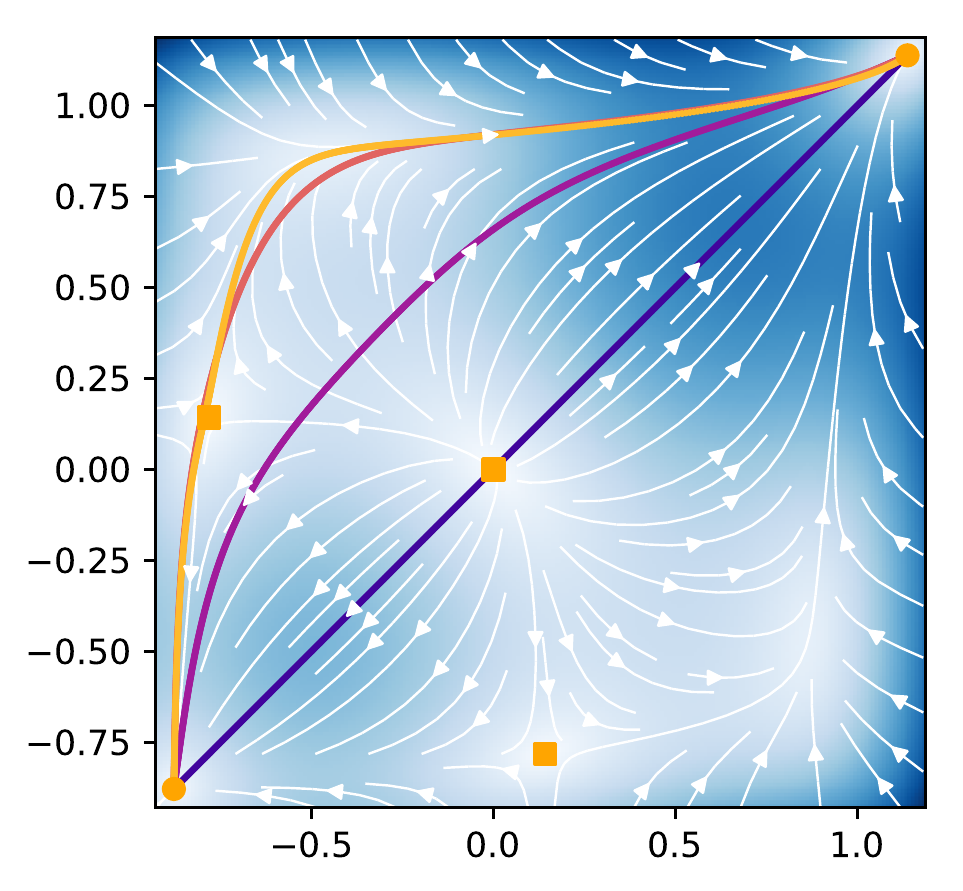}
    };
    \draw (-2.6,2.5) node[anchor=south west] {\bf a)};
  \end{tikzpicture}
  \hspace{-10pt}
  \begin{tikzpicture}
    \path (0,0) node {    \includegraphics[width=0.68\columnwidth]{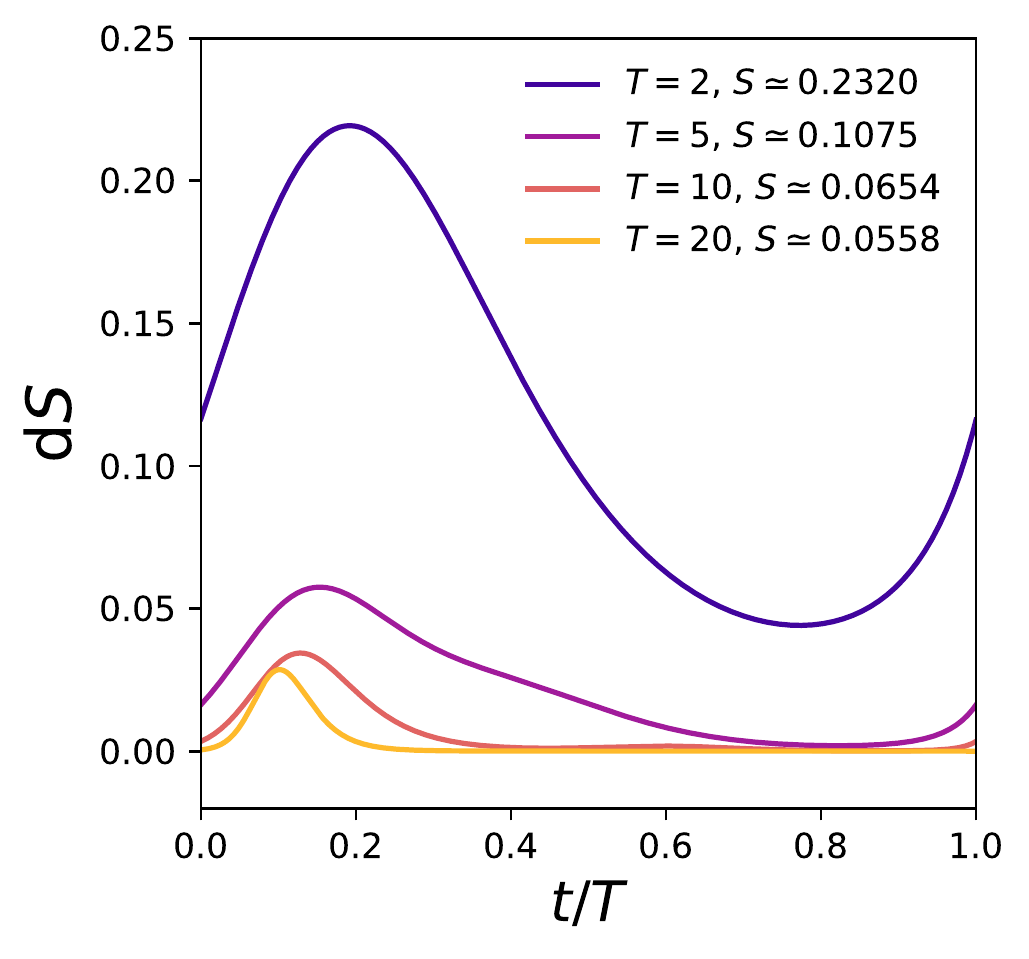}
    };
    \draw (-2.6,2.5) node[anchor=south west] {\bf b)};
  \end{tikzpicture}
  \hspace{-10pt}
  \begin{tikzpicture}
    \path (0,0) node {\includegraphics[width=0.68\columnwidth]{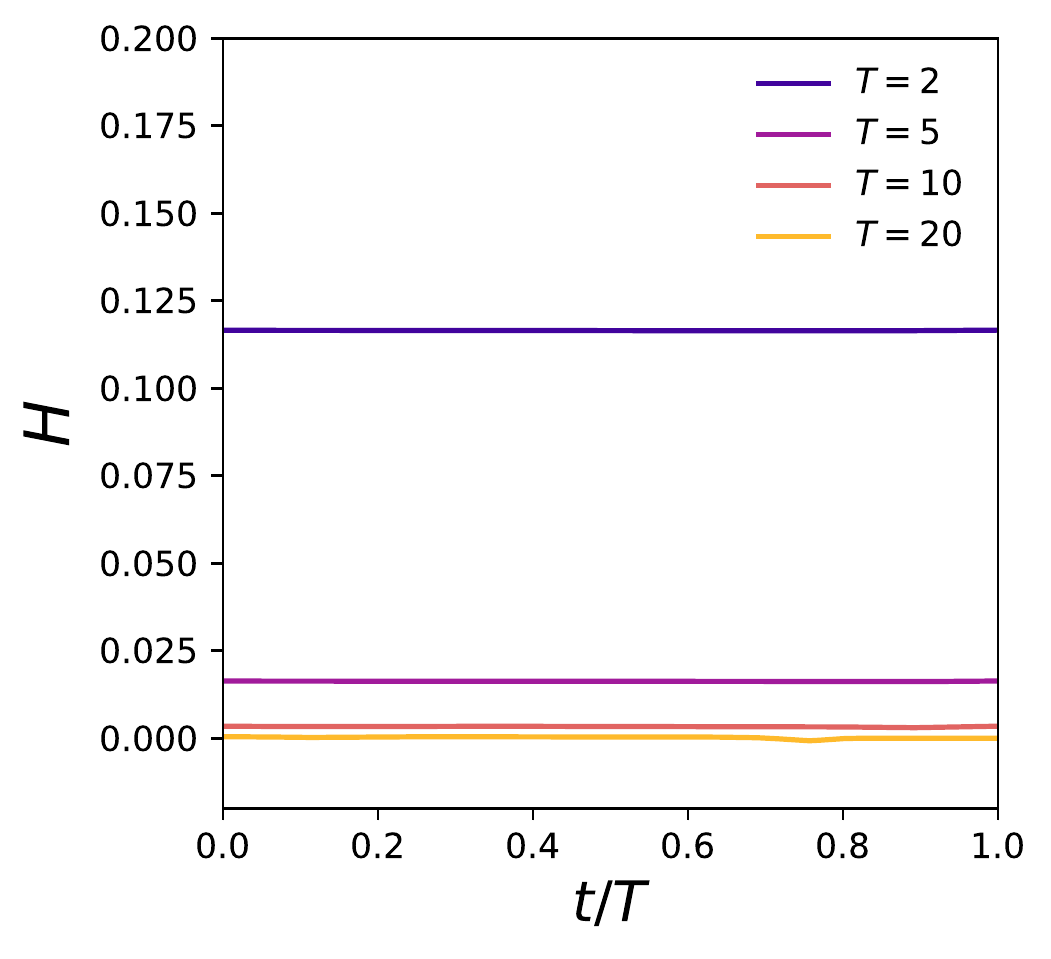}};
    \draw (-2.6,2.5) node[anchor=south west] {\bf c)};
  \end{tikzpicture}
    \caption{Comparison between paths in the low dimensional modified Ginzburg-Landau system for different values of the final time $T=2,5,10$ and $20$.  Minimum action paths (left), Lagrangian (middle) and Hamiltonian (right) along the paths.
    Orange squares (left): unstable fixed points. Orange disks: stable fixed points. Background color: intensity of the force field, darker encodes stronger force.
    As $T$ increases, the action decreases. For large $T$, the Lagrangian plateaus in the vicinity of critical points. We notice that the final Hamiltonian is constant along the trajectory. Parameters: $N=1200$, $D=0.03$, $\kappa=0.26$, $\Delta\tau=0.002$.}
    \label{fig:modifiedGL_dim2_finiteTime}
\end{figure*} 
To work in low dimensions, we discretize Eq.~\eqref{eq:spde_GL_extField} on two sites with periodic boundary conditions. Defining $x(t)$ and $y(t)$ the values of the field on site 1 and site 2, respectively, the system is now equivalent to studying a particle at position $(x,y)^T$ subjected to a nonequilibrium force.
The dynamics read
\begin{align}
    \begin{cases}
    \dot x = b_x(x,y) +\sqrt{2\epsilon}\eta_x\\
    \dot y = b_y(x,y)+ \sqrt{2\epsilon}\eta_y,
    \end{cases}
\end{align}
where, following Eq.~\eqref{eq:spde_GL_extField},  we set
\begin{align}
    b_x(x,y) = 8D(y-x)+x-x^3+\frac{\kappa}{2} (x^2+y^2),\\
    b_y(x,y) = 8D(x-y)+y-y^3+\frac{\kappa}{2} (x^2+y^2),
\end{align}
and where $\eta_x$ and $\eta_y$ are Gaussian white noise of variance unity. The term $\kappa(x^2+y^2)/2$ is the two-site analogue of $\kappa\int\rho^2$. For simplicity, we have taken $h=0$, such that the noiseless system has always two stable fixed points in $(x_1,y_1)=(q_1,q_1)$ and $(x_2,y_2)=(q_2,q_2)$, with $q_1=(-\kappa-\sqrt{\kappa^2+4})/2$ and $q_2=(-\kappa+\sqrt{\kappa^2+4})/2$, and one unstable fixed point at $(x_u,y_u)=(0,0)$. The system displays time-reversal symmetry for $\kappa=0$, and we therefore know that the dynamics follows a stochastic gradient descent on the landscape $V(x,y)=-x^2/2+x^4/4-y^2/2+y^4/4 + 4D(x-y)^2$, and that the minimum action path in this case corresponds to the minimum energy path, which simply follows the gradient ascent to escape the basin of attraction of one or the other stable fixed point.
For $\kappa\neq 0$ however, we need to resort to numerical algorithms to find the new minimum action paths between the stable fixed points. In this diffusive system, as presented in Sec.~\ref{sec:collect}, the Hamiltonian is explicitly given by
\begin{align}
    H=b_x(x,y)\theta_x + \theta_x^2 + b_y(x,y)\theta_y + \theta_y^2,
\end{align}
where $\theta_x$ and $\theta_y$ are the conjugated fields of $x$ and $y$, respectively. The solution to Hamilton's equation of motion is found via iteration of the gradient ascent-descent procedure introduced above.
\begin{figure*}
 \begin{tikzpicture}
    \path (0,0) node {    \includegraphics[width=0.68\columnwidth]{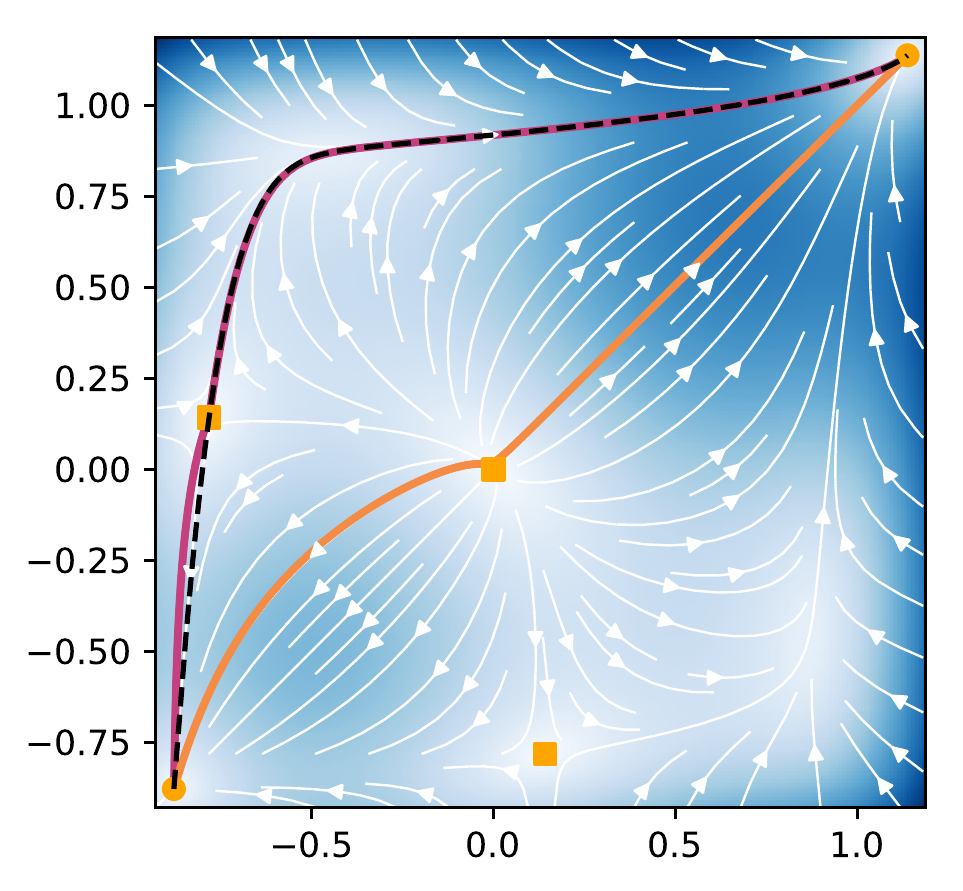}
    };
    \draw (-2.5,2.5) node[anchor=south west] {a)};
  \end{tikzpicture}
  \hspace{10pt}
  \begin{tikzpicture}
    \path (0,0) node {\includegraphics[width=1.10\columnwidth]{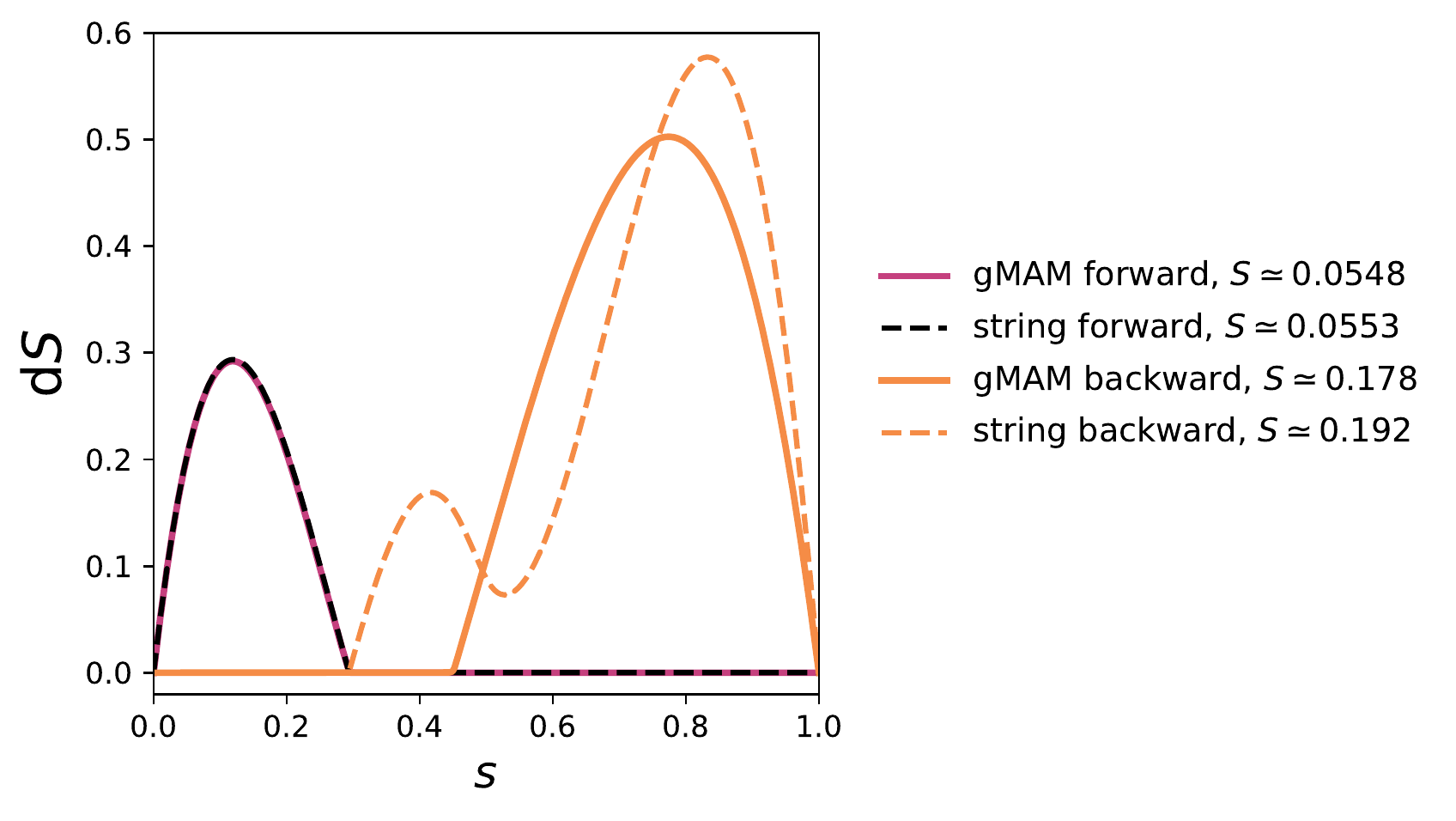}};
    \draw (-4,2.5) node[anchor=south west] {b)};
  \end{tikzpicture}
    \caption{Comparison between paths in the low dimensional modified Ginzburg-Landau system. Left panel: flow lines and paths. Right panel: action increment along the paths parametrized by normalized arclength. (a) Dashed line: heteroclinic orbit obtained with the string method. Purple line: forward minimizer obtained with the geometric algorithm. Orange line: backward minimizer obtained with the geometric algorithm.
    Squares: unstable fixed points. Disks: stable fixed points. Background color: intensity of the force field, darker encodes stronger force. Parameters: $N=600$, $D=0.03$, $\kappa=0.26$, $\Delta\tau=0.01$. }
    \label{fig:modifiedGL_dim2}
\end{figure*}
The results can be found in Fig.~\ref{fig:modifiedGL_dim2_finiteTime}.
We have checked that the scheme perfectly recovers the prediction of previous methods~\cite{weinan2004,grafke2017,kikuchi2020}. As expected, the action decreases when the final time $T$ increases, see Fig.~\ref{fig:modifiedGL_dim2_finiteTime}b). Also, the Hamiltonian $H$ is conserved along the path since the final path solves Hamilton's equation of motion. In addition we notice that $H\to 0$ when $T$ increases. 

Now, we know that the minimal action is reached when we allow $T\to \infty$, see Sec.~\ref{sec:minmax:G}. However, since the number of points along the trajectory must remain finite, increasing $T$ translates into an increase of the physical time step $\Delta t$ and decreases the resolution of the path.
To overcome this issue, we use the geometric parametrization of the path presented in Sec.~\ref{sec:minmax:G}. In Fig.~\ref{fig:modifiedGL_dim2}, we see that the geometric parametrization allows us to reach the minimum of the action. We have also checked that $H=0$ along the trajectory.

This example has illustrated the approach on low-dimensional systems subjected to an additive Gaussian noise. The method yields the optimal paths of finite duration but also the paths of infinite time length if the geometric formulation is used. Note that the numerical scheme can be adapted to determine transition paths for spatially extended fields: this is what we do in Sections~\ref{sec:modified_GL_mainSection} and~\ref{sec:schlogl}.

\section{Large $\om$ expansion and probability density}
\label{app:detail_largeDeviation_schlogl}

\begin{figure}
    \centering
    \includegraphics[width=0.99\columnwidth]{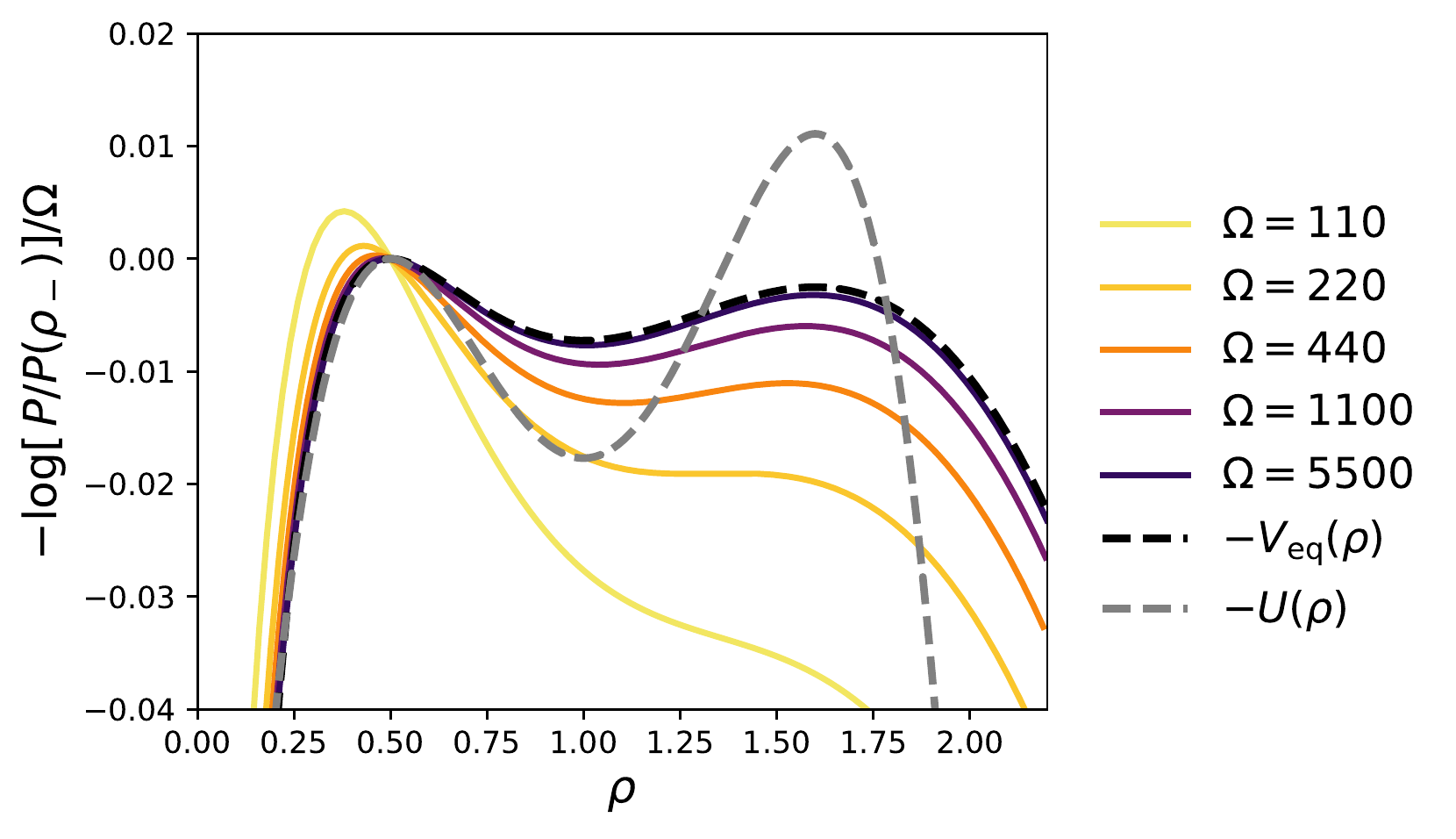}
    \caption{Rescaled and shifted logarithms of the probability density on a unique site in the Schl\"ogl model when the local number of particle $\om$ increases. The probability is explicitly known for this equilibrium case, and we show the convergence to the function $V_\mathrm{eq}(\rho)$ as the number of particles in the box increases. For the parameters chosen here,  $\rho_-$ is the stable phase, while looking at the naive Ginzburg-Landau potential $U(\rho)=-\lambda_0\rho-\lambda_2\rho^3/3+\lambda_1\rho^2/2+\lambda_3\rho^4/4$ would wrongly predict that $\rho_+$ is the stable phase. Parameters: $\lambda_0=0.8$, $\lambda_1=2.9$, $\lambda_2=3.1$, $\lambda_3=1$.}
    \label{fig:logP_continuousAction}
\end{figure}
In this section, we focus on the case of a unique well-stirred compartment of fixed volume. We work in the limit of a large number of particles, and we choose the microscopic rates $k_i$ such that the mean-field equation always displays two stable fixed points $n_-$ and $n_+$ that solve $0=k_0-k_1n+k_2n(n-1)-k_3n(n-1)(n-2)$, and where $n$ is the number of particles in the compartment. When all the particles coexist in the same compartment, the evolution of the probability $P(n)$ to find a number $n$ of particles $X$ at time $t$ is given by
\begin{align}
\begin{aligned}
    \partial_t P(n) =& W_+(n-1)P(n-1) + W_-(n+1)P(n+1)\\
    &-(W_+(n)+W_-(n))P(n),
\end{aligned}
\end{align}
where the rates are
\begin{align}
    W_+(n)&= k_0 + k_2 n(n-1)\\
    W_-(n)&=k_1 n +k_3 n(n-1)(n-2).
\end{align}
The stationary probability $P_\mathrm{eq}$ can be obtained explicitly
\begin{align}
    P_\mathrm{eq}(n)=K \prod_{i=1}^n \frac{W_+(i-1)}{W_-(i)}=K\exp\left( \sum_{i=1}^n \ln \frac{W_+(i-1)}{W_-(i)} \right)
\end{align}
where $K$ is a normalization constant. We are interested in the case of a large number of particles per compartment, such that we can extract a large deviation principle. 
We denote $\om\gg1$ the typical number of particles in the compartment, we define the rescaled number of particles $\rho=n/\om$, and we write the rates as
\begin{align}
    W_+(n)&=\om [w_+(\rho) +O(\om^{-1})]\\
    W_-(n)&=\om [w_-(\rho) +O(\om^{-1})],
\end{align}
with $w^+(\rho)=\lambda_0+\lambda_2\rho^2$ and $w^-(\rho)=\lambda_1\rho+\lambda_3\rho^3$, and where the $\lambda_i$ are now rescaled reaction rates verifying $\lambda_i=k_i\om^{i-1}$. This rescaling ensures that the deterministic mean-field dynamics 
\begin{align}
    \partial_t\rho=\lambda_0+\lambda_2\rho^2 -\lambda_1\rho-\lambda_3\rho^3
\end{align}
keeps the same fixed points $\rho_-$, $\rho_s$, $\rho_+$ when $\om\to\infty$, with $\rho_-$ and $\rho_+$ stable, while $\rho_s$ is unstable.
Following \cite{nicolis1979, dykman1994, tanase2012, grafke2019}, using the WKB (or eikonal) approximation and the continuum limit, the probability now becomes a probability density and can be cast into the following form
\begin{align}
    P_\mathrm{eq}(\rho)=K(\rho,\om)e^{-\om V_\mathrm{eq}(\rho)},
    \label{eq:large_deviation_eq_schlogl}
\end{align}
where the equilibrium potential is given by
\begin{align}
\label{eq:quasi_pot_schlogl_1d}
    V_\mathrm{eq}(\rho)=\int^\rho dy \ln \frac{w_+(y)}{w_-(y)}
\end{align}
and where $K(\rho,\om)$ is a function with the property that for any $\rho_1$, $\rho_2$, the fraction $K(\rho_1,\om)/K(\rho_2,\om)$ is bounded. As such, the ratio of the probabilities $P(\rho_1)/P(\rho_2)$ is completely determined by the difference $\Delta V\equiv V_\mathrm{eq}(\rho_1)-V_\mathrm{eq}(\rho_2)$, in the large $\om$ limit. 
We show in Fig.~\ref{fig:logP_continuousAction} that the potential $V_\mathrm{eq}$ and the naive Ginzburg-Landau potential $U(\rho)=-\lambda_0\rho-\lambda_2\rho^3/3+\lambda_1\rho^2/2+\lambda_3\rho^4/4$ display different maxima, thus different predictions for the relative stability of the states. In the large $\om$ limit, we check that the probability converges as expected to the probability density given by the function $V_\mathrm{eq}(\rho)$. We also notice that many particles per box may be needed to observe the convergence to the large deviation function.

To summarize, for a well-stirred and unique compartment, the analytical expression of the large deviation function can be obtained explicitly. This is no longer the case when spatial diffusion is taken into account and this is why one has to resort to other techniques to access the quasipotential.

\bibliographystyle{apsrev4-2}
\bibliography{biblio_phaseTransition}

\end{document}